# NANOEMULSION STABILITY ABOVE THE CRITICAL MICELLE CONCENTRATION: A CONTEST BETWEEN SOLUBILIZATION, FLOCCULATION AND KRAFFT PRECIPITATION

Kareem Rahn-Chique [1], Oriana Barrientos [2], German Urbina-Villalba [1]

[1] Instituto Venezolano de Investigaciones Científicas (IVIC), Centro de Estudios Interdisciplinarios de la Física (CEIF), Carretera Panamericana Km. 11, Aptdo. 20632, Caracas, Venezuela. Email: german.urbina@gmail.com

[2] Universidad Simón Bolívar, Departamento de Química, Valle de Sartenejas, Caracas, Venezuela.

**Abstract** The relative importance of micelle solubilization and Krafft temperature on the appraisal of the flocculation rate is studied using a dodecane-in-water nanoemulsion as a model system. In 0.5 mM solutions of sodium dodecylsulfate (SDS), neither the critical micelle concentration (CMC) nor the Krafft point of the surfactant are attained between 300 – 700 mM NaCl and $20 \leq T \leq 25$ °C. Hence, the addition of salt to a SDS-stabilized nanoemulsion only induces aggregation. Conversely, a surfactant concentration of 7.5 mM SDS promotes micelle solubilization or crystal precipitation depending on the physicochemical conditions. Solubilization decreases the absorbance of the system while flocculation and Krafft precipitation increase it. In this paper, the actual variation of the absorbance above the CMC was followed during five minutes for 300, 500 and 700 mM NaCl. The initial 60-second changes were used to determine apparent aggregation rates. The solubilization capacity of the systems was appraised using plots of absorbance vs. volume fraction of oil for different salt concentrations and observation times. The variation of the Krafft point as a function of the salt concentration was also studied using surfactant solutions and emulsions. Finally, small-scale simulations accounting for micelle solubilization were implemented in order to advance in the comprehension of the problem.

---

## 1. INTRODUCTION

Ionic surfactants are commonly used in the industry due to their ability stabilize oil-in-water emulsions by increasing the surface charge of the drops. The resulting electrostatic repulsion delays the flocculation of the particles which otherwise coagulate due to their van der Waals attraction and Brownian movement. Hence, the stability of an emulsion towards flocculation essentially relies on the adsorption equilibrium of the surfactant. This equilibrium is altered by the phase behavior of the amphiphile. In bulk solutions, surfactants may remain as monomers, aggregate into micelles, or precipitate in the form of crystals depending on temperature, the amount of salt, and its own concentration.

Aggregation can be induced screening of surface charges of the drops with electrolytes. This effect was formerly predicted by Debye-Hückel and formally incorporated in the theory of stability of lyophobic colloids by Derjaguin, Landau, Verwey and Overbeek ([Verwey, 1948]). According to the DLVO theory, the flocculation of the particles increases monotonically with the augment of the ionic strength of the solution. However, in the case of liquid/liquid dispersions, electrolytes also enhance the adsorption of ionic surfactants to the surface of the drops, and lower the threshold for micelle formation.

If the total surfactant concentration is low enough (0.5 mM SDS), the addition of salt (NaCl ≤ 1000 mM) is not sufficient to promote micellization (see Table 3 in [Urbina-Villalba, 2016]). Thus, small amounts of salt provoke an increase in the surface excess as well as the partial screening of this additional charge. As a result, a maximum electrostatic repulsion is achieved between 10 - 40 mM NaCl [Urbina-Villalba, 2013]. Higher salt concentrations screen the surface charge significantly, favoring the flocculation of the drops as predicted by DLVO. As shown by O'Brien and White [1978], and also by Ohshima [2005], a maximum of zeta potential vs. electrolyte concentration may also be observed as a result of the non-linear relation between the mobility of a particle and its zeta potential.

Sodium dodecylsulfate is commonly used as a detergent due to its ability to solubilize appreciable amounts of oil when used above its critical micelle concentration (CMC = 8.3 mM = 0.24 %wt [Ariyaprakai, 2007]). This property results from the strict hydrophobic nature of the micelle interior which conceals hydrocarbons from the aqueous solution, favoring molecular conformations typical of liquid alkanes [Lindman, 1979; Evans, 1994, Williams, 1954; Cui, 2008]. Spherical SDS micelles have an approximate radius of 2 nm [Cabane, 1985; Stigter, 1979] and are composed of 50-74 molecules [Reiss-Husson, 1964; Bales, 1998; Cabane, 1985; Israelachvilli, 2002]. As shown by Fig. 4 of Ref. [Urbina-Villalba, 2013], the CMC sensibly decreases with the addition of salt, reaching 0.98 mM SDS at 300 mM NaCl [Urbina-Villalba, 2016].

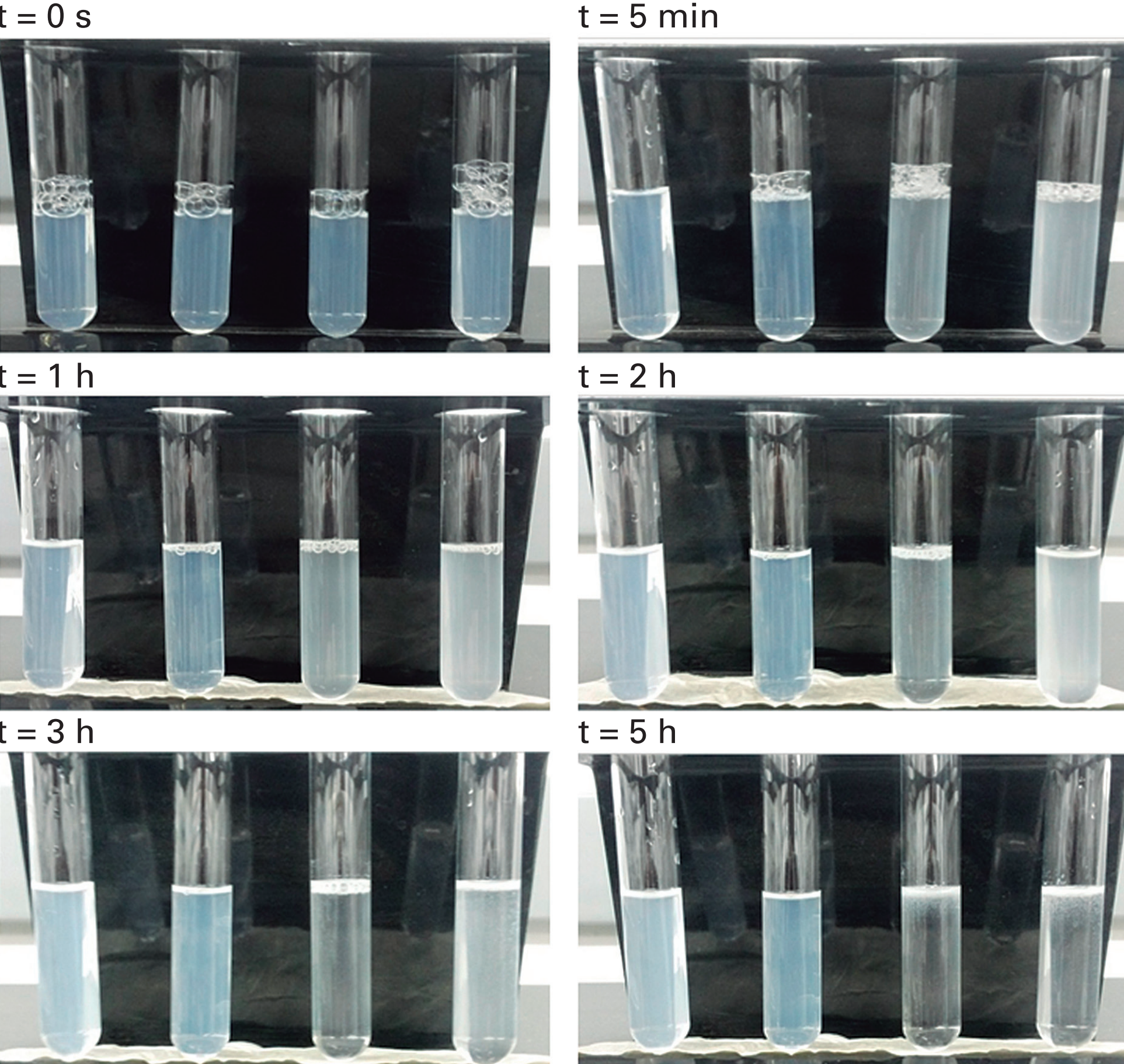

Fig. 1: Long time behavior of dodecane-in-water nanoemulsions stabilized with 0.5 mM SDS in 0, 300, 500, and 700 mM NaCl (T = 25°C).

Figure 1 illustrates the "long-time" behavior of a dodecane-in-water nanoemulsion stabilized with 0.5 mM sodium dodecyl sulfate (T = 25 °C) in 0, 300, 500 and 700 mM NaCl. Systems containing 0 or 300 mM NaCl do not appear to undergo significant changes even after 5 hours. Instead, the ones corresponding to 500 and 700 mM look opaque five minutes after the addition of salt. In 20 minutes the turbidity of the 700-mM system becomes higher than the one of 500-mM NaCl. Its maximum opacity is observed around 1 hour. After 2 hours the 500-mM system significantly clarifies and creams, evidencing the highest instability of the set. Clarification of the 700-mM dispersion occurs after 3 hours. At 5 hours, a clear solution with a substantial amount of cream is observed in both cases. Hence, the qualitative evolution of each system can be justified in terms of the DLVO theory, but the relative stability of the emulsions cannot.

Figure 2 shows the evolution of 7.5 mM SDS emulsions (T = 25 °C) in 0, 300, 500 and 700 mM NaCl. Again, the 0 mM emulsion is stable during 5 hours, but in this case, the 300 mM dispersion turns progressively clearer during the first 20 minutes, becoming completely translucent after 45 minutes. This behavior could possibly be justified by the partial solubilization of the drops into micelles of SDS. The systems corresponding to 500 and 700 mM NaCl turn opaque soon after the addition of salt (t = 0). As in the previous case, the turbidity of the 500-mM NaCl system increases first but then decreases. However, in this case, a slightly turbid solution remains after 5 hours. The gradient of particles observed at the top of the test tubes for the 0.5-SDS systems does not occur in the case of 7.5-mM SDS dispersions. The 500-mM tube becomes completely transparent after 72 hours. Instead, the turbidity of the 700-mM NaCl system increases more pronouncedly and in a few minutes small crystals begin to appear. These solids can be observed with the naked eye. The dispersion turns obscure until the size of the crystals causes their precipitation (45 min). The liquid above remains slightly unclear (not cloudy), and yet, it becomes significantly clearer than the one corresponding to 0 mM NaCl.

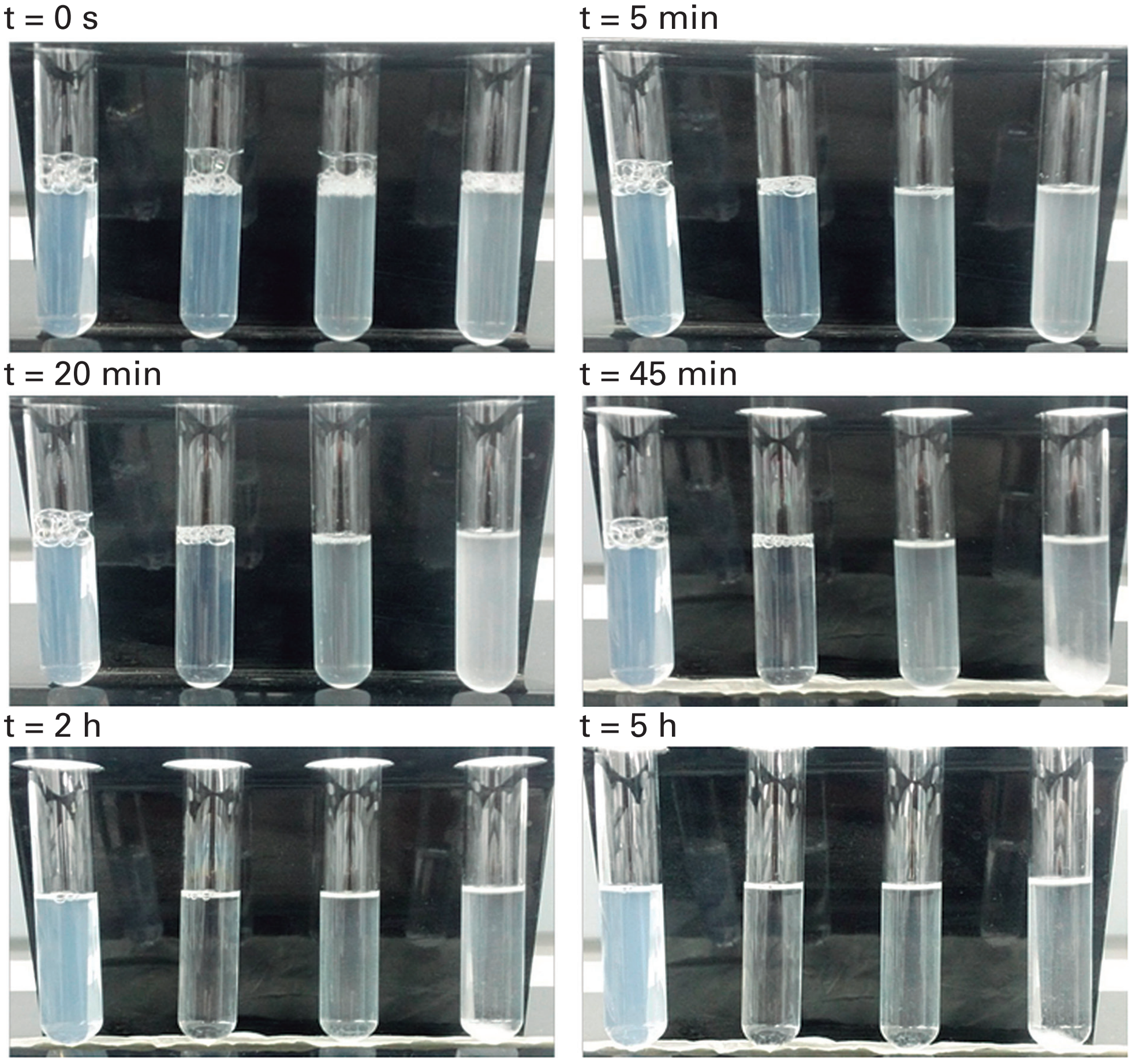

Fig. 2: Long time behavior of dodecane-in-water nanoemulsions stabilized with 7.5 mM SDS in 0, 300, 500, and 700 mM NaCl (T = 25°C).

It is evident from Figs. 1 and 2 that the role of the surfactant on the stability of alkane-in-water emulsion can be very involved. It seems impossible to predict the evolution of the average radius of a nanoemulsion in 7.5-mM SDS if the phase behavior of the surfactant is not considered. Furthermore, the higher instability of the 500-mM NaCl emulsion in 0.5 mM SDS is difficult to justify.

During the last few years our group developed a theoretical formalism suitable for the appraisal of a mixed flocculation-coalescence rate ($k_{FC}$) from turbidity measurements [Rahn-Chique, 2012a]. We also incorporated a computational code (Emulsion Stability Simulations) adequate for predicting the temporal evolution of oil-in-water dispersions (see [Urbina-Villalba, 2016; Osorio, 2009; Urbina-Villalba, 2000] and references therein). The agreement between theory and experiment in the case of hexadecane-in-water nanoemulsions stabilized with 0.5 mM SDS allowed a non-trivial interpretation of the measurements [Urbina-Villalba, 2015; Urbina-Villalba, 2016]. The monotonous increase of the flocculation rate in terms of the ionic strength of the solution is due to the reversible aggregation of drops in the secondary minimum of the interaction potential. This reversibility decreases as the depth of the minimum increases, favoring the cohesion of the aggregates. Drops do not cross the potential energy barrier existing between the secondary and the primary minimum. As a result, the concept of "stability ratio" [Fuchs, 1936; Lozsán, 2006] generally used for the quantification of the Derjaguin-Landau-Verwey-Overbeek (DLVO) theory [Verwey, 1947; Derjaguin, 1967], fails.

While the behavior of $k_{FC}$ can be reasonably reproduced for 0.5 mM SDS emulsions, the one of 7.5 mM SDS systems cannot be justified, not even if the deformability of the drops is included in the calculations besides flocculation, coalescence, creaming and Ostwald ripening [Urbina-Villalba, 2016]. Experiments show that $k_{FC}$ increases up to 500 mM NaCl, reaching a small plateau, and then slightly decreases until 1 M NaCl. According to several empirical equations (see Table 3 in [Urbina-Villalba, 2016] and Fig. 4 in [Urbina-Villalba, 2013]), a surfactant concentration of 7.5 mM SDS is substantially higher than the CMC in the range of salinities studied: $300 < \text{NaCl} < 1000$ mM. Nonetheless the simulations did not consider the possible solubilization of the drops.

Up to now, the flocculation rate of dodecane-in-water nanoemulsions has only been studied above the CMC (8 mM SDS, $380 \leq [\text{NaCl}] \leq 600$ mM, see Table 5 in Ref. [Rahn-Chique, 2012a]). As expected, some anomalies were found. The highest value of $k_{FC}$ corresponds to 420 mM NaCl and not to the maximum salinity. Moreover, the flocculation rate of the 600-mM NaCl emulsion is substantially lower than the ones observed for 500 and 550 mM NaCl. The error bars cannot justify the difference.

Experimental aggregation rates of dilute nanoemulsions are now routinely obtained in our lab from the fitting of an algebraic equation to the temporal change of the turbidity. However, the turbidity also changes with the solubilization of drops and the formation of crystals. The purpose of the paper is to ascertain the influence of micelle solubilization and Krafft precipitation on the behavior of a dodecane-in-water nanoemulsion during a period of 60 seconds. The question we wish to address is: Is it possible to establish a connection between the anomalous performance of the flocculation rate at 7.5 mM SDS and the referred phenomena.

## THEORETICAL BACKGROUND

Two major mechanisms of micelle solubilization are found in the literature. The first (surface reaction) envisages the transference of oil through a transient attachment of the micelles to the surface of the drops. The second (bulk reaction) assumes the capture of the molecules of oil already dissolved, from the aqueous phase. In the case of ionic emulsions, only the second mechanism is usually considered due to the electrostatic repulsion between the micelles and the drops.

Todorov *et al*. [Todorov, 2002] observed the dissolution of large drops of oil (radius $R > 10$ μm) in SDS using a microscope and a thermostated capillary. Plots of $R$ vs. t were found to be linear for decane ($C_{eq} = 1.11 \times 10^{-6}$ M) and non-linear for benzene ($C_{eq} = 2.28 \times 10^{-2}$ M). To interpret the data, a kinetic model of bulk solubilization was developed. It is based on the molecular diffusion of the oil from the surface of the drop to a region of width $\kappa^{-1}$, followed by micellar uptake. In the most general case, molecular dissolution and micellar solubilization occur simultaneously. The rate of solubilization ($dR/dt$) is nonlinear and depends on: a mass transfer coefficient ($\alpha$), the magnitude of $\kappa$, and a parameter $\beta$ determined by the properties of the oil. If $\kappa R \gg 1$, a linear dependence of emerges. However, if $\alpha\kappa^{-1} \gg 1$ two limiting cases may occur: For $\kappa R \gg 1$ (case a): the oil dissolution is dominated by micellar solubilization, and $|dR/dt| \propto \kappa$; For $\kappa R \ll 1$ (case b): the rate of diminishing is determined by the molecular dissolution of the oil. This induces a non linear dependence of the dissolution rate with time: $|dR/dt| \propto R(t)^{-1}$. If only molecular solubilization occurs, $\kappa \approx 0$, and $|dR/dt| \propto (t_0 - t)^{-½}$, where $t_0$ is the time required for complete dissolution of the drop. This simple equation fits the solubilization of a drop of benzene below the CMC (0.007 mM SDS) and approaches its qualitative behavior at higher surfactant concentrations.

In 0.025 - 0.25 mM SDS solutions, apparent solubilization rates of $|dR/dt| < 3$ and $> 70$ nm/s were found for decane and benzene, respectively [Todorov, 2002]. However, using the same experimental data, a reverse order is obtained for the reaction: {micelle} + {oil molecule} = {swollen micelle}: $k+ = 1.0 \times 10^{-13}$ $cm^3$/s n-decane, and $k+ = 2.0 \times 10^{-19}$ $cm^3$/s for benzene. According to the authors, benzene molecules interact more favorably with the water molecules than decane molecules. Therefore they exhibit a larger molecular solubility but a higher activation barrier for micelle solubilization. However, since $|dR/dt|$ depends on both k+ and $C_{eq}$, it turns out that the oil which exhibits a greater solubility in pure water dissolves faster, irrespective of its smaller solubilization rate. Moreover: the close parallelism between the solubilization formalism and the theory of aggregation [Smoluchowski, 1917] allowed the authors to define a diffusion-limited rate constant: $(k+)_{fast} = 4\pi D_{oil} R_m = 2.4 \times 10^{-11}$ $cm^3$/s (where $D_{oil}$ is the diffusion constant of a molecule of oil, and $R_m$ the radius of a micelle). This theoretical limit is substantially quicker than the actual rates. Hence, it was concluded that there exists an energetic barrier for the micellar uptake of even just one molecule.

According to Ariyaprakai and Dungan [2007; 2008], the solubilization of monodispersed hexadecane emulsions (0.3-0.8 μm) conforms to a first order equation:

$$\frac{dR_i}{dt} = \frac{k_{eff}}{\rho_{oil}} \left( C_{aq} - C^{eq}_{mic} \right) \tag{1}$$

where $k_{eff}$ is an effective mass transfer coefficient, $\rho_{oil}$ is the density of oil, $R_i$ the average radius of emulsion drops, and $C_{aq}$ the aqueous concentration of oil at time t. The maximum solubilization capacity of SDS ($C^{eq}_{mic}$) can be de-termined adding emulsions of different oil concentration to a fixed volume of a very concentrated (12-g) surfactant solution. The use of monodisperse emulsions leads to a simpler analysis, equivalent to assuming $C_{aq} = 0$ in Eq. (1). In this case $|dR/dt| = 9.9 \times 10^{-4}$ μm/h for hexadecane and 0.021μm/h for tetradecane. SDS solutions of 1, 2, 3 and 4 wt% can solubilize: 0.007, 0.036, 0.067 and 0.109 wt% of hexadecane after 700 hours [Ariyaprakai, 2007]. These values suggest an approximate average of **two molecules of hexadecane per SDS micelle** [Ariyaprakai, 2007]. Additionally, it was confirmed that the solubilization rates are independent of the initial average radius of the drops, and proportional to the total surfactant concentration.

The approximate time required for the solubilization of a drop depends on its composition:

$$\tau = \frac{R_0}{|dR/dt|_{t=0}} \quad (2)$$

Using the values outlined above, it takes either: 202 h, 95 h, 66 s, or 2.9 s to solubilize a 200-nm drop of: hexadecane, tetradecane, decane or benzene, respectively.

The thermodynamic stability of swollen micellar solutions is strikingly different from the kinetic stability of drops of nanometer size. This confirms the irreversibility of the solubilization process, and outlines the free-energy advantages of the swollen micelle structure as compared to those of conventional drops. The possible ability of micelles to transport oil between drops has been the subject of a long debate [Dungan, 2003; Kabalnov, 1994]. However, their capability to destabilize nanoemulsions through solubilization is much less evident. Recently, Rao and McClements [Rao, 2012] studied the solubilization of lemon oil nanoemulsions in sucrose monopalmitate and Tween 80 solutions. The turbidity of the nanoemulsions was followed as a function of time for various oil concentrations. It was found that lemon oil is transferred from nanoemulsion droplets to microemulsion micelles until a critical saturation concentration ($C_{sat}$) is reached. Beyond this point, the excess oil remains in the form of droplets. Hence, after a convenient dissolution time, plots of the turbidity of the dispersion versus the oil concentration show a negligible horizontal slope (due to complete solubilization of the drops), followed by a steep increase (due to the scattering of the remaining emulsion).

It is noteworthy that the synthesis of nanoemulsions by low-energy methods usually involves a very high surfactant concentration, between 5 to 10 times the CMC [Maestro 2008; Izquierdo, 2002]. A high concentration decreases the interfacial tension, favoring the creation of small drops during emulsification. However, in order to preserve the resulting drop size distribution, the system has to be quickly diluted due to its high instability. That instability is often associated with flocculation, but it is likely to be related to solubilization. A strong dilution is able to stop micelle solubilization by decreasing the aqueous surfactant concentration below the CMC.

Recent experiments showed that the capacity of sodium dodecylbenzene sulfonate (SDBS) to solubilize hexadecane is stronger at concentrations below the CMC than above it [Zhong, 2015]. The apparent solubility increased linearly with the total surfactant concentration, exhibiting slopes of 0.84 and 0.16, respectively. Up to our knowledge, a similar study has not been conducted for the case of SDS.

Temperature is also a key parameter in the subtle equilibrium between micellar and monomer surfactant solutions. Below the Krafft temperature, monomers act as typical solutes, maintaining equilibrium with a precipitate of hydrated surfactant crystals. At the Krafft temperature the solubility of the surfactant reaches its CMC and increases pronouncedly afterwards [Lindman, 1979; Evans, 1994]. As a result, crystals disappear when micelles are formed. In the case of SDS, the Krafft temperature changes between 20 and 25 °C, depending on the surfactant concentration and the ionic strength of the solution. These are typical working conditions in the laboratory as well as in most applications. Hence, the appearance of crystals or micelles might depend on factors as bizarre as the air conditioning system. The formation of crystals in the bulk of the external phase increases the total absorbance of the dispersion. These crystals are expected to settle at the bottom of the container clarifying the solutions at long times. It is uncertain if the precipitation of the surfactant is able to withdraw a substantial amount of molecules from the surface of the drops. This is not expected in the case of micelles which only form after the saturation of the oil/water interface.

## EXPERIMENTAL DETAILS

### 3.1 Absorbance of Aqueous Surfactant Solutions

In order to discard any inherent surfactant contribution to the absorbance of the emulsions, aliquots of brine were added to aqueous surfactant solutions of 0.5 and 7.5 mM SDS. The absorbance of the solutions with 0, 300, 500, and 700 mM NaCl was observed during 5 minutes at T = 20 and T = 25 °C. Complementary measurements at 350 and 375 mM NaCl were included in order to illustrate the competition between aggregation and solubilization (see below).

## 3.2 Absorbance of Dodecane-in-water nanoemulsions

### 3.2.1 Abs vs. $\phi$

Dodecane-in-water nanoemulsions with an average radius of 68.8 nm (61.0 -76.5 [LS230, Coulter]) 89.4 nm (82.0 – 96.0 [BI-200SM, Brookhaven]) were prepared following the procedure described in reference: [Rahn-Chique, 2012a]. The resulting stock nanoemulsion contained 0.024 M SDS, 0.02 M NaCl, 0.46 %wt isopentanol. Different aliquots of the stock emulsion were added to NaCl + SDS solutions in order to yield diluter emulsions with a fixed chemical composition (7.5 mM SDS, [NaCl] = 0, 300, 350, 375, 500, and 700 mM) but distinct volume fractions of oil $5 \times 10^{-5} \leq \phi \leq 3 \times 10^{-3}$. The absorbance (Abs) of these samples at $\lambda$ = 400 nm was measured periodically using a Shimadzu 1800 UV spectrophotometer. This wavelength does not match any absorption frequency of emulsion components. Hence, the values measured basically appraise the amount of light scattered by the suspended particles. Plots of Abs vs. $\phi$ were drawn at several observation times: 10 min, 2h, 4h, 24h, 48h, 72h, 96h, and 7 days.

Below the CMC the absorbance increases linearly as a function of the drop concentration, passing through the origin (Abs = 0 for $\phi$ = 0) [Rao, 2012]. If the surfactant concentration is higher than the CMC, the solubilization of the oil decreases the turbidity of the dispersion almost to zero until a critical volume fraction of oil: $\phi = \phi_c$ is reached. From that point on, the turbidity increases linearly due to the scattering of the remaining drops. Hence, $\phi_c$ corresponds to the amount of oil required for the saturation of the micellar solution.

If salt is added to an emulsion stabilized with SDS, the CMC of the surfactant decreases, reaching values significantly lower than 8.3 mM. In these circumstances the number of micelles increases, as well as $\phi_c$. Hence, the horizontal segment of the Abs vs. $\phi$ plot should increase as a function of the NaCl concentration. However, the addition of salt also destabilizes the emulsion promoting flocculation, coalescence and creaming. As a result the curves of Abs vs. $\phi$ are non linear, but in most cases still show an abrupt change of slope near the saturation of the micellar solution. Whenever possible, a linear fit was calculated using the data corresponding to the inclined (non-horizontal) part of the curve. The value of $\phi_c$ was taken as the intercept of the fitting line with the x axis ($\phi$(Abs = 0)). If the variation of the absorbance was totally non-monotonic, $\phi_c$ was approximated as lowest volume fraction for which a substantial increase in the absorbance of the emulsion suddenly occurred.

### 3.2.2 Abs vs. t

The evolution of the absorbance was also continuously monitored during 5 minutes after the addition of salt (0, 300, 350, 375, 500, and 700 mM NaCl, T = 20, 25 °C, [SDS] = 0.5, 7.5 mM , $\phi = 3.2 \times 10^{-4}$). These measurements allow a direct comparison between the absorbance of emulsions and surfactant solutions (section 3.1).

### 3.2.3 Abs vs. t. Emulsions with Squalene

In order to appraise the influence of Ostwald ripening in the behavior of the absorbance of emulsions at 300 and 375 mM NaCl (T = 25 °C, [SDS] = 7.5 mM) an additional set of emulsions was prepared using a combination of dodecane (90%wt) and squalene (10%wt) as the oil. Squalene lowers the vapor pressure of the drops decreasing the Oswald ripening rate to a minimum [Mendoza, 2013].

## 3.3 Measurements of $k_{FC}$

In a previous work our group developed a theoretical expression for the temporal variation of the turbidity ($\tau$) as a result of flocculation and coalescence [Rahn-Chique, 2012a]:

$$\tau = n_1 \sigma_1 + x_a \sum_{k=2}^{k_{max}} n_k \sigma_{k,a} + (1 - x_a) \sum_{k=2}^{k_{max}} n_k \sigma_{k,s} \tag{3}$$

Here: $\sigma_1$, $\sigma_{k,a}$ and $\sigma_{k,s}$ represent the optical cross sections of primary drops, aggregates of size $k$ , and spherical drops of size $k$ ($R_k = \sqrt[3]{k}\, R_0$). According to the theory of Smoluchowski for irreversible flocculation [Smoluchowski,

1917], the number of aggregates per unit volume composed of $k$ primary particles is equal to:

$$n_k = \frac{n_0 (k_{FC} n_0 t)^{k-1}}{(1 + k_{FC} n_0 t)^{k+1}} \tag{4}$$

where:

$$n = \sum n_k \tag{5}$$

$$n_0 = n(t = 0) \tag{6}$$

Fitting of Eq. (3) to the experimental variation of the turbidity (or absorbance) allows the calculation of $k_{FC}$, and $x_a$. In the case of ionic nano-emulsions, aggregation is induced by injecting a high concentration of salt into the sample vessel. This practice screens the surface charge of the drops inducing their flocculation. The method for determining the initial aggregation time is nicely illustrated in Fig. 2 of Ref. [Rahn-Chique, 2012b].

The procedure outlined above is efficient and generally requires a very short sampling time (60-100 s). Up to this work it was uncertain if the absorbance of the emulsion was perturbed by micellar solubilization or crystallization of the surfactant solution. The absorbance of these dispersions in 0, 300, 500, and 700 mM NaCl ([SDS] = 0.5 and 7.5 mM SDS, T = 20, 25 °C) was monitored during 60 s. Equation (3) was used to obtain approximate values of $k_{FC}$.

## 3.4 Evolution of the average radius.

### 3.4.1 Experimental measurement

With our current equipments (LS230 Coulter; Goniometer BI-200SM, Brookhaven) it is not possible to measure the variation of the average radius of the emulsions during the first 80 s after the addition of salt. Therefore it is not possible to ascertain how the drop size distribution changes during the collecting of the flocculation data. On the one hand, the sample is too dilute and the size of the particles is too small for the LS230 equipment. On the other hand, the system evolves too fast for the calculation of a stable correlation function required by the BI-200SM. Despite these limitations, the size of the drops was measured at relatively long times using the Goniometer for a total lapse of 500 s. Salt was added to each emulsion 40 seconds after the start of the first measurement.

### 3.4.2 Theoretical prediction of the average radius

It is clear from Figs. 1 and 2 that the behavior of the emulsion at relatively long times (5 hours) can be very different depending on temperature and salt concentration. However, solubilization and Krafft precipitation take a finite time, and it is uncertain if they influence the evaluation of $k_{FC}$ which takes at most 60 seconds. It is also unknown if the values of $k_{FC}$ deduced thereby have any relevance, that is: Can they describe the behavior of the system during the time of measurements and afterwards? In order to answer this question the simplest theoretical expression which accounts for flocculation and coalescence was used to approximate the average radius [Rahn-Chique, 2015]:

$$R_{FC} = \left[\frac{n_1}{n}\right] R_1 + \sum_{k=2}^{k_{\max}} \left[\frac{n_k}{n}\right] [x_a R_{k,a} + (1 - x_a) R_{k,s}] \tag{7}$$

where: $R_{k,a}$ corresponds to the average radius of the aggregates with $k$ primary particles, and $R_{k,s}$ to the radius of a drop resulting from the coalescence of $k$ primary particles ($R_{k,s} = \sqrt[3]{k}\, R_0$). In the case in which $R_{k,a} \approx R_{k,s}$, Eq. (7) can be easily evaluated:

$$R_{FC} = \left[\frac{n_1}{n}\right] R_1 + \sum_{k=2}^{k_{\max}} \left[\frac{n_k}{n}\right] R_{k,a} [x_a + (1 - x_a)] = \\ = \sum_{k=1}^{k_{\max}} \left[\frac{n_k}{n}\right] R_k \tag{8}$$

Notice that $R_{k,a} \approx R_{k,s}$ is a rather strong approximation. It means that the hydrodynamic radius of an aggregate of size $k$ is equal to the one of a big drop with the same volume (composed of $k$ primary drops). Nevertheless, this approximation is usually incorporated in the software of the light scattering equipment. These apparatus supply the size of an "equivalent" sphere despite the actual shape of the particle.

In general the dependence of $R_{k,a}$ on $k$ is unknown due to the variety of conformations available for each aggregate size. Alternative approximate expressions can be formulated based on Eq. (3) and on the connection between the optical cross section of a spherical particle ($\sigma$) and its radius ($\sigma = Q_s \pi R^2$) [Rahn-Chique, 2015]:

$$R_{FC} = n_1(\sigma_1/Q_s\pi)^{1/2} + \sum_{k=2}^{k_{max}} n_k[x_a(\sigma_{k,a}/Q_a\pi)^{1/2} + (1-x_a)(\sigma_{k,s}/Q_s\pi)^{1/2}] \quad (9)$$

where $Q$ is the scattering coefficient [Gregory, 2009]. According Ref. [Rahn-Chique, 2015] the predictions of Eq. (9) tend to be lower than the actual values. Although Eq. (9) could probably constitute a better approximation to the average radius determined by light scattering, it depends on two additional parameters. A further attempt to approximate the average radius was made assuming $Q = Q_s = Q_a$ with $Q = 0.003$ and $0.11$ [Rahn-Chique, 2015].

### 3.5 Effect of Gravity

As will be evident from the results, the absorbance of several systems decreases at long times. This can be the consequence of numerous processes, including solubilization and creaming. In the case of nanoemulsions, creaming is not expected due to the small size of the drops. Disregarding hydrodynamic effects, the creaming of a particle can be roughly estimated using Eq. (10):

$$V_g = \frac{2R_i^2 \Delta\rho g}{9\eta} \quad (10)$$

where $V_g$ is the creaming rate, $\Delta\rho$ the density difference between the external and the internal phase, g is the gravity force (9.8 m/s$^2$), and $\eta$ is the viscosity of the water phase (8.9 x 10$^{-4}$ Pas). The time required for the creaming of a dispersion can be estimated dividing the height of the container (aprox 4 cm) by the creaming velocity. For nanometric particles this time is of the order of months, and does not change too sensibly if the particles aggregate or coalesce. Despite this fact, we used a Quickscan turbimeter (Formulaction) in order to evaluate the effect of creaming on the behavior of the absorbance at salinities of 350 and 375 mM NaCl, (where the most pronounced clarification of the system was observed).

### 3.6 Refractive Index of surfactant solutions

Using a 13-947 Refractometer (Fisher Scientific), the refractive index of the surfactants solutions at 100, 300, 500, 700, 900 and 1000 mM NaCl was measured, in order to confirm, that it differs considerably from the one of dodecane under the working conditions.

## 4. SIMULATIONS

For a complete description of Emulsion Stability Simulations the reader is referred to [Urbina-Villalba, 2009; Toro-Mendoza, 2010; Osorio, 2011; Urbina-Villalba, 2015] and references therein. In ESS the displacement of a particle: $\Delta r = \left|\vec{r}_i(t+\Delta t) - \vec{r}_i(t)\right|$ during a time step $\Delta$t is the product of the potential of interaction between the particles: ($F_{ij} = -dV/dr_{ij}$) and the Brownian drift induced by the solvent molecules $\sqrt{2D_{eff,i}\,\Delta t}\ [\vec{G}auss]$:

$$\vec{r}_i(t+\Delta t) = \vec{r}_i(t) + \sum_j (\vec{F}_{ji} D_{eff,i}/k_B T)\,\Delta t + \sqrt{2D_{eff,i}\,\Delta t}\ [\vec{G}auss] \quad (11)$$

Here "Gauss" is a three component vector which stands for a set of random variables characterized by a Gaussian distribution (with zero mean and unit variance), and $D_{eff,i}$ is the effective diffusion constant of particle *i*.

The effects of solubilization were incorporated in a new routine. Its algorithm is very simple:

1. The equation of Corrin and Harkins is used to estimate an approximate CMC for the given salt concentration [Urbina-Villalba, 2015; 2016]. Solubilization proceeds if the surfactant concentration is higher than the value of the CMC.
2. The number of surfactants adsorbed to the interface of the drops is subtracted from the total surfactant population.
3. The rest of the surfactant is used to calculate the number of available micelles. A typical aggregation number is an input of the simulation. Sixty molecules/micelle were considered in the case of SDS.
4. Depending on the experimental conditions the molecular solubility of the oil in water can be subtracted or not from the total number of molecules susceptible of dissolution. If the emulsions are prepared in advance prior to their destabilization with brine, it is likely that the water solution is already saturated with oil at sub-CMC concentrations. In this scenario it seems reasonable to subtract the molecular solubility of the oil from the total population of oil molecules prior to the process of solubilization. If instead the mixture of components is pre-pared in situ, the total number of molecules of oil should be susceptible of dissolution.

Table 1: Efective charge of the surfactant molecule predicted by T-I and T-II parameterization.

| [NaCl] (M) | [SDS] (M) | Ionic Strength (M) | Average Radius (nm) | Exp. Pot. (mV) | Adjusted Pot. (mV) | Method | Surfactant Charge q (Coul) |
|---|---|---|---|---|---|---|---|
| 5.0 x $10^{-3}$ | 8.0 x $10^{-3}$ | 1.3 x $10^{-2}$ | 217.5 | 83.8 | 83.8 | T - I | 0.088 |
| 9.5 x $10^{-5}$ | 8.0 x $10^{-3}$ | 8.95 x $10^{-3}$ | 72.5 | 81.7 | 81.73 | T - I | 0.075 |
| 2.0 x $10^{-3}$ | 2.96 x $10^{-3}$ | 4.96 x $10^{-3}$ | 73.3 | 85.3 | 85.34 | T - I | 0.072 |
| 5.0 x $10^{-3}$ | 8.0 x $10^{-3}$ | 1.3 x $10^{-2}$ | 217.5 | 83.8 | 83.82 | T - II | 0.3945 |
| 5.0 x $10^{-3}$ | 2.96 x $10^{-3}$ | 7.96 x $10^{-3}$ | 73.3 | 94.6 | 94.62 | T - II | 0.392 |
| 8.0 x $10^{-3}$ | 2.96 x $10^{-3}$ | 1.10 x $10^{-2}$ | 73.3 | 102.4 | 102.36 | T - II | 0.395 |

5. The maximum number of molecules than can be solubilized in a micelle is an input of the calculation ($N^{oil}_{max}$). In the case of hexadecane and SDS, this corresponds approximately to two hexadecane molecules per spherical micelle [Ariyaprakai, 2007].
6. The solubilization rate is given by Eq. (1) with $C_{aq}$ = 0. Thus, the decrease of the radius of each drop (and its corresponding volume) can be calculated. Hence, the number of molecules of oil that can be dissolved at each time step is known ($N^{oil}_{step}$).
7. The solubilization of each drop occurs only if the number of molecules already dissolved does not exceed the maximum solubility of the oil ($N^{oil}_{step} + N^{oil}_{dis} < N^{oil}_{max}$). In this case, the radius of the drops diminishes. Otherwise it remains unchanged.

If the amount of oil is small or the surfactant concentration is very high, the solubilization occurs until the particles completely dissolve ($N^{oil}_{dis} \ll N^{oil}_{max}$). In the usual scenario, the particles only dissolve until the micellar solution is saturated.

The solubilization routine was included in between the exchange molecules of oil due to Ostwald ripening, and the coalescence "check". Hence, the particles move due to their interaction forces and the thermal exchange with the solvent, exchange molecules due to Ostwald ripening, dissolve due to micellar solubilization, and may or may not coalesce depending on their final position and size.

### 4.1 Twenty five particle simulations

In these calculations the simulations boxes only contained 25 drops (R = 72.5 nm [Rahn-Chique, 2012a]. On the one hand, this small number of particles allowed following the evolution of the system for 300 s at a time step 7.45 x $10^{-8}$ s. Thus, the qualitative behavior of the drops can be directly observed. On the other hand, the flocculation rate ($k_{FC}$) cannot be evaluated due to the lack of statistics (significant information).

The simulated systems comprise:

a) hexadecane-in-water (h/w) emulsions in 300, 350, 400, 500, 700 and 900 mM NaCl. This system was extensively studied by our group during the last four years [Urbina-Villalba, 2013; 2015; 2016]. The potential of interaction used in the present calculations is the one that allowed reproducing the experimental variation of $k_{FC}$ vs. [NaCl] for 0.5 mM SDS [Urbina-Villalba, 2015]. It employs an adsorption isotherm which assumes a logarithmic dependence of the surface surfactant concentration on the *electrolyte concentration* (Model 4 in Ref. [Urbina-Villalba, 2013] and calculations Type II (T-II) in Ref. [Urbina-Villalba, 2015]).

b) dodecane-in-water (d/w) emulsions in 300 or 500 mM NaCl. As in Ref. [Urbina-Villalba, 2015], two different procedures for calculating the interaction potential were tested. A Type I calculation corresponds to the use of "macroscopic" adsorption isotherms which depend logarithmically on the *surfactant concentration*. A Type II parameterization corresponds to the procedure referred above for h/w calculations. This is a kind of "nanoscopic" adsorption isotherm deduced from the variation of the electrostatic surface potential of the drops as a function of the salt concentration.

The charge of the surfactant molecule required for d/w calculations was obtained by adjusting the theoretical value of the surface potential of a drop to its zeta potential $\zeta$ (Table 1), using an interfacial area of 50 Å$^2$. Average values of q = 0.0783 and 0.3938 Coul were obtained for Type I (T-I) and Type II (T-II) parameterizations, respectively. Figure 3 shows the change of $|\zeta|$ vs. [NaCl] predicted by the two procedures. T-I parameterization predicts lower absolute values of $\zeta$ which decrease monotonically with the ionic strength. Instead, T-II parameterization shows a peak around 20 mM NaCl, and a much subtle decrease. As a consequence, the potential of interaction between two spherical drops shows low repulsive barriers in the former case (Figure 4) and high, insurmountable barriers in the latter case. Figure 5 illustrates the shape of the secondary minimum corresponding to Type II parameterization.

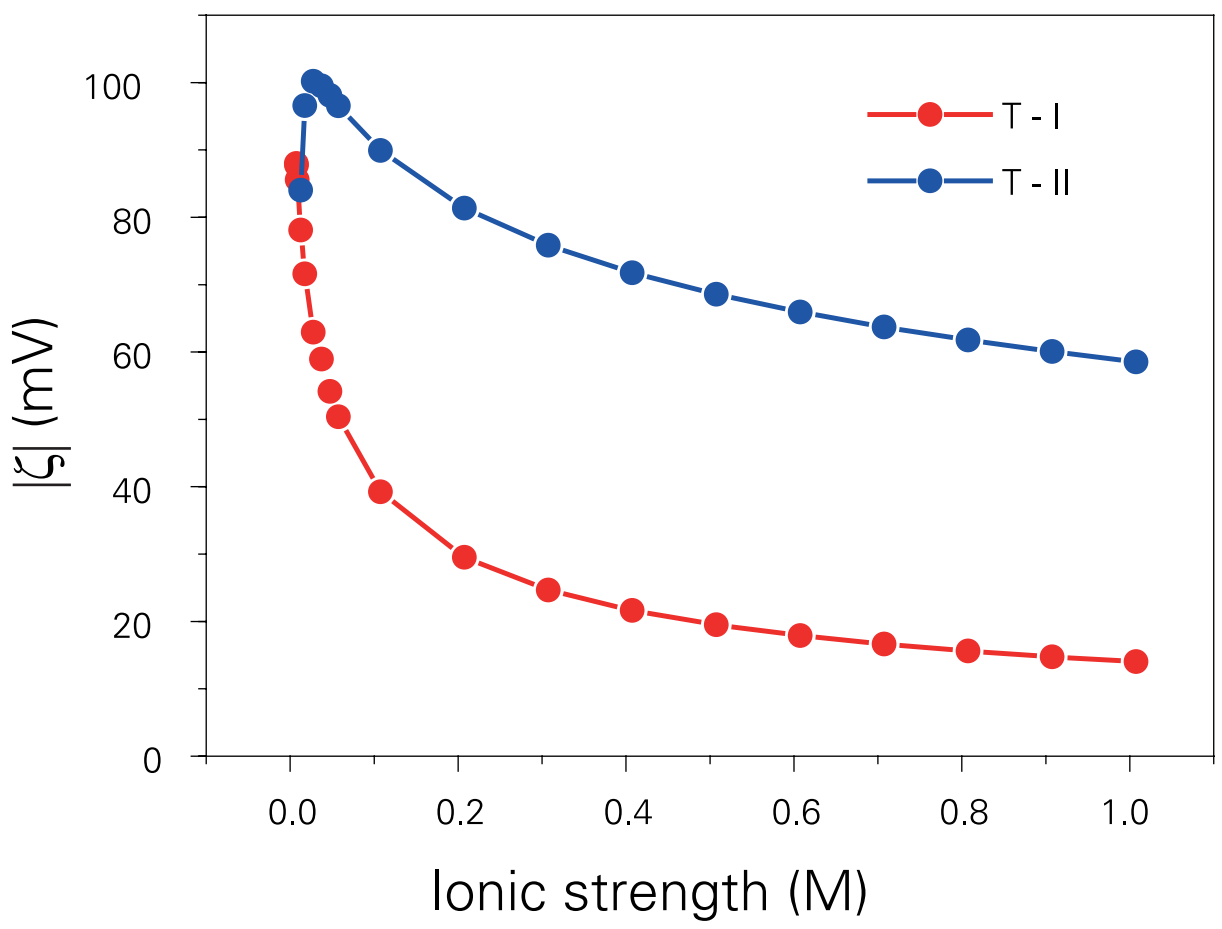

Fig. 3: Variation of the electrostatic surface potential (ζ) of a drop as a function of the ionic strength of the solution, for T-I and T-II parameterization.

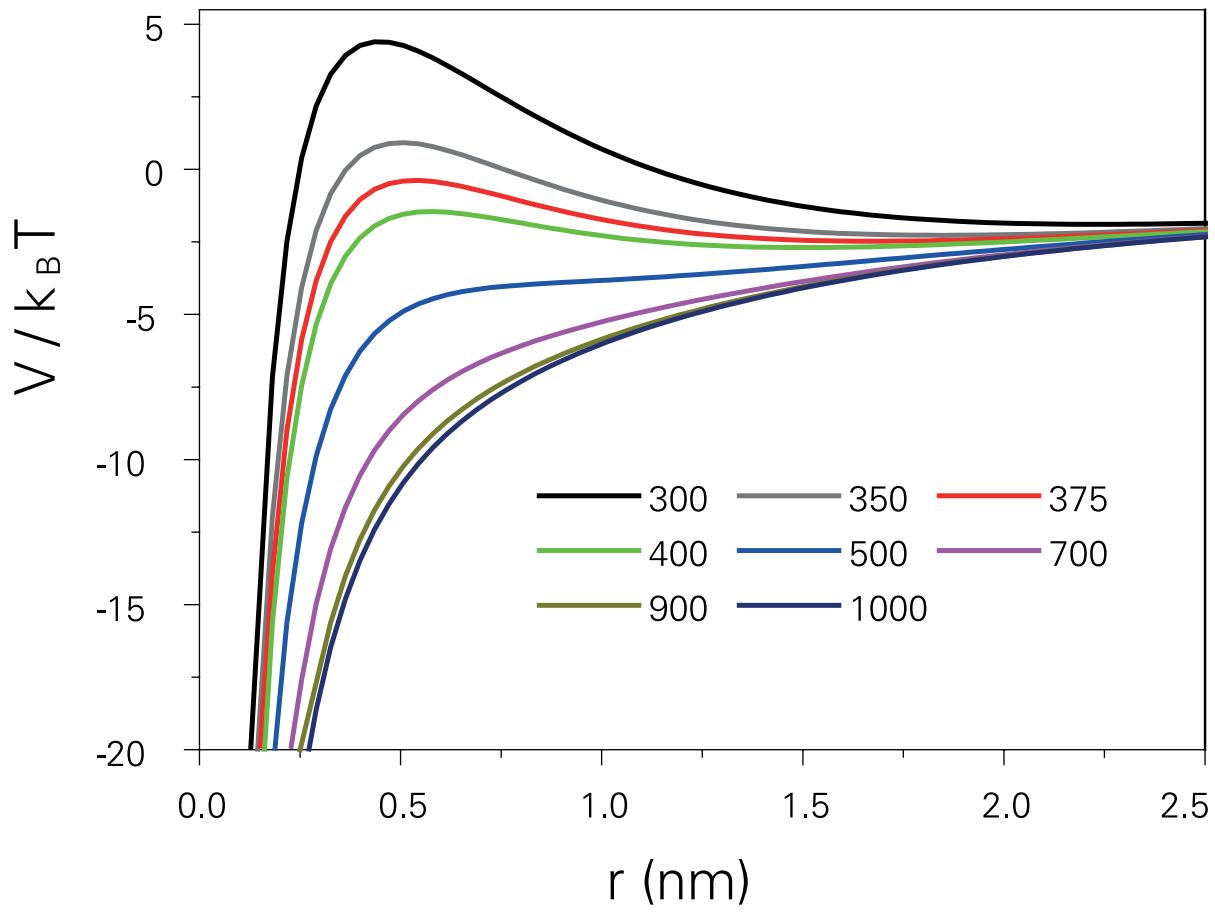

Fig. 4: Interaction potential between two drops of dodecane (R = 72.5 nm) in water according T-I parameterization.

### 4.2 Doublets

In order to approximate the qualitative behavior of $k_{FC}$ as a function of the salt concentration, the methodology employed in Ref. [Urbina-Villalba, 2016] was used. It consists on the calculation of one many-particle simulation along with a set of stability ratios ($W_{11}$). Using this procedure the actual value of the rate under distinct experimental conditions can be approximated as:

$$k_F^{slow} = k_F^{fast} / W_{11} \tag{12}$$

The values of $W_{11}$ are obtained from the average flocculation/coalescence time predicted by one hundred binary calculations ($N_{max} = 100$, $t_f = 0.01$ s). Small cubic boxes with edges of $L = 5\, R_0$ were used. The initial distance of approach depends on the system and interaction potential. In the case of dodecane $d_{floc}$ = 7 nm was sufficient except for 0 mM NaCl and van der Waals computations ($d_{floc}$ = 40 nm (T-I), 20 nm (T-II)). In the case of hexadecane, both spherical and deformable drops were tested, besides Vrij coalescence

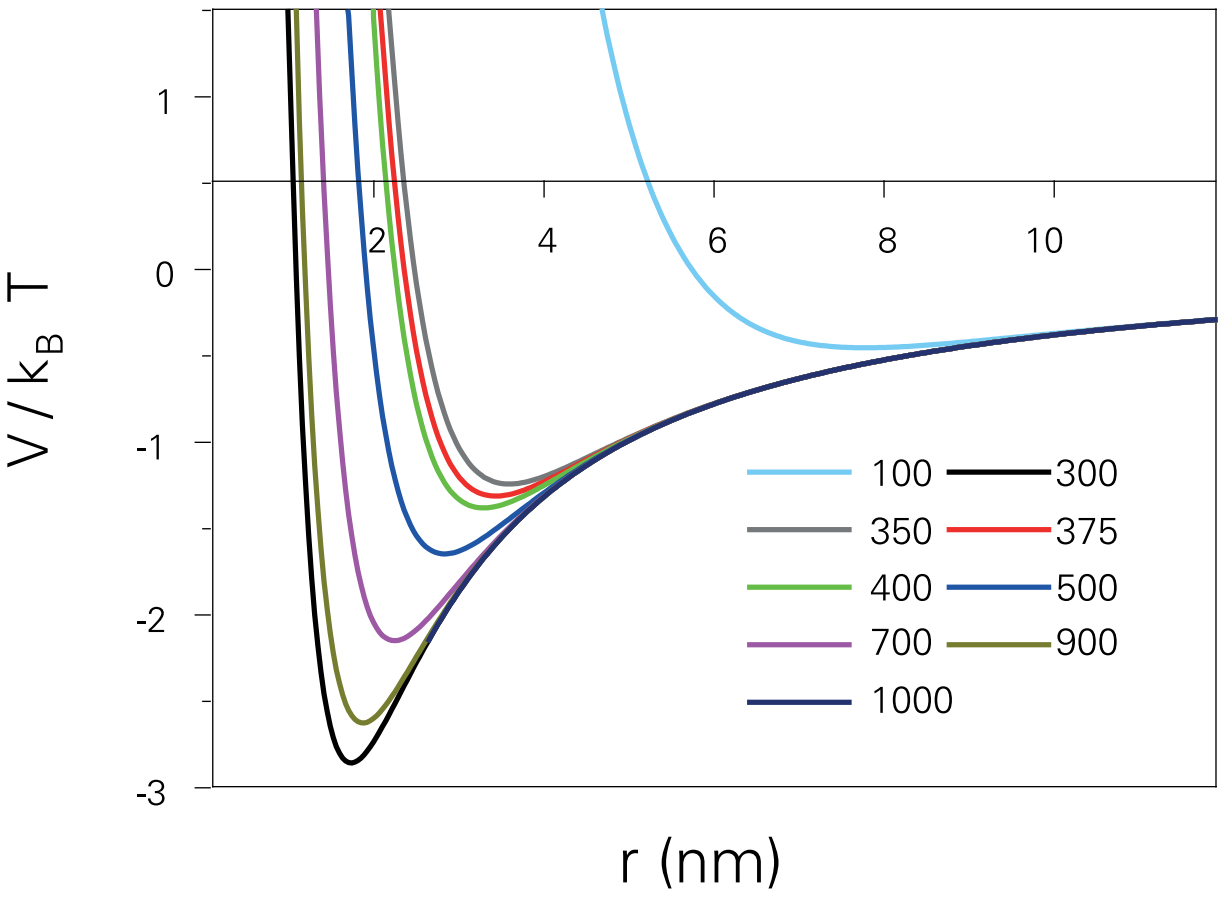

Fig. 5: Interaction potential between two drops of dodecane (R = 72.5 nm) in water according T-II parameterization.

mechanism. Further details regarding these algorithms can be found in: [Urbina-Villalba, 2016].

## 6. RESULTS AND DISCUSSION

### 6.1 Experimental Evidence

Figure 6 illustrates the behavior of the absorbance (Abs) of an aqueous surfactant solution when it is diluted to yield 0.5 mM SDS. This system was prepared adding 0.6 ml of pure water to 2.4 ml of a stock solution. The absorbance is stable until water is added. The process of insertion implies the opening of the cell chamber, the injection, the consequent dilution and perturbation of the sample, and the closing of the chamber's door. Hence the behavior of Abs is erratic for at least 15 s. The signal alternates showing three sets of sharp decreases and sudden recoveries, until it reaches a final maximum and then progressively decreases towards a stable value (in approximately 1.5 min). A similar pattern is observed in most cases when an aliquot of salty water is injected to yield 300 – 700 mM NaCl solutions with 0.5 or 7.5 mM SDS (Figure 7). Notice that the peak of absorbance does not surpass 0.2 a.u. (arbitrary units of absorbance). Moreover, the (asymptotic) stable signal of the solutions does not exceed 0.002 a.u. except in the case of 7.5 mM SDS and 700 mM NaCl at 20 °C. In this system the absorbance passes through a minimum around t = 1 min, and then increases linearly reaching a value of 0.1 a.u. after 5 minutes (Figure 8). In this regard, Aveyard [Aveyard, 1988] found that the absorbance of an AOT solution in the presence of NaCl increases as a function of the surfactant content. However, significant differences were observed when brine was added to the surfactant solution or vice versa. They suggested that the method of mixing has a profound effect in the size of the micelles formed, and therefore, on the conduct of the absorbance. In any event, macroscopic crystals of SDS are observed at longer times (not shown). Crystals grow substantially faster at 20 than at 25 °C as exposed by the curves of $R$ vs. $t$ (see below).

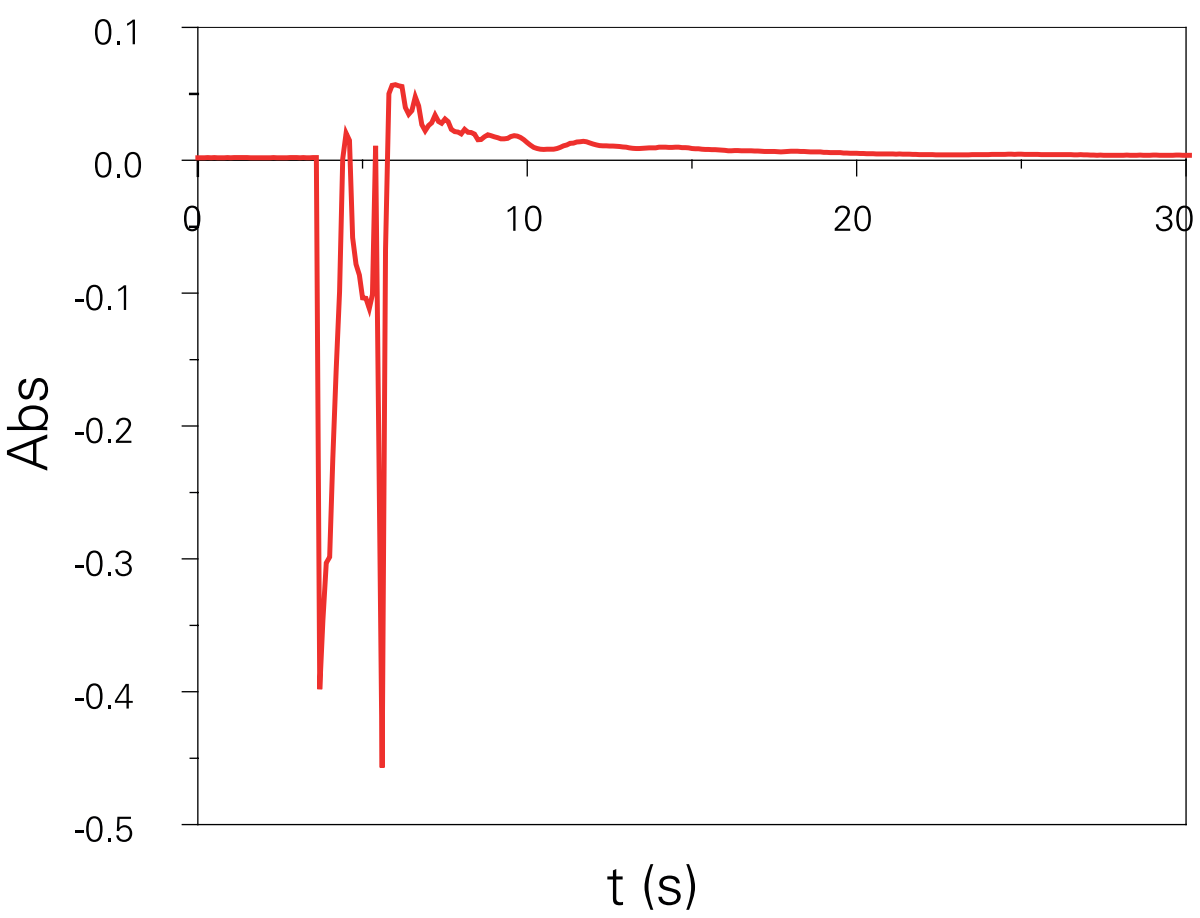

Fig. 6: Change of the absorbance (Abs) of a stock SDS solution (2.4 ml) after its dilution with water (0.6 ml) to yield 0.5 mM SDS.

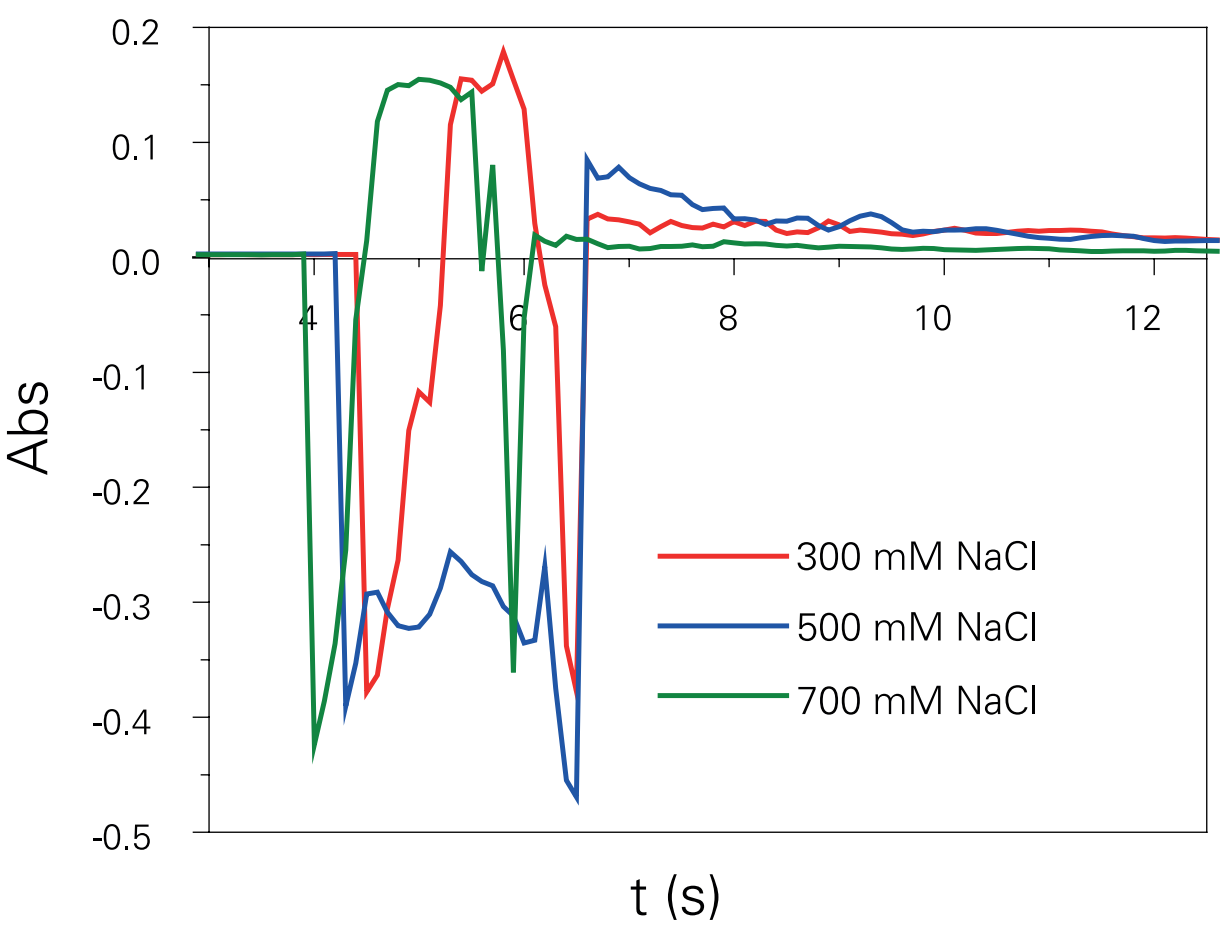

Fig. 7: Change of the absorbance of a stock SDS solution after the addition of an aliquot of brine and SDS to yield salt concentrations between 300 – 700 mM NaCl and 0.5 mM SDS.

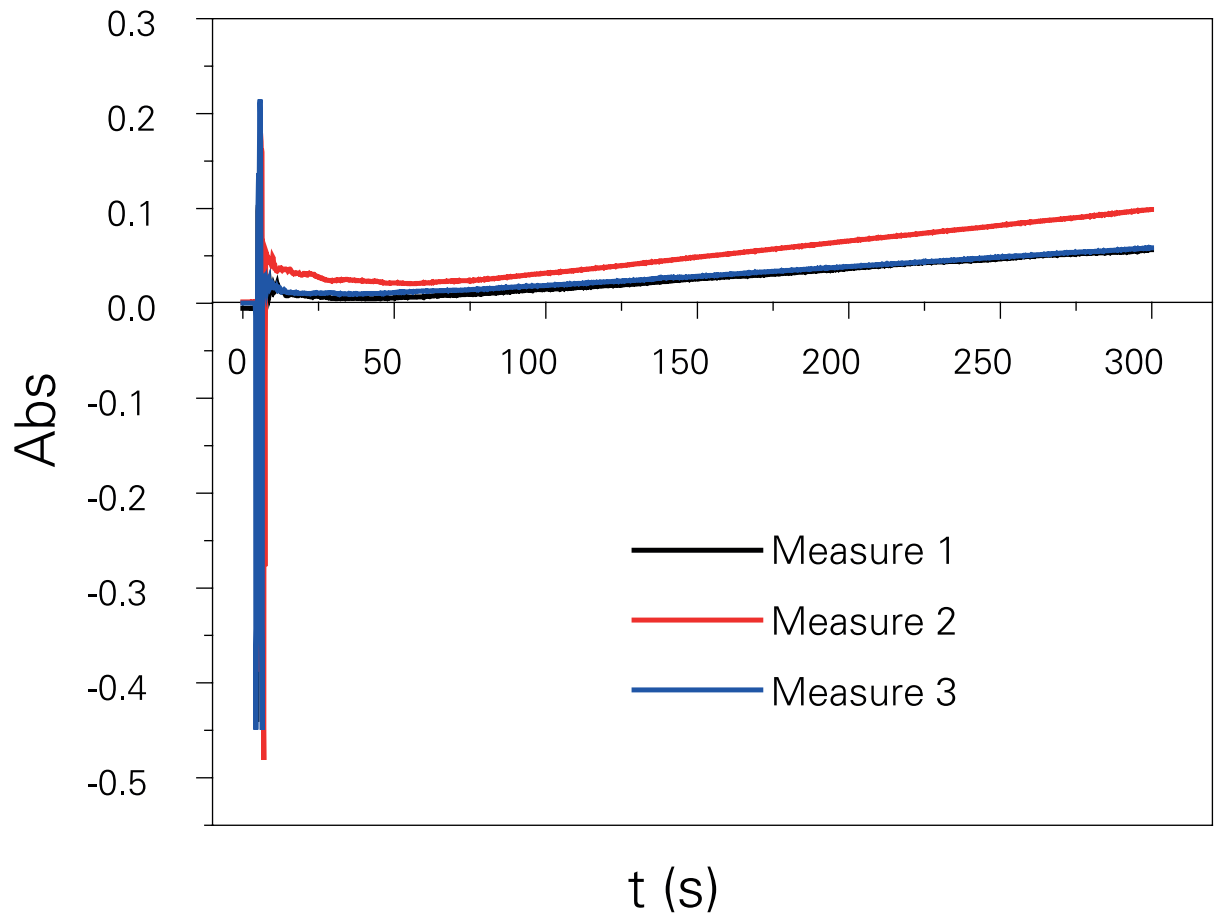

Fig. 8: Change of the absorbance of a stock SDS solution after the addition of an aliquot of brine and SDS to yield 700 mM NaCl and 7.5 mM SDS (T = 20 °C). Results from three different measurements.

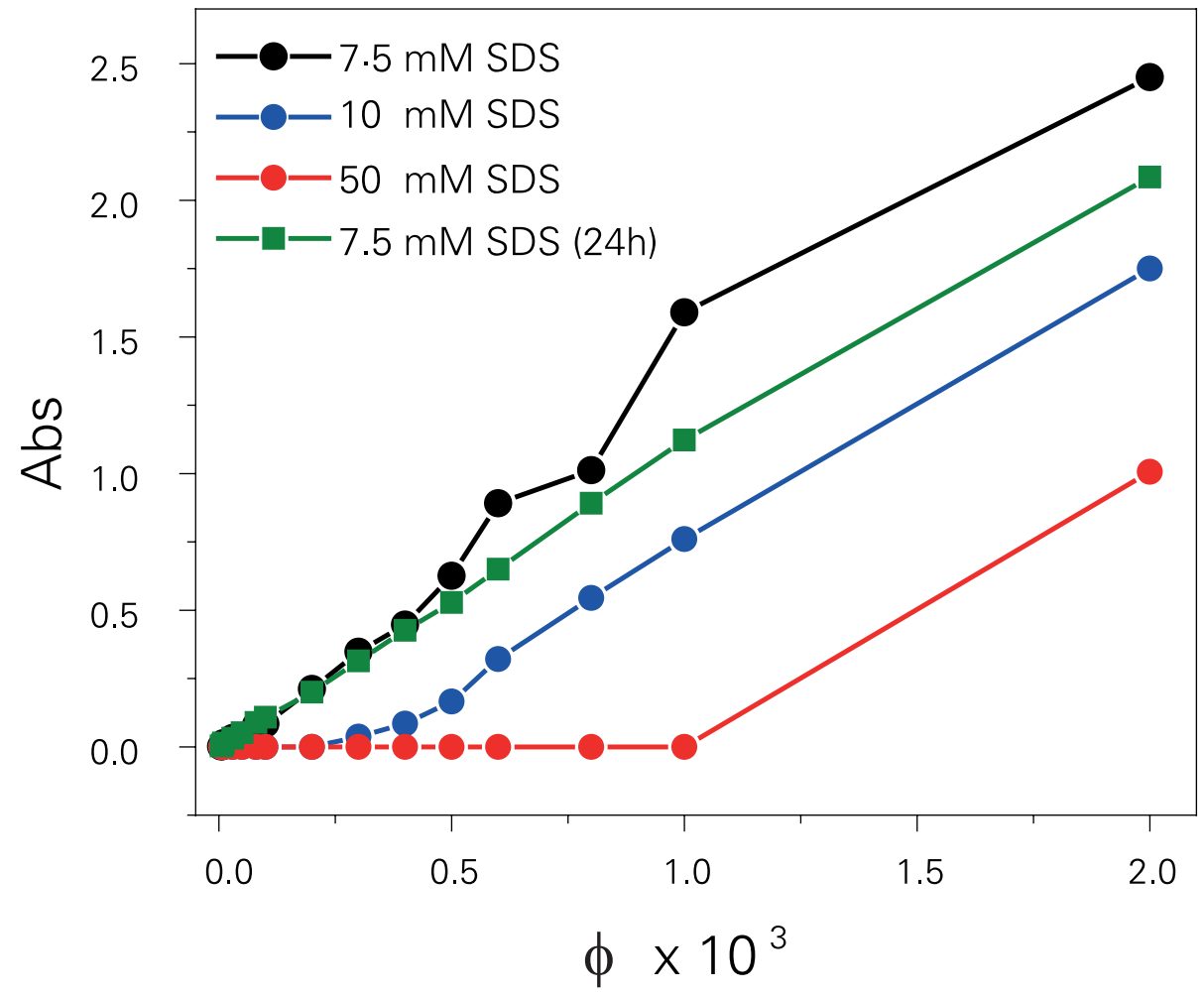

Fig. 9: Absorbance of a dodecane-in-water emulsion stabilized with 7.5, 10.0 and 50.0 mM SDS as a function of the volume fraction of oil (T = 25 C, t = 72h). The green line corresponds to 7.5 mM SDS after 24 hours.

Figure 9 illustrates the behavior of the absorbance of dodecane-in-water (d/w) emulsions containing 7.5, 10.0 and 50.0 mM NaCl after 72 h. As shown by Rao (Fig. 3a in [Rao, 2012]), the absorbance of a stable emulsion increases linearly as a function of the volume fraction of oil. This is the case of our initial, 7.5 mM SDS emulsion at 24 hours: Abs = 1099.6 $\phi$ ($r^2$ = 0.9988). As the number of drops increases, the amount of light scattered by the drops is higher. Hence, the intensity of light that reaches the detector is lower (absorbance increases). However, if the surfactant concentration happens to be higher than the CMC (10, 50 mM SDS), a certain amount of oil is solubilized by the micelles. In this case, the curve of Abs vs. $\phi$ does not pass through the origin [Rao, 2012]. Instead, the absorbance presents an initial horizontal slope close to the x-axis (~ 0) due to the complete dissolution of the drops. At a certain concentration (critical volume fraction $\phi_c$), the micelle solution saturates and the drops in excess promote a sudden increase of the absorbance. For $\phi > \phi_c$ , the plot of Abs vs. t is linear (for instance, for 10 mM SDS at t = 72 h, Abs = 1083.2 $\phi$ – 0.3309, $r^2$ = 0.9894). The slope is similar to the one previously found for the original 7.5-mM SDS emulsion. The linearity of the plots progressively declines during 72 h, indicating that the emulsions are only temporarily stable.

As expected, the value of $\phi_c$ for 50 mM SDS is 5 times higher than the one of 10 mM SDS: $\phi_c$ (50 mM) = 5.01 $\phi_c$ (10 mM). This proportionality is almost exact although it is known that micelles change their shape as a function of the surfactant concentration [Mishic, 1990].

If the value of $\phi_c$ for 50 or 10 mM SDS is used to extrapolate the one of 7.5 mM SDS, the amount of oil that could be solubilized by this surfactant solution *under the presence of micelles* is obtained: $\phi_{hc} = 2.3 \times 10^{-4}$ (where h stands for a "hypothetical" critical fraction). This value is substantially lower than the $\phi_c$ observed for 7.5 mM SDS in 100 – 300 mM NaCl (at 25 °C) and 100 mM NaCl (at 20 °C): $\phi_c = (1.0 - 1.5) \times 10^{-3}$. Counter intuitively, it appears that a small salt concentration has a synergetic effect on solubilization. This situation is reversed at higher salt concentrations where either the solubilization is smaller or a value of $\phi_c$ is not observed (Figs. 10-11).

When NaCl is added to a surfactant solution its CMC lowers considerably. Hence, a curve similar to the ones of 10 and 50 mM SDS would be expected for 7.5 mM SDS if it weren't for the fact that this emulsion is also destabilized by the screening of its surface charges. Therefore, plots of Abs vs. $\phi$ at several salt concentrations are in general non linear. In order to understand the behavior of these systems, it is useful to draw curves of Abs = 1402.6 $\phi$ (stable emulsion), and Abs = 1402.6 ($\phi$ - $\phi_{hc}$) (remaining emulsion after a hypothetical micelle solubilization) upon the graphs. On the one hand, a curve of Abs vs. $\phi$ that lies below these guide-lines will indicate partial solubilization of the emulsion. On the other hand, curves above the guidelines suggest flocculation of the drops, since the scattering of light increases with the size of the scattering objects.

According to Eq. (2), a drop of 72.5 nm will dissolve in 26.9 min (at $dR/dt = 4.49 \times 10^{-11}$ m/s [Ariyaprakai, 2008]). However, it is found that a salt concentration as low as 100 mM NaCl lowers the solubilization **rate** considerably even at T = 25 °C (Figure 10). Larger times are required for complete solubilization of the emulsion. The value of $\phi_c$ increases progressively with time, reaching a magnitude of $10^{-3}$ after 7 days. This fact strongly suggests that the micelles remove the molecules of oil from the aqueous solution and not directly from the drops. The addition of salt lowers the molecular solubility of the oil hindering their solubilization. According to the literature [Verschueren, 2001] the solubility of dodecane decreases from 0.0037 mg/l to 0.0029 mg/l in sea water (aprox 600 mM NaCl) at T = 25 °C. Hence, the rate of solubilization should be slower than in the absence of salt, although the total amount of oil solubilized by the micelles at long times is higher.

At T = 25 °C, the value of $\phi_c$ slightly increases for 300 mM NaCl, but then decreases at higher ionic strengths, vanishing for [NaCl] > 700 mM NaCl. In these cases, the curves of Abs vs. $\phi$ do not touch the x-axis during 7 days. In most cases the absorbance decreases as a function of time except at very low fractions of oil, where small maxima are observed (notice the difference between the systems of 500 and 700 mM NaCl). The partial solubilization of the drops can be appraised, comparing the absorbance of each curve to the guide-lines previously drawn. As shown by Figure 10, most curves lie below the guidelines but above the horizontal axis, indicating a certain degree of solubilization. The process of flocculation prevails at high salt concentrations.

The main features of the absorbance of emulsions at T = 25 °C are reproduced at 20 °C (Figure 11). However, the evolution of the systems is different. It depends strongly on the volume fraction of oil. In principle, the molecular solubility of the oil should decrease at a lower temperature, and therefore, the rate of solubilization by the micelles should diminish.

At a low salinity of 100 mM NaCl, the curves resemble the ones of 25 °C, and a critical volume fraction of oil can be ascribed. At 300 mM the absorbance shows some erratic peaks at low volume fractions of oil, and then reaches the x-axis around $\phi = 7.5 \times 10^{-4}$. Higher salt concentrations evidence a substantial destabilization of the systems basically due to flocculation and more importantly: to the formation of crystals. These crystals of SDS can be observed by visual inspection of the samples. As expected, the absorbance changes irregularly when crystals appear in the solution.

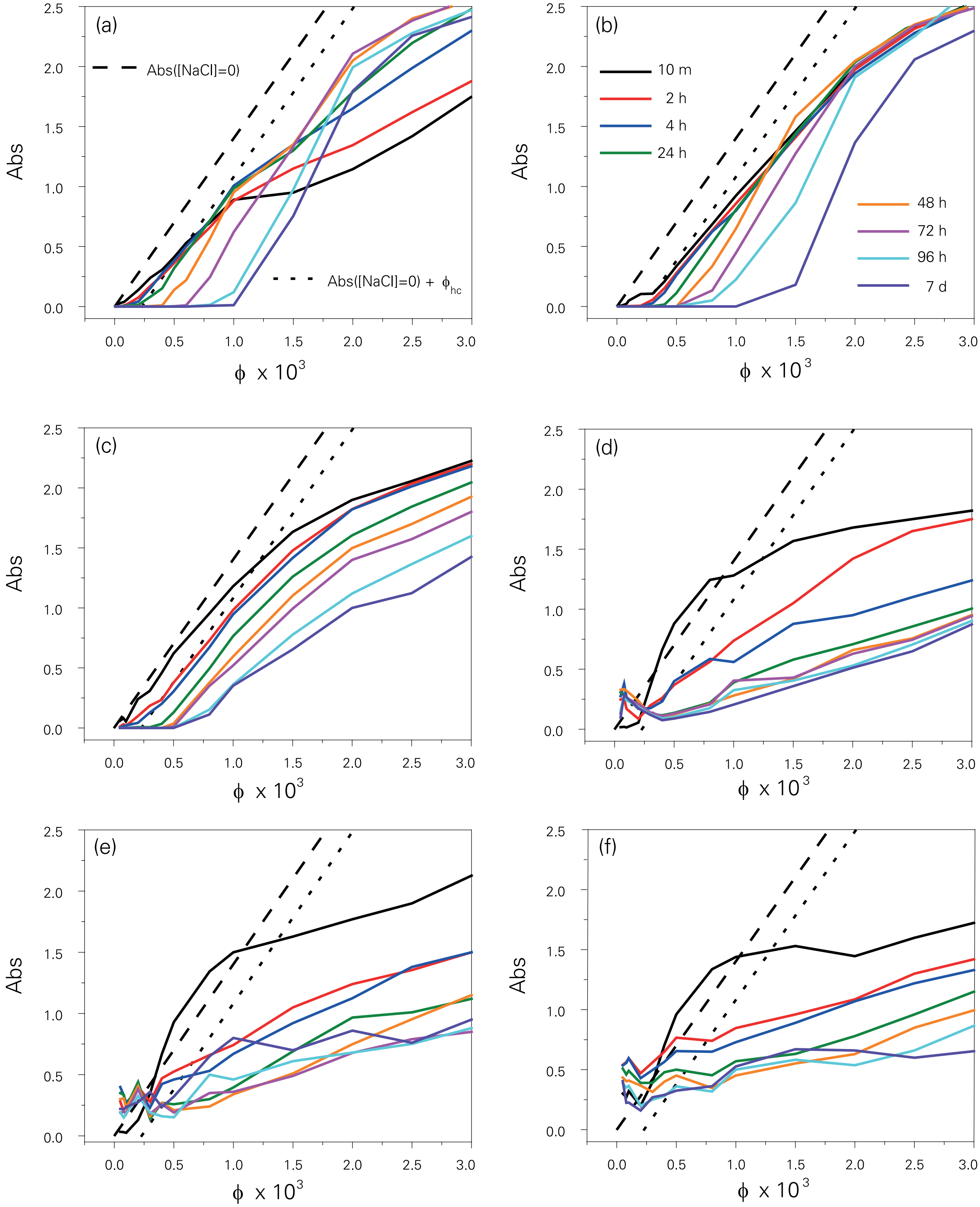

Fig. 10: Variation of Abs vs. $\phi$ for a d/w emulsion stabilized with 7.5 mM SDS at several times of observation. (T = 25 C). (a) 100, (b) 300, (c) 500, (d) 700, (e) 900, (f) 1000 mM NaCl.

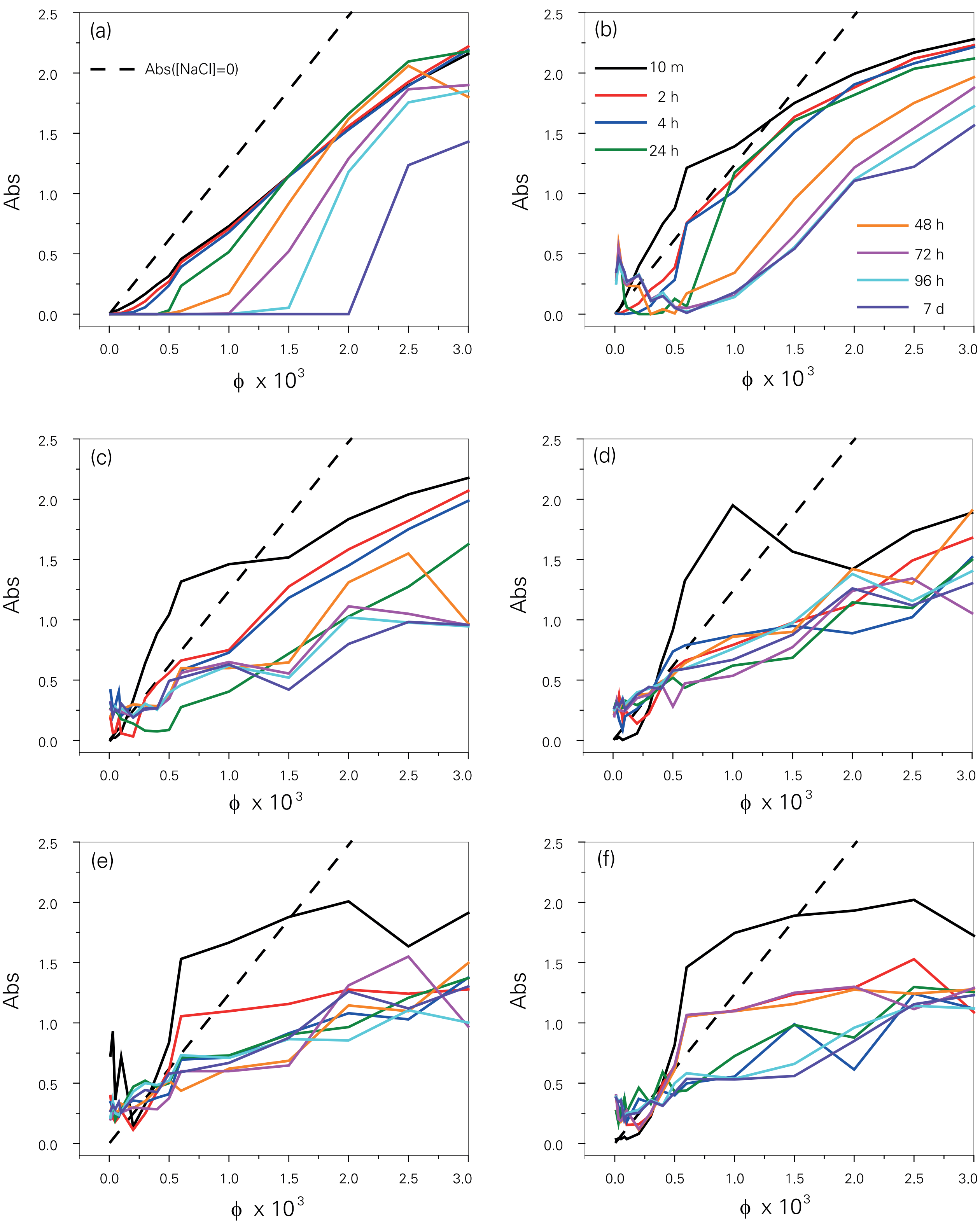

Fig. 11: Variation of Abs vs. $\phi$ for a d/w emulsion stabilized with 7.5 mM SDS at several times of observation. (T = 20 C). (a) 100, (b) 300, (c) 500, (d) 700, (e) 900, (f) 1000 mM NaCl.

Tables 2 and 3 shows approximate values of $\phi_c$. These were estimated in each case from the intercept of a linear regression of the points corresponding to the absorbance increase. In general, the value of $\phi_c$ increases with the observation time until the complete saturation of the micelles is attained.

The absorbance of emulsions was also studied continuously during 5 minutes. The initial value of Abs is approximately 0.4 units. This is more than twice the value of the highest peak observed in surfactant solutions. The relaxation time after the dilution with brine is substantially shorter: approximately 4 seconds. The absorbance presents initially a "W" shape: it drops abruptly, goes up, backs down again, and finally rises, either to stabilize at a certain value or to keep growing up to a higher level.

Figure 12 shows the 300-second variation of the absorbance for 500 and 700 mM NaCl (7.5 mM SDS, T = 20 and 25 °C). Three of these systems show the appearance of crystals in test tubes after 10 minutes. However, the possible existence of microscopic solids at very short times is unknown. In all cases the absorbance grows monotonically during approximately 100 s. Notice that 60 s are required for the appraisal of the flocculation rate. In fact, the behavior of the curves is typical of flocculation, except for the terminal decrease of the 500 mM NaCl system at T = 25 °C. A close look at the rest of the systems reveals an almost imperceptible drop of the absorbance at long times (changes occur in the second

Table 2: Critical volume fraction of solubilization for 7.5 mM SDS at T= 25°C. The increase of the absorbance after the critical value $\phi_c$, is characterized by the average slope of a linear fit. The number of points used in the computation is also specified.

| | | 100 mM NaCl | | T=25°C | |
|---|---|---|---|---|---|
| Time | $\phi_c$ | Slope (Abs vs. $\phi$) | $r^2$ | N° data points | Observations |
| 10 min | 3,27 x $10^{-5}$ | 910 | 0,9975 | 11 | ---- |
| 2 h | 7,53 x $10^{-5}$ | 858 | 0,9876 | 10 | ---- |
| 4 h | 7,67 x $10^{-5}$ | 919 | 0,9837 | 11 | ---- |
| 24 h | 1,42 x $10^{-4}$ | 960 | 0,8830 | 9 | ---- |
| 48 h | 3,75 x $10^{-4}$ | 1266 | 0,9887 | 7 | ---- |
| 72 h | 6,20 x $10^{-4}$ | 1527 | 0,9987 | 4 | ---- |
| 96 h | 9,51 x $10^{-4}$ | 1874 | 0,9975 | 3 | ---- |
| 7 d | 9,75 x $10^{-4}$ | 1555 | 0,9810 | 4 | ---- |
| | | 300 mM NaCl | | T=25°C | |
| 10 min | 1,13 x $10^{-4}$ | 1009 | 0,9937 | 11 | ---- |
| 2 h | 2,15 x $10^{-4}$ | 1057 | 0,9957 | 9 | ---- |
| 4 h | 2,57 x $10^{-4}$ | 1119 | 0,9974 | 8 | ---- |
| 24 h | 3,89 x $10^{-4}$ | 1278 | 0,9992 | 6 | ---- |
| 48 h | 5,19 x $10^{-4}$ | 1439 | 0,9855 | 5 | ---- |
| 72 h | 7,10 x $10^{-4}$ | 1566 | 0,9991 | 4 | ---- |
| 96 h | 7,48 x $10^{-4}$ | 1263 | 0,9740 | 6 | ---- |
| 7 d | 1,36 x $10^{-3}$ | 1877 | 0,9775 | 3 | ---- |
| | | 500 mM NaCl | | T=25°C | |
| 10 min | 2,71 x $10^{-5}$ | 1103 | 0,9929 | 7 | Cream $\phi$ > 0.002 |
| 2 h | 1,04 x $10^{-4}$ | 1007 | 0,9915 | 8 | Cream $\phi$ > 0.002 |
| 4 h | 1,45 x $10^{-4}$ | 1010 | 0,9931 | 9 | Cream $\phi$ > 0.001 |
| 24 h | 2,51 x $10^{-4}$ | 885 | 0,9808 | 9 | Cream $\phi$ > 0.001 |
| 48 h | 4,22 x $10^{-4}$ | 977 | 0,9955 | 5 | Cream $\phi$ > 0.001 |
| 72 h | 4,48 x $10^{-4}$ | 920 | 0,9969 | 5 | Cream $\phi$ > 0.001 |
| 96 h | 4,59 x $10^{-4}$ | 665 | 0,9866 | 7 | Cream $\phi$ > 0.001 |
| 7 d | 4,60 x $10^{-4}$ | 577 | 0,9833 | 7 | Cream $\phi$ > 0.001 |
| | | 700 mM NaCl | | T=25°C | |
| 10 min | 1,74 x $10^{-4}$ | 2777 | 0,9940 | 4 | Cream $\phi$ > 0.002 |
| 2 h | 3,60 x $10^{-5}$ | 734 | 0,9950 | 7 | Crystals |
| 4 h | -8,67 x $10^{-4}$ | 329 | 0,9689 | 7 | Crystals |
| 24 h | 3,03 x $10^{-5}$ | 350 | 0,9853 | 8 | Crystals |
| 48 h | 1,56 x $10^{-4}$ | 328 | 0,9943 | 8 | Crystals |
| 72 h | 2,18 x $10^{-5}$ | 312 | 0,9786 | 8 | Crystals |
| 96 h | 1,35 x $10^{-4}$ | 305 | 0,9878 | 8 | Crystals |
| 7 d | 2,62 x $10^{-4}$ | 303 | 0,9910 | 8 | Crystals |
| | | 900 mM NaCl | | T=25°C | |
| 10 min | 9,80 x $10^{-5}$ | 2010 | 0,9738 | 7 | Crystals |
| 2 h | -4,09 x $10^{-4}$ | 492 | 0,9948 | 6 | Crystals |
| 4 h | -4,29 x $10^{-4}$ | 457 | 0,9858 | 9 | Crystals |
| 24 h | -1,31 x $10^{-3}$ | 458 | 0,9628 | 7 | Crystals |
| 48 h | 9,52 x $10^{-5}$ | 388 | 0,9870 | 6 | Crystals |
| 72 h | -2,51 x $10^{-4}$ | 291 | 0,9849 | 6 | Crystals |
| 96 h | -8,76 x $10^{-5}$ | 412 | 0,8475 | 6 | Crystals |
| 7 d | 1,57 x $10^{-4}$ | 969 | 0,9946 | 4 | Crystals |
| | | 1000 mM NaCl | | T=25°C | |
| 10 min | 6,51 x $10^{-5}$ | 1921 | 0,9612 | 5 | Crystals |
| 2 h | -1,61 x $10^{-3}$ | 310 | 0,9702 | 10 | Crystals |
| 4 h | -1,35 x $10^{-3}$ | 313 | 0,9864 | 10 | Crystals |
| 24 h | -7,25 x $10^{-4}$ | 300 | 0,9805 | 6 | Crystals |
| 48 h | -5,60 x $10^{-4}$ | 267 | 0,9625 | 5 | Crystals |
| 72 h | -5,77 x $10^{-4}$ | 285 | 0,8995 | 7 | Crystals |
| 96 h | -4,37 x $10^{-4}$ | 343 | 0,9568 | 6 | Crystals |
| 7 d | -5,11 x $10^{-4}$ | 351 | 0,9470 | 6 | Crystals |

Table 3: Critical volume fraction of solubilization for 7.5 mM SDS at T= 20°C. The increase of the absorbance after the critical value $\phi_c$, is characterized by the average slope of a linear fit. The number of points used in the computation is also specified.

| | | 100 mM NaCl | | T=20°C | |
|---|---|---|---|---|---|
| Time | $\phi_c$ | Slope (Abs vs. $\phi$) | $r^2$ | N° data points | Observations |
| 10 min | $3{,}04 \times 10^{-5}$ | 753 | 0,9975 | 12 | ---- |
| 2 h | $1{,}33 \times 10^{-4}$ | 824 | 0,9983 | 10 | ---- |
| 4 h | $1{,}92 \times 10^{-4}$ | 838 | 0,9977 | 9 | ---- |
| 24 h | $4{,}32 \times 10^{-4}$ | 1033 | 0,9954 | 6 | ---- |
| 48 h | $8{,}13 \times 10^{-4}$ | 1272 | 0,9876 | 4 | ---- |
| 72 h | $1{,}02 \times 10^{-3}$ | 1270 | 0,9947 | 4 | ---- |
| 96 h | $1{,}41 \times 10^{-3}$ | 1703 | 0,9663 | 3 | ---- |
| 7 d | $1{,}88 \times 10^{-3}$ | 1430 | 0,8501 | 3 | ---- |
| | | 300 mM NaCl | | T=20°C | |
| 10 min | $-3{,}20 \times 10^{-4}$ | 802 | 0,9029 | 15 | ---- |
| 2 h | $9{,}69 \times 10^{-5}$ | 1194 | 0,9800 | 9 | ---- |
| 4 h | $1{,}49 \times 10^{-4}$ | 1085 | 0,9671 | 8 | ---- |
| 24 h | $4{,}45 \times 10^{-4}$ | 1631 | 0,9305 | 4 | Crystals |
| 48 h | $4{,}80 \times 10^{-4}$ | 896 | 0,9877 | 6 | Crystals |
| 72 h | $7{,}47 \times 10^{-4}$ | 868 | 0,9858 | 5 | Crystals |
| 96 h | $7{,}69 \times 10^{-4}$ | 806 | 0,9851 | 5 | Crystals |
| 7 d | $6{,}62 \times 10^{-4}$ | 690 | 0,9662 | 5 | Crystals |
| | | 500 mM NaCl | | T=20°C | |
| 10 min | $-3{,}78 \times 10^{-4}$ | 748 | 0,8451 | 15 | Crystals |
| 2 h | $-4{,}10 \times 10^{-4}$ | 624 | 0,9841 | 7 | Crystals |
| 4 h | $-9{,}73 \times 10^{-5}$ | 705 | 0,9762 | 5 | Crystals |
| 24 h | $2{,}63 \times 10^{-4}$ | 587 | 0,9824 | 5 | Crystals |
| 48 h | $2{,}89 \times 10^{-4}$ | 703 | 0,9067 | 4 | Crystals |
| 72 h | $-8{,}52 \times 10^{-4}$ | 338 | 0,5985 | 4 | Crystals |
| 96 h | $4{,}75 \times 10^{-4}$ | 369 | 0,7958 | 6 | Crystals |
| 7 d | $-1{,}65 \times 10^{-3}$ | 197 | 0,4689 | 6 | Crystals |
| | | 700 mM NaCl | | T=20°C | |
| 10 min | $1{,}48 \times 10^{-4}$ | 2436 | 0,9654 | 6 | Crystals |
| 2 h | $-7{,}33 \times 10^{-4}$ | 446 | 0,9826 | 8 | Crystals |
| 4 h | $-1{,}31 \times 10^{-3}$ | 317 | 0,7948 | 10 | Crystals |
| 24 h | $-6{,}58 \times 10^{-4}$ | 388 | 0,8696 | 6 | Crystals |
| 48 h | $-4{,}80 \times 10^{-4}$ | 540 | 0,9649 | 13 | Crystals |
| 72 h | $-6{,}77 \times 10^{-4}$ | 364 | 0,8840 | 15 | Crystals |
| 96 h | $-7{,}87 \times 10^{-4}$ | 399 | 0,9422 | 15 | Crystals |
| 7 d | $-8{,}55 \times 10^{-4}$ | 363 | 0,9487 | 15 | Crystals |
| | | 900 mM NaCl | | T=20°C | |
| 10 min | $-1{,}12 \times 10^{-3}$ | 542 | 0,6703 | 15 | Crystals |
| 2 h | $-7{,}46 \times 10^{-4}$ | 418 | 0,7628 | 15 | Crystals |
| 4 h | $-7{,}81 \times 10^{-4}$ | 369 | 0,9441 | 15 | Crystals |
| 24 h | $8{,}50 \times 10^{-4}$ | 364 | 0,9454 | 15 | Crystals |
| 48 h | $-6{,}76 \times 10^{-4}$ | 382 | 0,9625 | 15 | Crystals |
| 72 h | $-5{,}77 \times 10^{-4}$ | 389 | 0,8149 | 15 | Crystals |
| 96 h | $-1{,}23 \times 10^{-3}$ | 277 | 0,8577 | 15 | Crystals |
| 7 d | $-8{,}55 \times 10^{-4}$ | 363 | 0,9487 | 15 | Crystals |
| | | 1000 mM NaCl | | T=20°C | |
| 10 min | $-3{,}10 \times 10^{-4}$ | 742 | 0,7425 | 15 | Crystals |
| 2 h | $-7{,}19 \times 10^{-4}$ | 432 | 0,7293 | 15 | Crystals |
| 4 h | $-9{,}40 \times 10^{-4}$ | 304 | 0,8461 | 15 | Crystals |
| 24 h | $-8{,}00 \times 10^{-4}$ | 358 | 0,9229 | 15 | Crystals |
| 48 h | $-7{,}41 \times 10^{-4}$ | 418 | 0,7628 | 15 | Crystals |
| 72 h | $-8{,}00 \times 10^{-4}$ | 407 | 0,7272 | 15 | Crystals |
| 96 h | $-9{,}38 \times 10^{-4}$ | 304 | 0,9305 | 15 | Crystals |
| 7 d | $-8{,}57 \times 10^{-4}$ | 354 | 0,8856 | 15 | Crystals |

decimal place). For these salinities, the behavior of the 0.5 SDS systems is very similar (not shown).

Figure 13 illustrates the performance of the 300 mM NaCl systems. The 0.5 mM SDS emulsions are stable for at least 300 s, reaching a stable signal after the injection of brine. However, for 7.5 mM SDS, the absorbance decreases monotonically during the whole observation time. The decline is not pronounced but sustained. The absorbance of the original emulsions does not change significantly during 300 s after its dilution with pure water (0.6 ml), but it increases monotonically at significantly longer times (Figure 14). Notice the difference in the scale of the x-axis between Figs. 13 and 14. Since the addition of salt induces the formation of micelles, the results suggest that micelle solubilization predominates at 300 mM NaCl. The flocculation and the ripening rates are too slow to produce an absorbance increase (see below).

If a low salt concentration favors solubilization and a high salt concentration promotes aggregation, there should be intermediate cases comprising a significant contribution of both processes. This induced us to study the behavior of the emulsions at additional salinities (350, 375 mM SDS). Figure 15 shows the change of the absorbance for 350 mM NaCl. At T = 20 °C the absorbance shows a substantial increase with a minor terminal decrease. But at T = 25 °C, the absorbance rises first and then decreases after 40-50 seconds. These results were confirmed using an additional set of fresh emulsions, and at least five different injections for each salt concentration. It was observed that the outcome of the 375 mM NaCl systems was more reproducible, but still, a maximum of absorbance was observed in the majority of the injections at 350 mM NaCl.

The results of the previous paragraph suggest the initial preponderance of floc-culation, followed by the prevalence of solubilization. Surprisingly, the same phenomenon occurs not only for 7.5 mM SDS, but also for 0.5 mM ! This was totally unexpected since at 350 mM NaCl, the CMC is predicted to be 0.9 mM SDS [Urbina-Villalba, 2013; 2015].

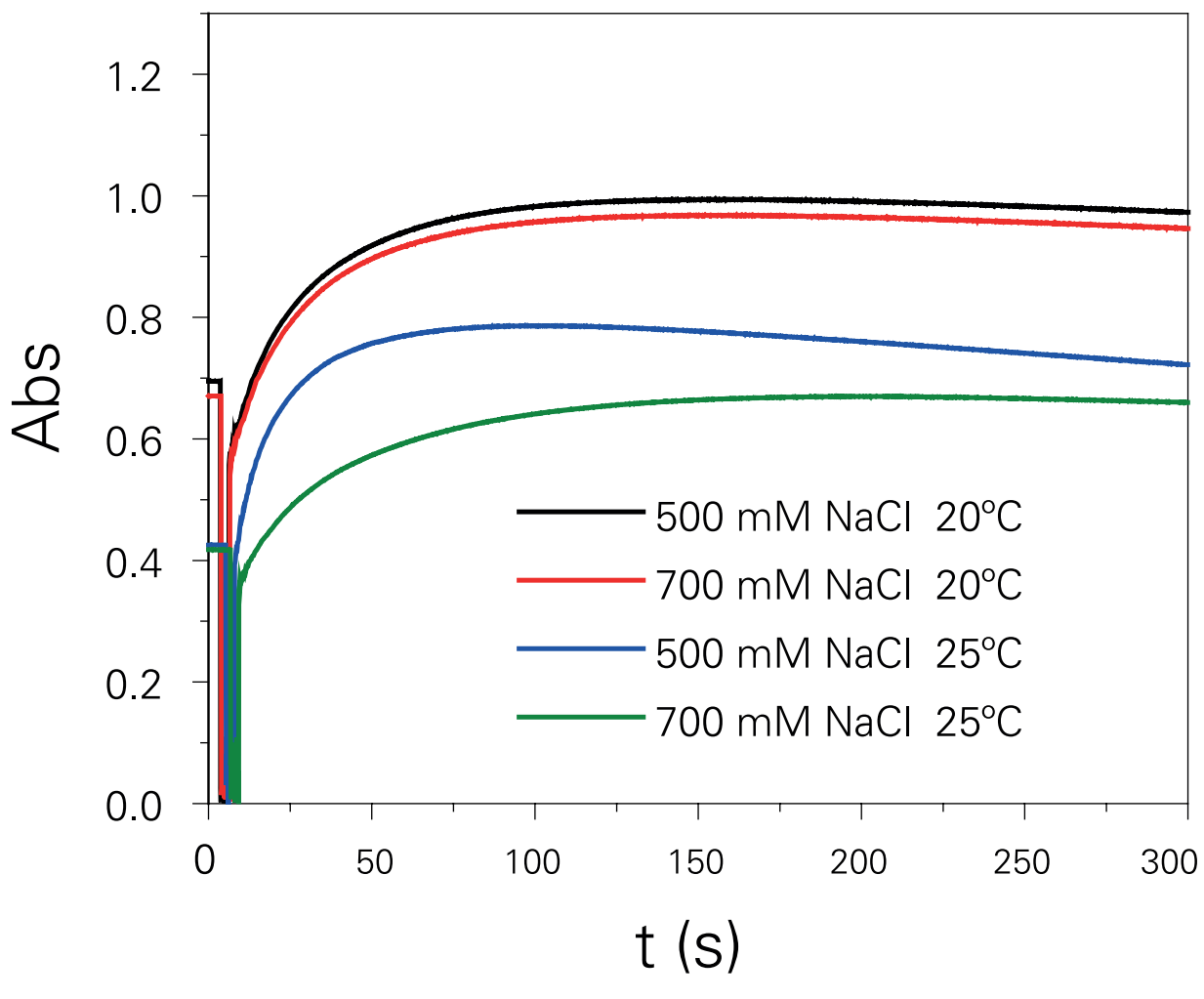

Fig. 12: 300-second variation of the absorbance for d/w emulsions of 500 and 700 mM NaCl (7.5 mM SDS, T = 20 and 25 °C).

Fig. 14: Absorbance of the neat d/w emulsions (0.5, 7.5 mM SDS, T = 20 and 25 °C) in the absence of salt (long time behavior).

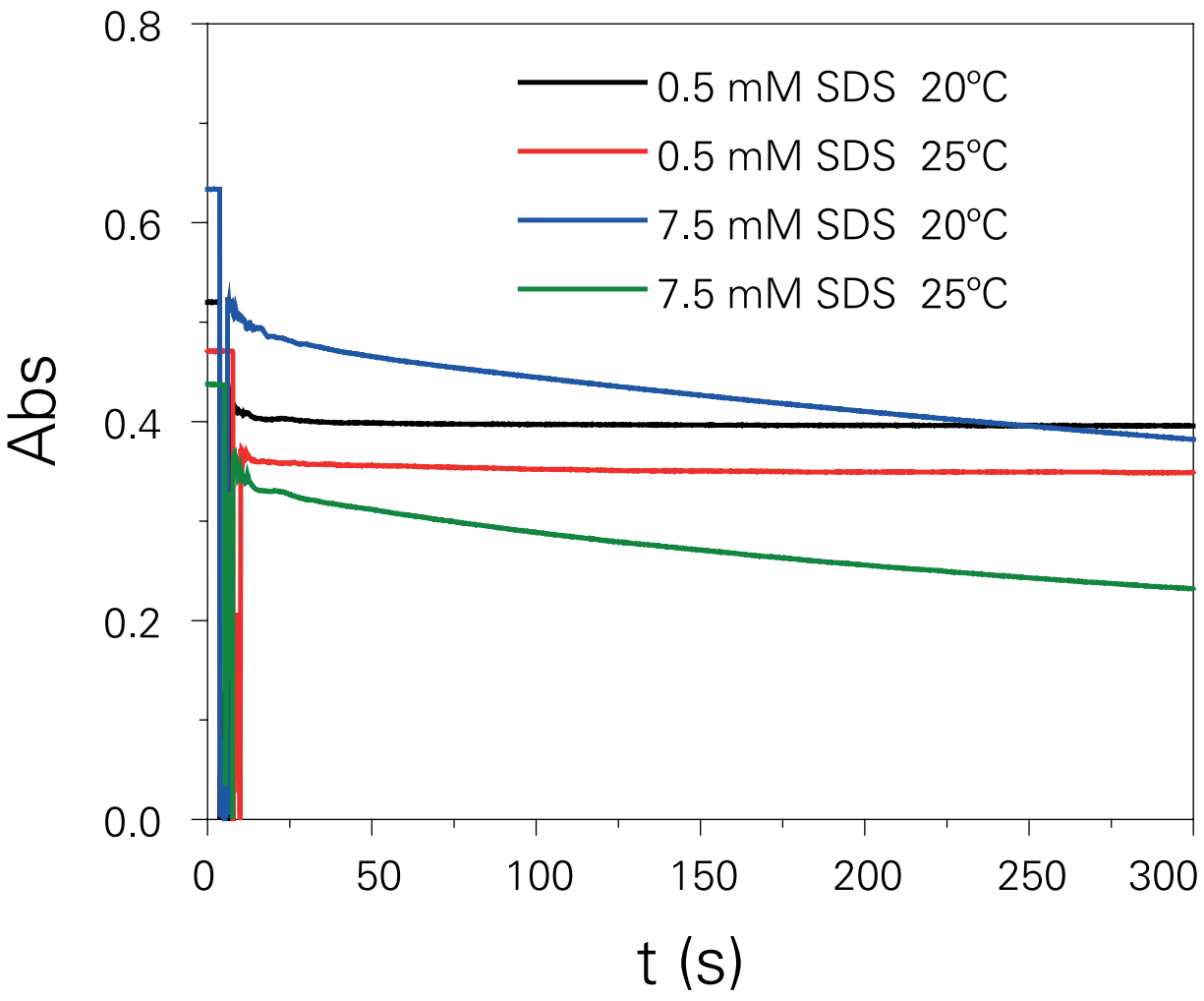

Fig. 13: 300-second variation of the absorbance for d/w emulsions at 300 mM NaCl (0.5, 7.5 mM SDS, T = 20 and 25 °C).

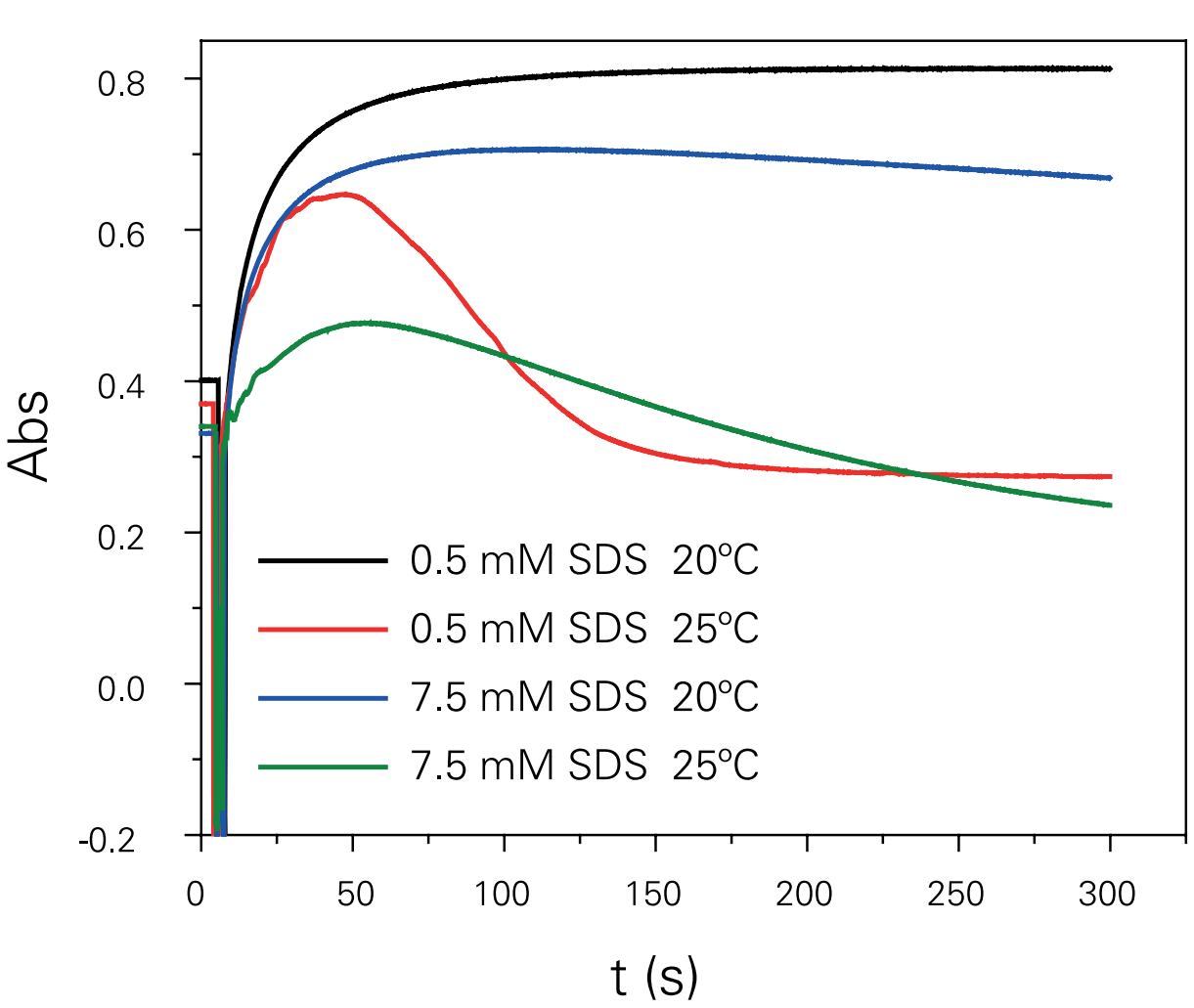

Fig. 15: Change of the absorbance for d/w emulsions at 350 mM NaCl (0.5, 7.5 mM SDS, T = 20 and 25 °C).

A significant amount of experimental effort was made in order to understand the origin of the absorbance maximum observed. In regard to creaming and despite the predictions of Eq. (10), supplementary measurements of the absorbance as a function of height (at 300 – 350 mM NaCl) were made using a Quickscan (Figure 16). The results confirmed the theoretical estimations: the velocity of creaming is too low to justify a significant clarification of the samples during 300 s. The curves are parallel to the x-axis at all times, independently of the salt concentration. Additionally, the refractive index of the micelle solutions was measured in order to discard the possible matching of this property between the aqueous solution (1.35) and dodecane (1.43). However, the index of 7.5-mM SDS solutions barely changes with the ionic strength, going from 1.35 (at 100 – 300 mM NaCl) to 1.36 (for 500 – 1000 mM NaCl). Finally, new emulsions were prepared using 10 %wt of squalene and 90 %wt of dodecane as the oil phase.

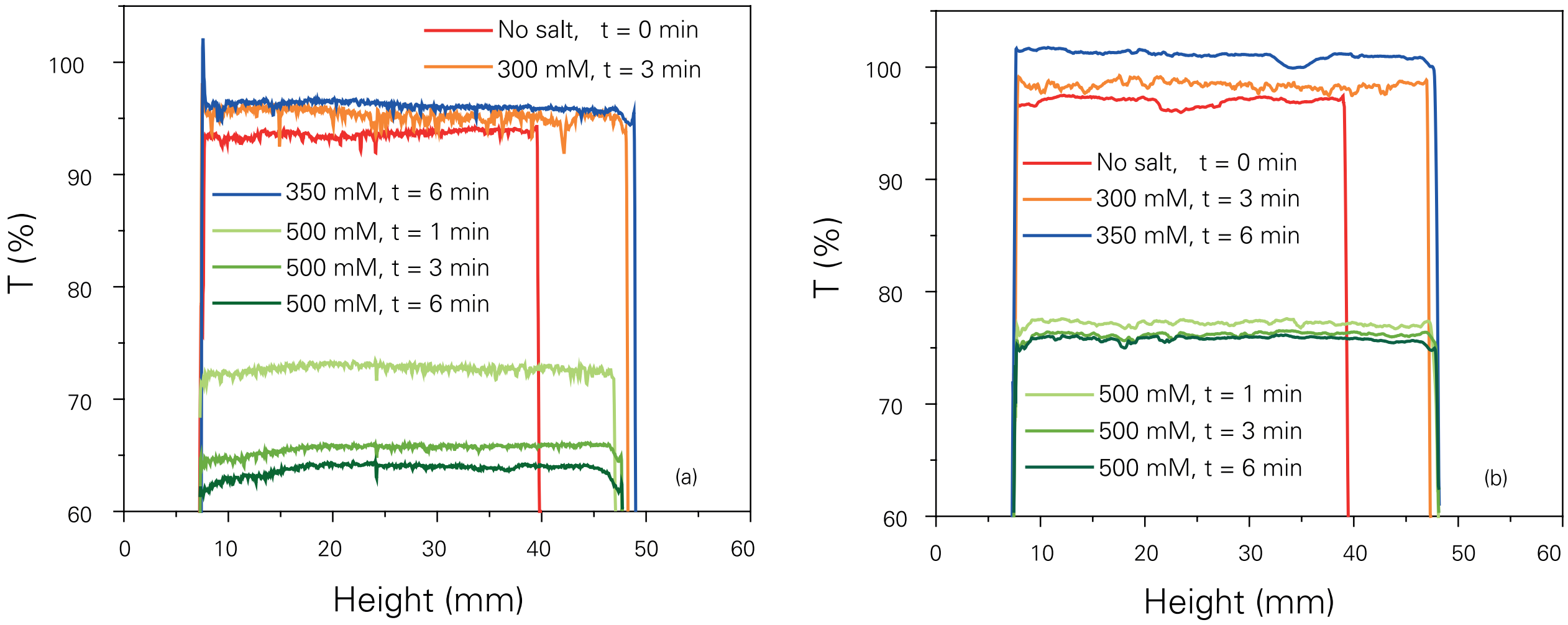

Fig. 16: Change of the absorbance as a function of height for d/w emulsions in 300, 350 and 500 mM NaCl (T = 25 °C). a) 0.5 mM SDS, b) 7.5 mM SDS.

This mixture was purposely prepared to eliminate a possible contribution of Ostwald ripening [Mendoza, 2013]. As Figure 17 shows the elimination of ripening does not prevent either the appearance of a maximum or the decrease of the absorbance beyond 100 s.

The possible occurrence of a mixed process between ripening and solubilization leads to several fundamental questions which, up to our knowledge, do not have a definite answer: Should swollen micelles be considered for the calculation of the average radius of the emulsion? , or in other words: What is the minimum size of a drop? At what size does it develop a macroscopic interfacial tension?

In any event, the present results suggest that either micellization also occurs in 0.5 mM SDS solutions for salt concentrations above 300 mM NaCl, or, a significant molecular solubilization of dodecane takes place. Based on the good agreement previously found between experimental flocculation rates (derived from the absorbance), and theoretical rates predicted by ESS for the case of 0.5 mM SDS emulsions [Urbina-Villalba, 2015], the second hypothesis seems more plausible.

Table 4 shows the flocculation rates evaluated from the 60-second increase of the absorbance. The values in red correspond to the rates of the original emulsions without salt. The data **follows the theoretical model (Eq. 3)**, suggesting that the most relevant processes of destabilization during the period of observation are flocculation and coalescence. Notice that the decrease of the drop size due to solubilization is not contemplated in Eq. (3). However, Eq. (3) has demonstrated to be flexible enough to reproduce the data of a continuous distribution of sizes with only a few points. Hence, if solubilization occurs, its effects might be concealed within the fitting coefficients. It is unlikely that crystal nucleation could be fitted with Eq. (3). This would require at least different analytical expressions for $n_k(t)$ and $R_k(t)$.

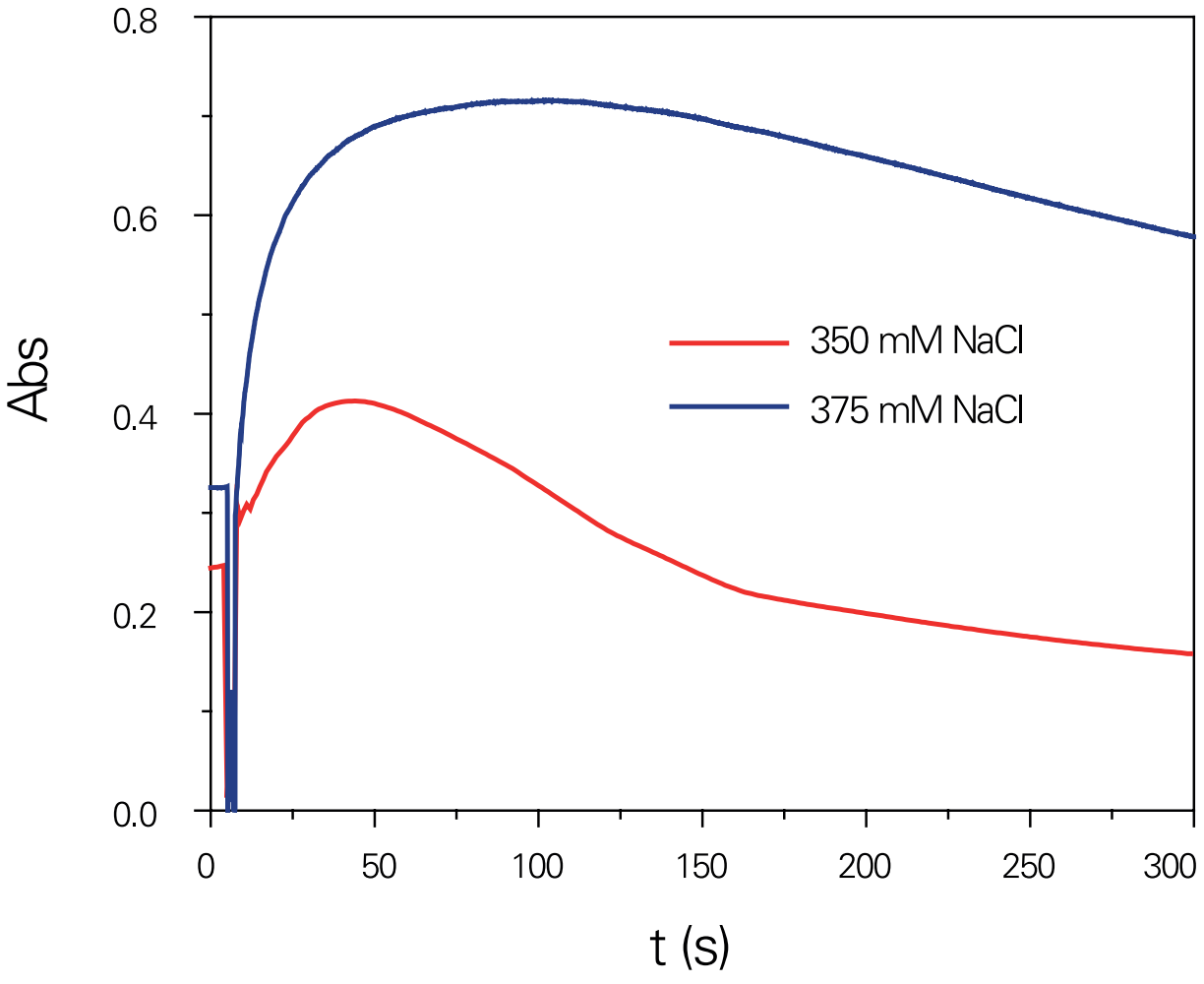

Fig. 17: Behavior of the absorbance for mixed (dodecane + squalene) emulsions at 350 and 375 mM NaCl (T = 25 C).

Despite the successful fittings, some inconsistencies of the rates stand out. According to theoretical considerations the rates are expected to: 1) be directly proportional to the temperature ($k_F = 4\,\pi\, D\, R$). The values of the rates for T = 25°C are higher than the ones of T = 20°C in 5 out of 8 cases (5/8); 2) increase with the salt concentration according to DLVO (7/14); decrease with the surfactant concentration (5/8). In any event, the values of $k_{FC}$ at 500 and 700 mM NaCl decrease when going from 0.5 to 7.5 mM SDS, as experimentally found for the case of hexadecane-in-water emulsions [Urbina-Villalba, 2015].

Table 4: Flocculation rates evaluated from the 60-s variation of the absorbance as a function of time.

| [NaCl] (mM) | [SDS] (mM) | Temperature (°C) | $x_a$ | $k_{FC}$ ($m^3/s$) |
|---|---|---|---|---|
| 0 | 0.5 | 25 | 0.60 | (7.6 ± 0.5) x $10^{-22}$ |
| 350 | 0.5 | 25 | 0.79 | (1.5 ± 0.5) x $10^{-18}$ |
| 375 | 0.5 | 25 | 0.80 | (1.04 ± 0.07) x $10^{-18}$ |
| 500 | 0.5 | 25 | 0.88 | (1.0 ± 0.3) x $10^{-18}$ |
| 700 | 0.5 | 25 | 0.81 | (1.0 ± 0.2) x $10^{-18}$ |
| 0 | 7.5 | 25 | 0.03 | (2.9 ± 0.5) x $10^{-22}$ |
| 350 | 7.5 | 25 | 1.22 | (3 ± 2) x $10^{-19}$ |
| 375 | 7.5 | 25 | 0.80 | (1.27 ± 0.01) x $10^{-18}$ |
| 500 | 7.5 | 25 | 0.89 | (8 ± 2) x $10^{-19}$ |
| 700 | 7.5 | 25 | 0.90 | (1.9 ± 0.4) x $10^{-19}$ |
| 0 | 0.5 | 20 | 0.59 | (5.1 ± 0.5) x $10^{-22}$ |
| 350 | 0.5 | 20 | 0.83 | (1.3 ± 0.2) x $10^{-18}$ |
| 500 | 0.5 | 20 | 0.87 | (1.1 ± 0.1) x $10^{-18}$ |
| 700 | 0.5 | 20 | 0.85 | (8.0 ± 0.2) x $10^{-19}$ |
| 0 | 7.5 | 20 | 1.09 | (2.6 ± 0.5) x $10^{-22}$ |
| 350 | 7.5 | 20 | 0.79 | (1.56 ± 0.07) x $10^{-18}$ |
| 500 | 7.5 | 20 | 0.93 | (4.4 ± 0.5) x $10^{-19}$ |
| 700 | 7.5 | 20 | 0.92 | (4.5 ± 0.4) x $10^{-19}$ |

### 6.2 Simulations

The algorithm employed to mimic the solubilization of the drops assumes that this process only occurs above the CMC. The influence of salt on the diminution of the CMC is considered, as well as the reduction of the electrostatic potential due to the screening of the surface charges. However, the solubilization rates used in the simulations are the ones previously reported in the bibliography for d/w and h/w emulsions *in the absence of salt*. Hence they do not take into account the observed decrease of the solubilization rate.

According to previous estimations a surfactant concentration of 0.5 mM SDS is lower than the CMC even at 1 M NaCl. Hence, only the highest surfactant concentration of the experiments was considered (7.5 mM SDS) for the computations. Moreover, Figure 2 shows that a salt concentration of 700 mM NaCl already induces the formation of crystals at 25 °C, but this additional process cannot be easily incorporated into the ESS code. Therefore, 25-particle simulations were limited to 300 and 500 mM NaCl. In all calculations the initial position of the drops is the same.

According to the experiment (Fig. 2), it requires between 30 and 45 minutes to completely solubilize a d/w emulsion of $\phi = 3.2 \times 10^{-4}$ at 300 mM NaCl. Curiously, the same process takes between 2 and 5 hours in 500 mM NaCl. Figure 18a illustrates the effect of micelle solubilization on a d/w emulsion using *fixed* particle positions (immobile drops). A large time step of $7.8 \times 10^{-4}$ s can be used in this case in order to reach 30 minutes of simulation. If it is assumed that only two molecules of oil can be dissolved per micelle, the radii of the drops diminish from 72.5 to 67.3 within the first two minutes, and then remain constant due to the saturation of the micellar solution (Figure 19a). Moreover: if the number of surfactant molecules per micelle decreases, the solubilization capacity of the solution increases as well as the solubilization time. However, it was shown above that the addition of salt increases the solubilization capacity of the micellar solution. If it is assumed that complete solubilization occurs (*at the same solubilization rate*), the radii of the drops progressively decreases and the emulsion totally disappear after 29 minutes (Fig. 18b).

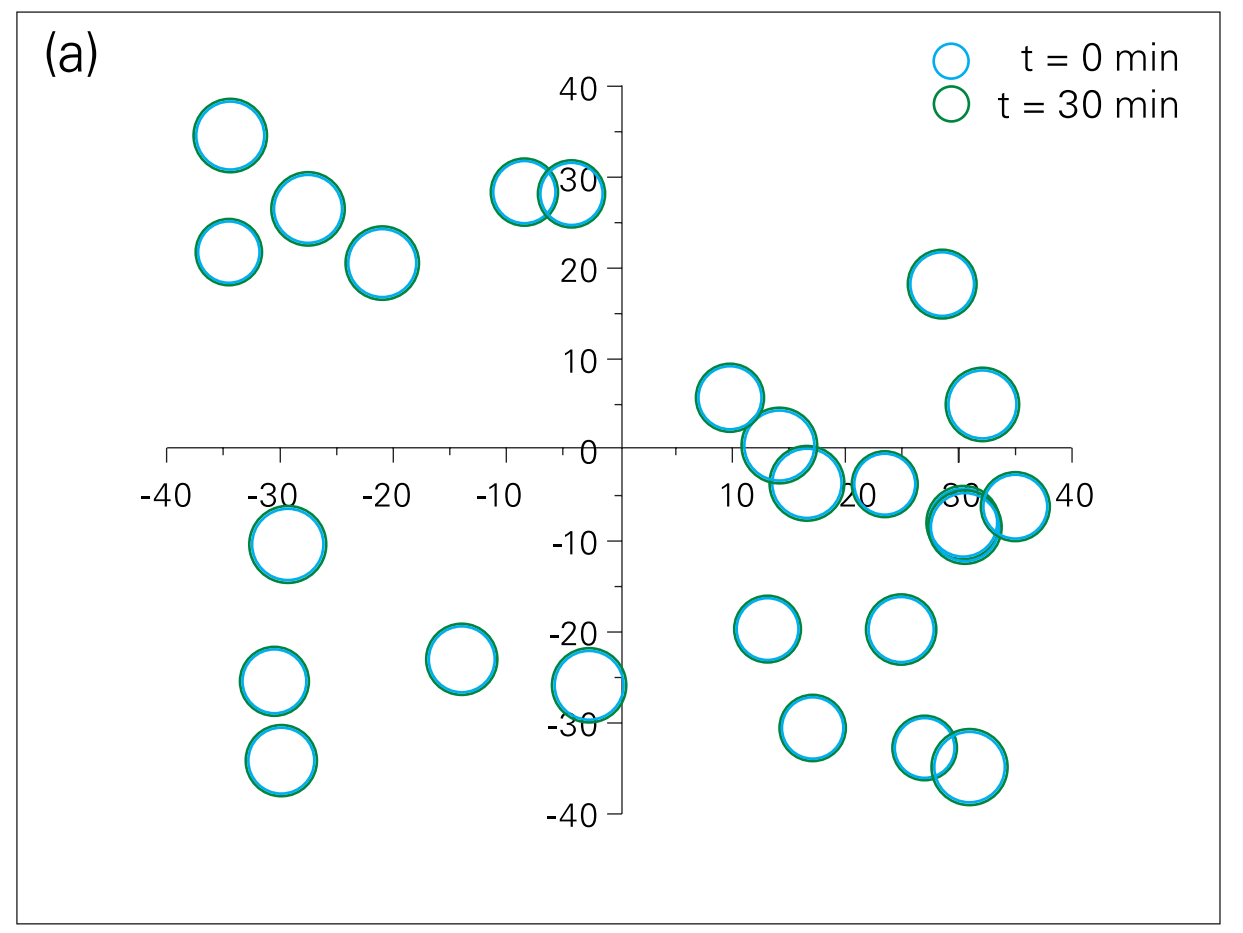

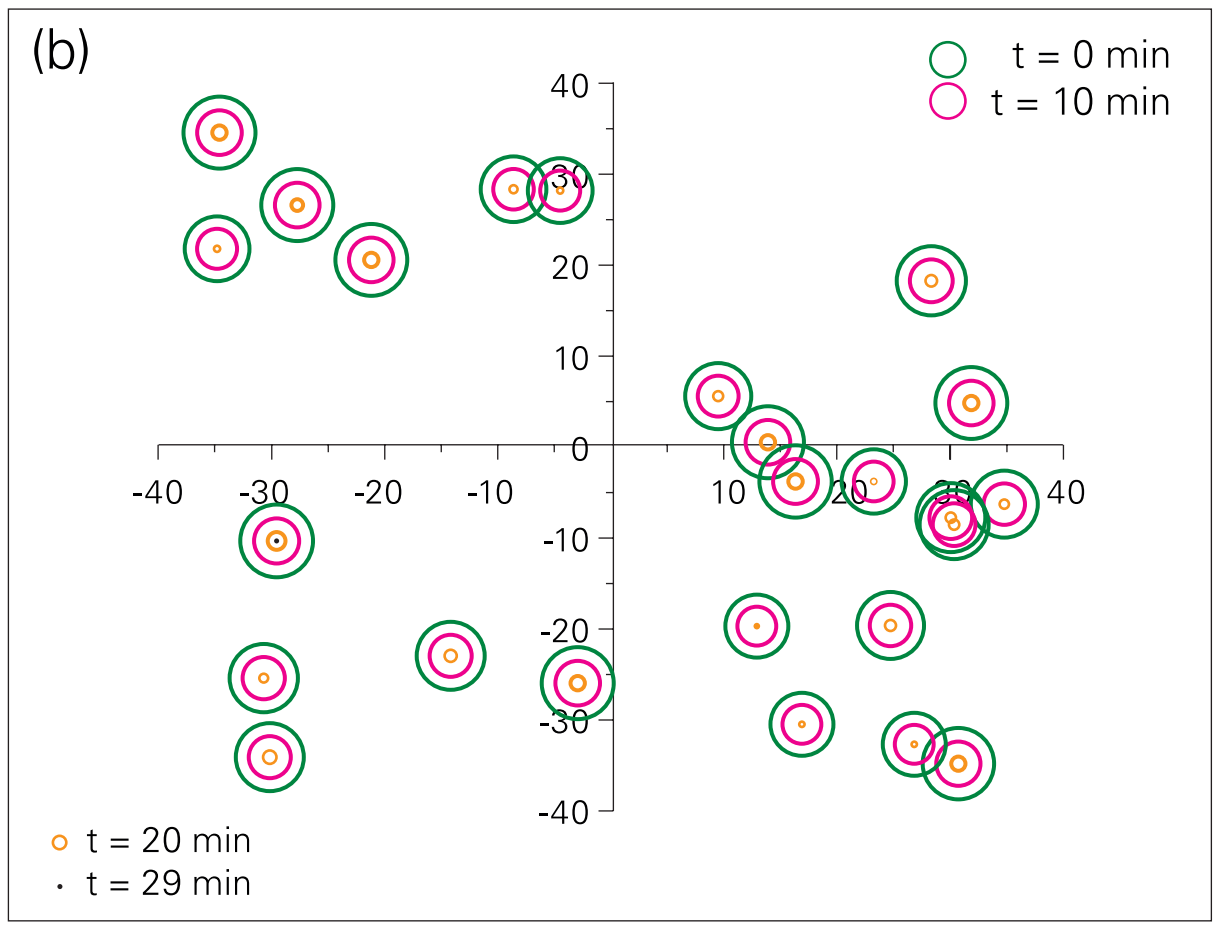

Fig. 18: Solubilization of a dodecane-in-water emulsion: (a) saturation of the micellar solution; (b) infinite solubility (complete solubilization).

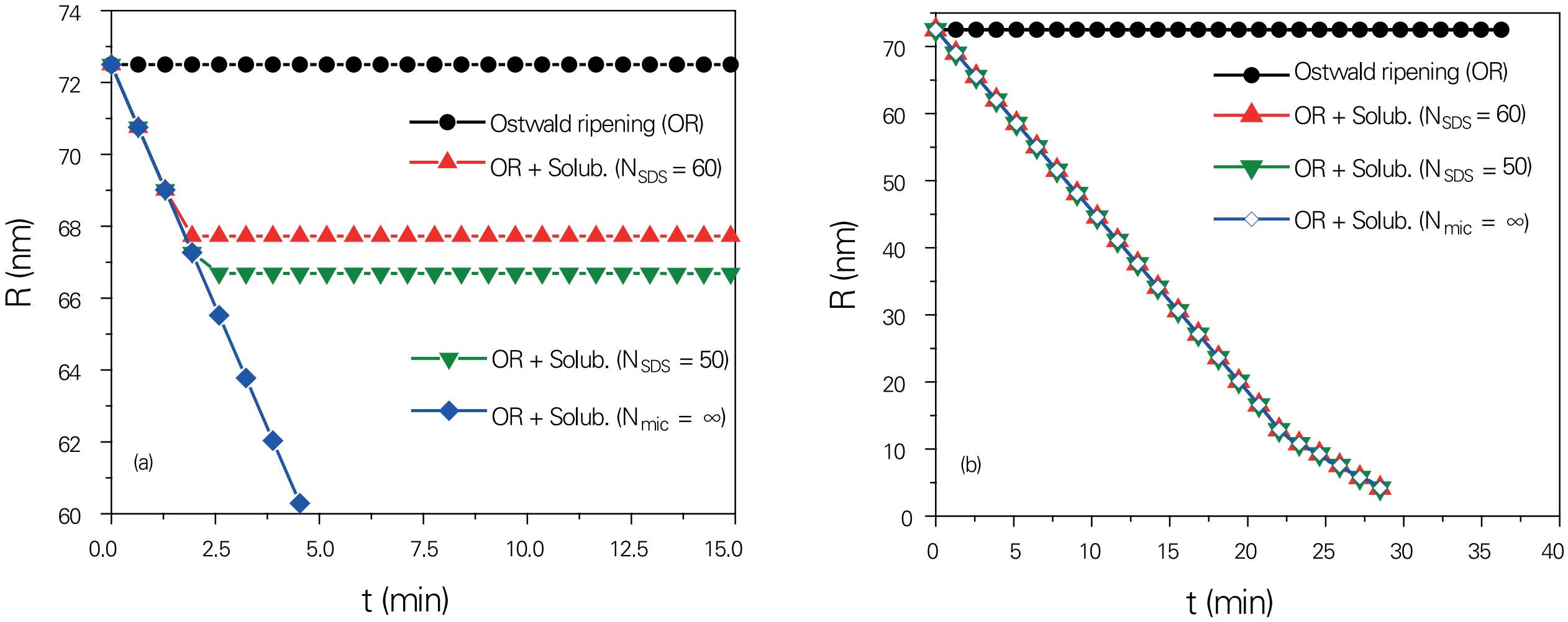

Fig. 19: Predictions of *R vs. t* for d/w emulsions resulting from the simulation of 25 immobile drops. In these calculations the drops are fixed at their initial positions. $N_{SDS}$ stands for the number of surfactant molecules in a micelle. $N_{mic}$ is equal to the number micelles. In order to simulate an infinite solubility, an extremely high number of micelles was implemented. (a) It is assumed that two (2) molecules of dodecane are dissolved per micelle as in the case of hexadecane. (b) 25 molecules dissolved per micelle.

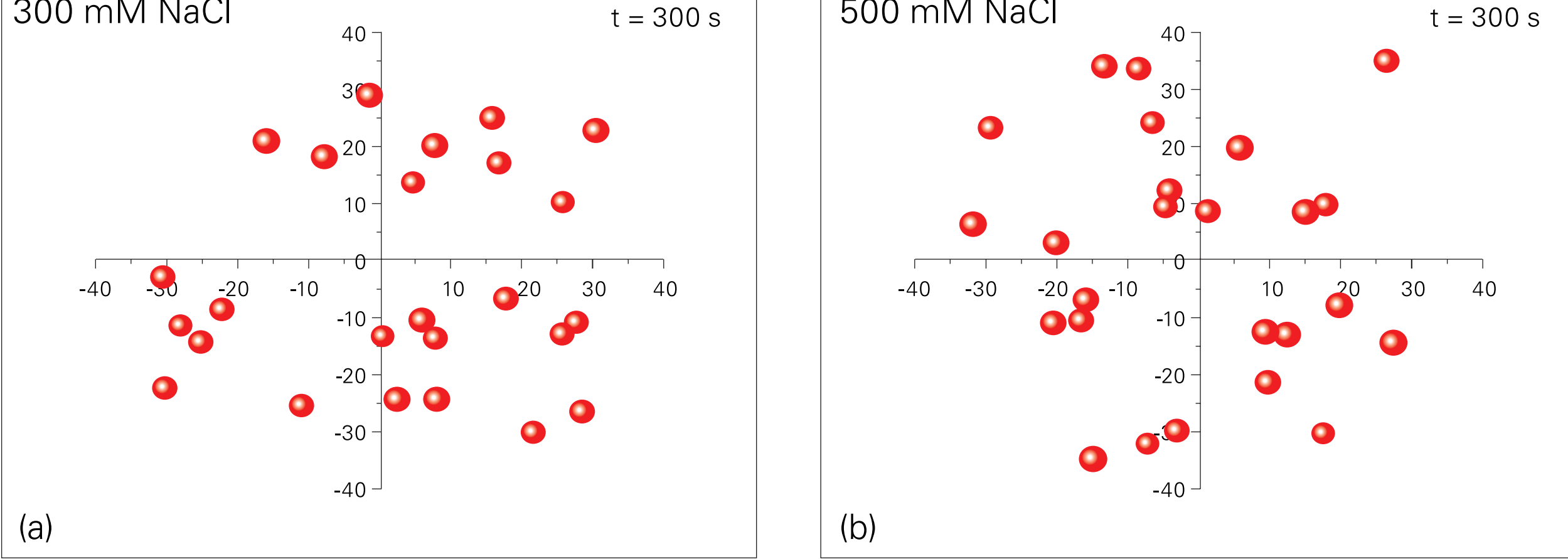

Fig. 20: Final state of the simulations corresponding to a d/w emulsion with 7.5 mM SDS in (a) 300 and (b) 500 mM NaCl. T-II interaction potential.

Complete solubilization may occur if the volume fraction of oil is lower than $\phi_c$ or the number of oil molecules per micelle ($N^{oil}_{max}$) is higher. According to Ariyaprakai [2008], this number changes in the case of dodecane between 25 and 54 for SDS concentrations between 35 and 139 mM, respectively. Figure 19b illustrates the predictions of the simulations for $N^{oil}_{max} = 25$: the emulsion is completely solubilized after 29 minutes. The change of slope at t ~ 22 min marks the beginning of a sensible contribution of Ostwald ripening. At shorter times, Ostwald ripening decreases the average radius of the emulsion in a negligible amount (the negative slope cannot

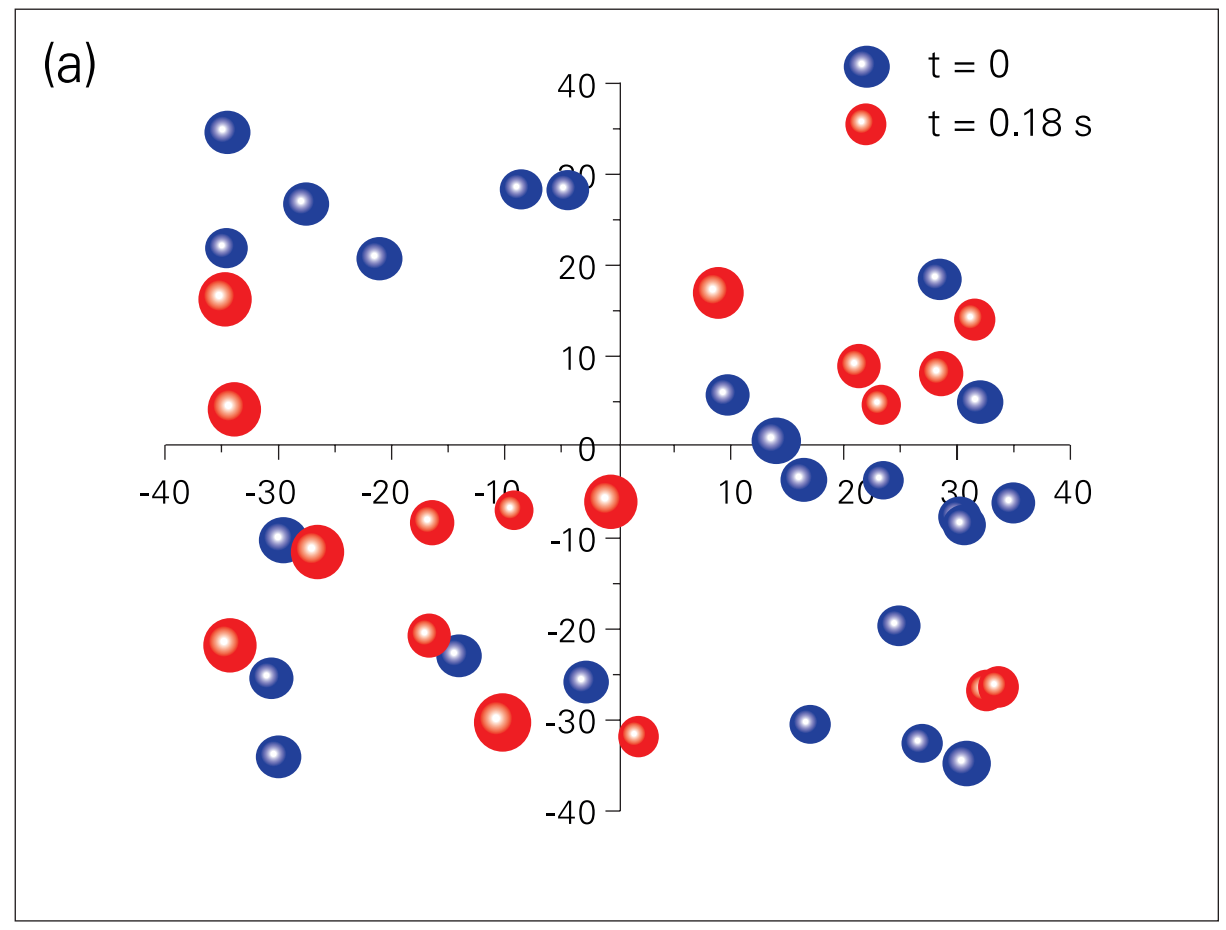

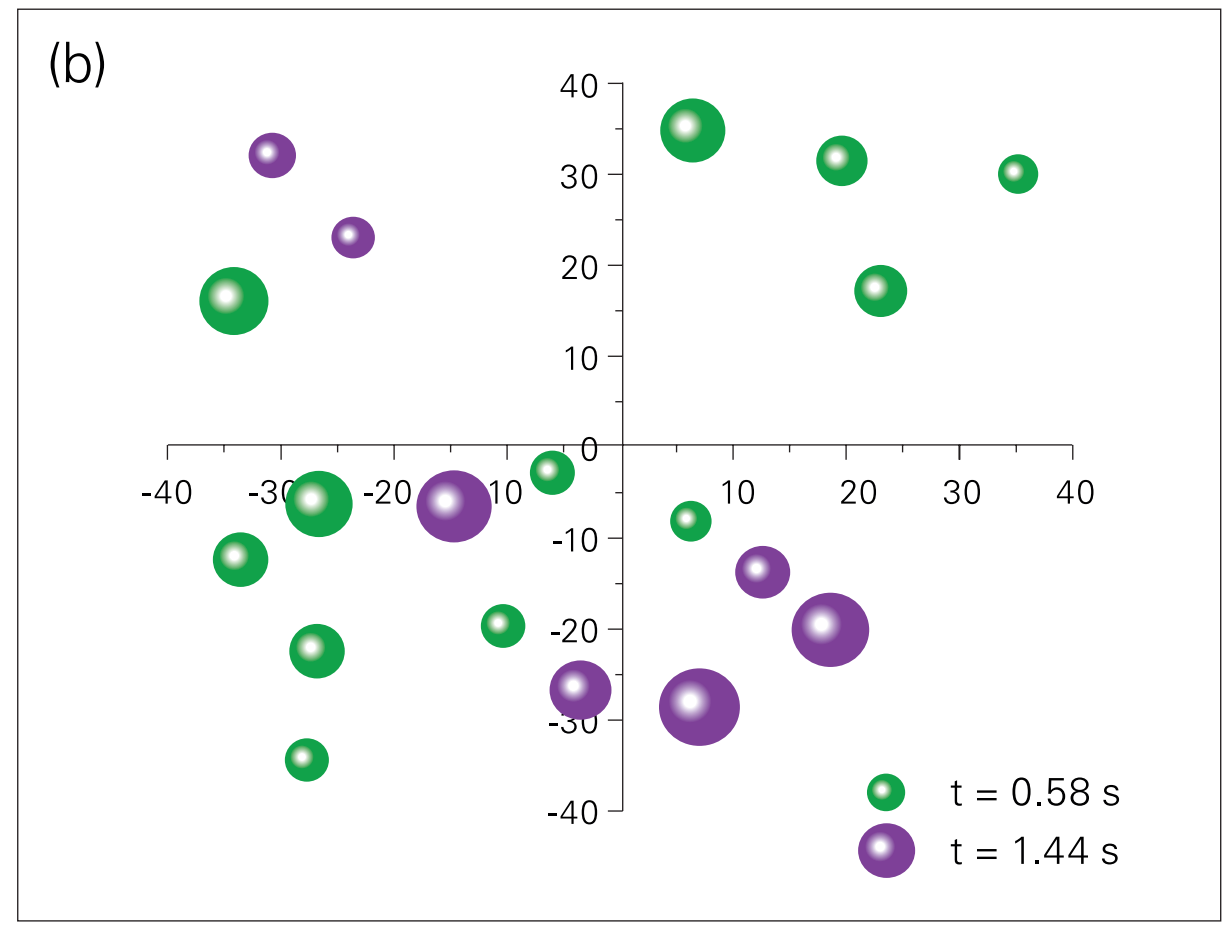

Fig. 21: T-I prediction for the short time evolution of a d/w emulsion during (a) 0, 0.18 s, (b) 0.58, 1.44 s, after the addition of brine (500 mM NaCl, 7.5 mM SDS).

be appreciated from inspection of the top curve in Figures 19a and 19b).

Since the solubilization rate is constant for each oil, the use of 2 or 25 molecules per micelle does not change drastically the outcome of the simulations during the first 5 minutes. In the former case, the radius of the drops decreases to 68 nm, and in the latter case it reaches 59 nm (Figure 19). Therefore, a value of $N^{oil}_{max} = 2$ was implemented in all calculations in order to highlight the effect of the solubilization *rate*, which differs significantly between dodecane and hexadecane. However, the long time prediction of stability of dodecane-in-water emulsions severely changes if $N^{oil}_{max}$ is equal to 2 or 25, due to the appreciable solubility of the oil in the micellar solution. In the first case the emulsion remains stable in 300 mM NaCl (at 7.5 mM SDS). In the second case, it is completely solubilized as experimentally observed: the drops disappear and the emulsion becomes transparent between 20 and 45 minutes (Figure 2).

Figure 20 exemplify the interplay between particle movement and micelle solubilization. When particles move the effect of the interaction potential on flocculation and coalescence is decisive. High repulsive barriers (as predicted by Type II parameterization) preserve the stability of the system at both 300 and 500 mM NaCl (Figures 20ab). The radii of the drops slightly decrease at these concentrations but no significant aggregation is observed. Hence, the decrease of the absorbance found for 300 mM NaCl during 300 s could possibly be justified, but its increase for 500 mM NaCl cannot. On the other hand, low repulsive barriers (Type I parameterization) induce flocculation and coalescence. In this case, the number of drops per unit volume diminishes. The average radius of the remaining drops increases due to coalescence but decreases due to solubilization. According to the simulations, coalescence predominates at short times for 500 mM NaCl (Figure 21ab). For 0, 0.58, 1.44, 3.49 and 17.1 s, the average radius (in units of $R_0$ = 72.5 nm) augments from 1.0 (at t = 0) to: 1.23, 1.45, 1.75, and 2.92. Only one drop remains after 17 s (38 s in the case of 300 mM NaCl), Figure 22. The decrease of the radii due to the solubilization of the drops is concealed by the coalescence process. Thus, the behavior of the absorbance depends on the prevalence of two opposite phenomena. The absorbance decreases due to a reduction in the number of drops per unit volume, but it also increases due to the larger size of the remaining drops. Additionally, it decreases further due to the solubilization of the drops. This explanation justifies the existence of a maximum absorption for 350 and 375 mM NaCl. More importantly: it also rationalizes the existence of the maximum at sub-micellar concentrations (0.5 mM SDS).

Notice that the number of particles does not change with the aggregation of the drops. The competition between aggregation and solubilization might also produce a maximum of absorbance, but an enhancement of solubilization in the absence of micelles (at 0.5 mM SDS) is difficult to justify. Hence, the occurrence of coalescence seems to be a necessary condition for the appearance of the maximum in this case. Unfortunately, this hypothesis is difficult to prove.

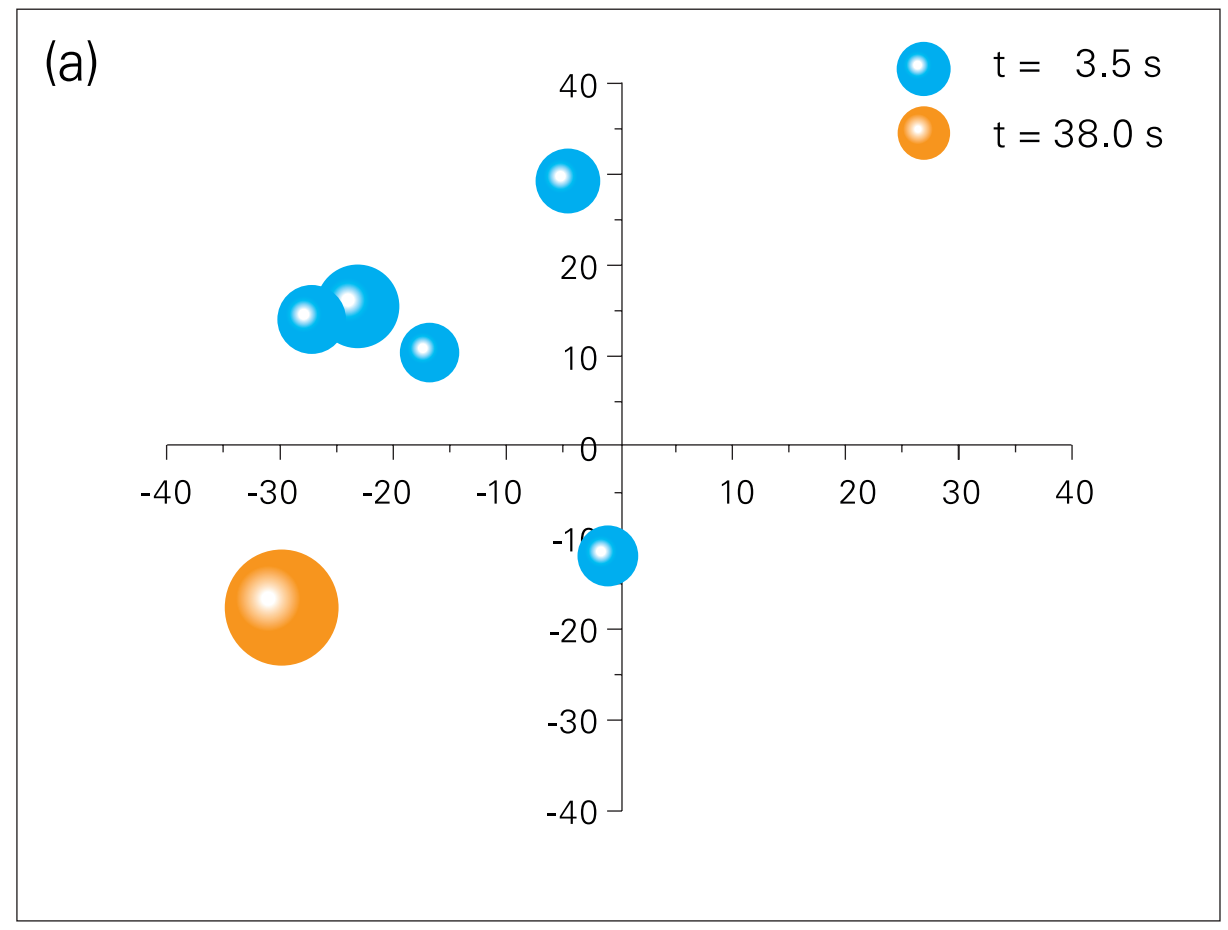

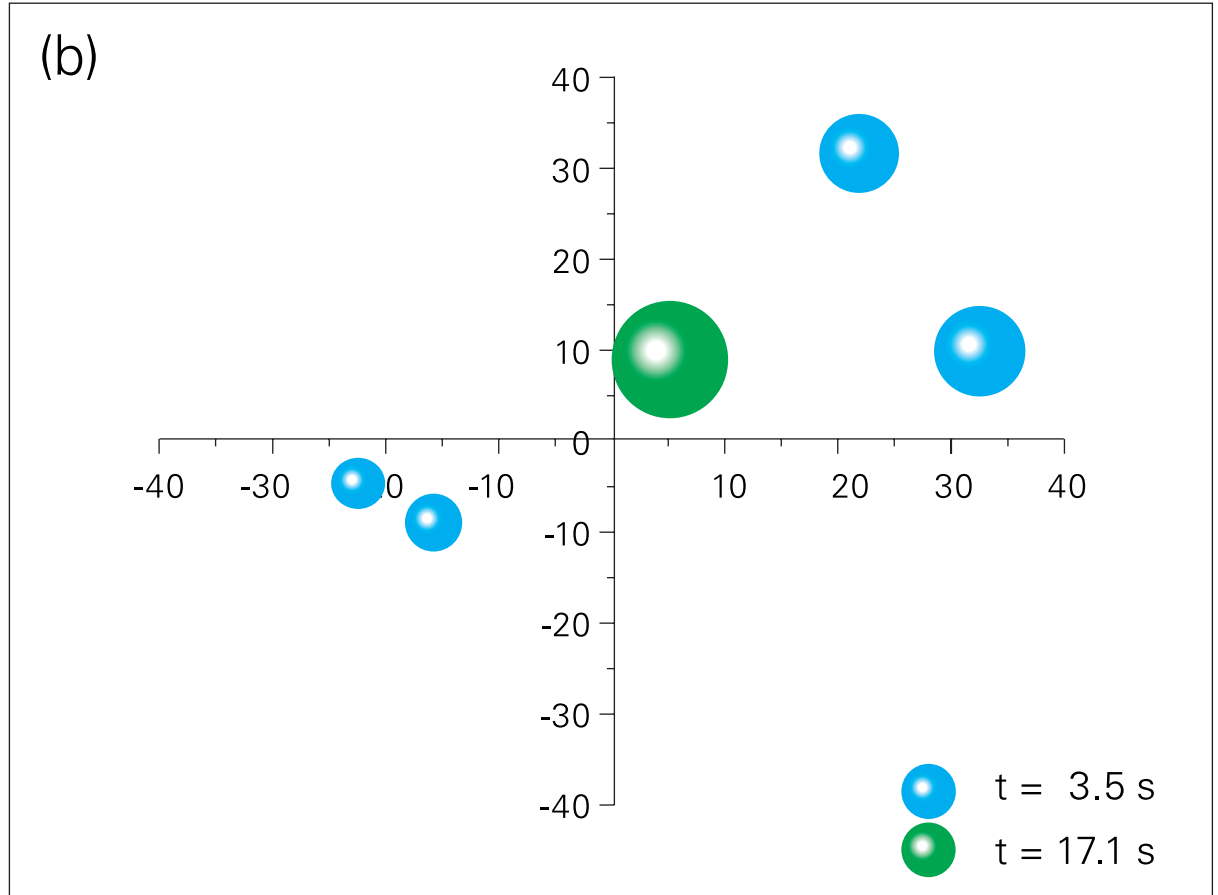

Fig. 22: Final state of the d/w emulsions according to T-I potential for 300 and 500 mM NaCl (7.5 mM SDS).

Most modifications of the system perturb several variables at once. Possibly the variation of the volume fraction of oil to increase the number of particles, or the addition of a non ionic surfactant to prevent coalescence, could be of help.

Figure 23 shows the results of using a set of doublets and one long simulation in order to approximate the qualitative behavior of $k_{FC}$ vs. [NaCl] using T-II potential (Eq. (12)). The figure outlines a technical problem of this type of approach: when the barriers are very high it is uncertain if $k_{FC}^{fast}$ should be approximated using only the van der Waals interaction (lower curves in the graph, parallel to the x axis), or the complete DLVO potential at very high salt concentration. According to Eq. (12), $k_{FC}^{fast}$ sets up the order of magnitude of $k_{FC}$, while the values of W establish the relative behavior of the rest of the simulations with respect to the reference system (the one chosen to appraise $k_{FC}^{fast}$ ). When the repulsive barrier between the drops is very high, a large salt concentration does not eliminate it. Hence, the flocculation rate obtained is an average aggregation rate for reversible secondary minimum flocculation ($k_{FC}^{fast}$ = 2.1 x $10^{-18}$ $m^3$/s, $r^2$ = 0.95578). The one calculated with the van der Waals potential is always substantially faster ($k_{FC}^{fast}$ = 7.1 x $10^{-18}$ $m^3$/s, $r^2$ = 0.9971). It is a theoretical limit predicted for primary minimum flocculation. According to Figure 23 only the values deduced for [NaCl] = 1000 mM approximate the experimental data. In any event, the main features of the experimental curve are not reproduced. The situation does not improve if a larger number of sampling events, $N_{max}$, is increased from 100 to 1000 (see [Urbina-Villalba, 2016] for details). It does not get better either if instead of using the viscosity of water, the viscosity of the micellar solution in brine is employed (9.4466 x $10^{-4}$ Pas at 120 mM NaCl, T = 25 °C [Kushner, 1952], see also [Gamboa, 1986]): $k_{FC}^{fast}$ = 6.9 x $10^{-18}$ $m^3$/s, $r^2$ = 0.9983. The possible generation of a depletion potential due to the presence of micelles [Danov, 1994] was disregarded after plotting the interaction potential, due to its negligible contribution at these dilute concentrations. Finally, the time step of the simulation was lowered from 7.9 x $10^{-8}$ s to 4.9 x $10^{-9}$ s in order to appraise its effect on the evaluation of $k_{FC}^{fast}$. A faster (even more unfavorable) value was obtained: 7.92 x $10^{-18}$ $m^3$/s ($r^2$ = 0.9978).

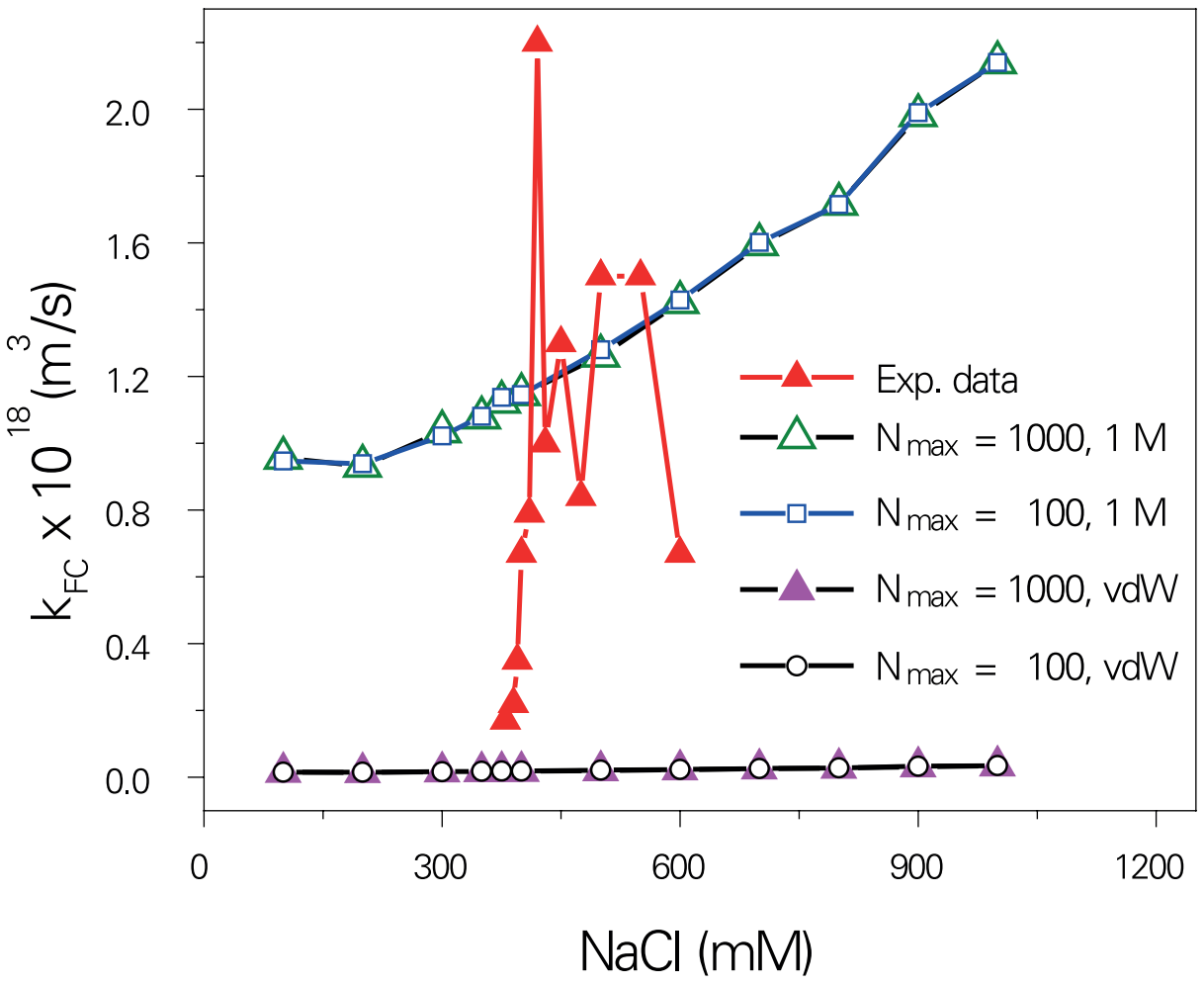

Fig. 23: Prediction of $k_{FC}$ vs. NaCl according to potential Type II in the case of non-deformable drops (see text for details).

Figure 24 presents the results of the calculation of doublets for the evaluation of $k_{FC}$ vs. [NaCl] using T-I potential

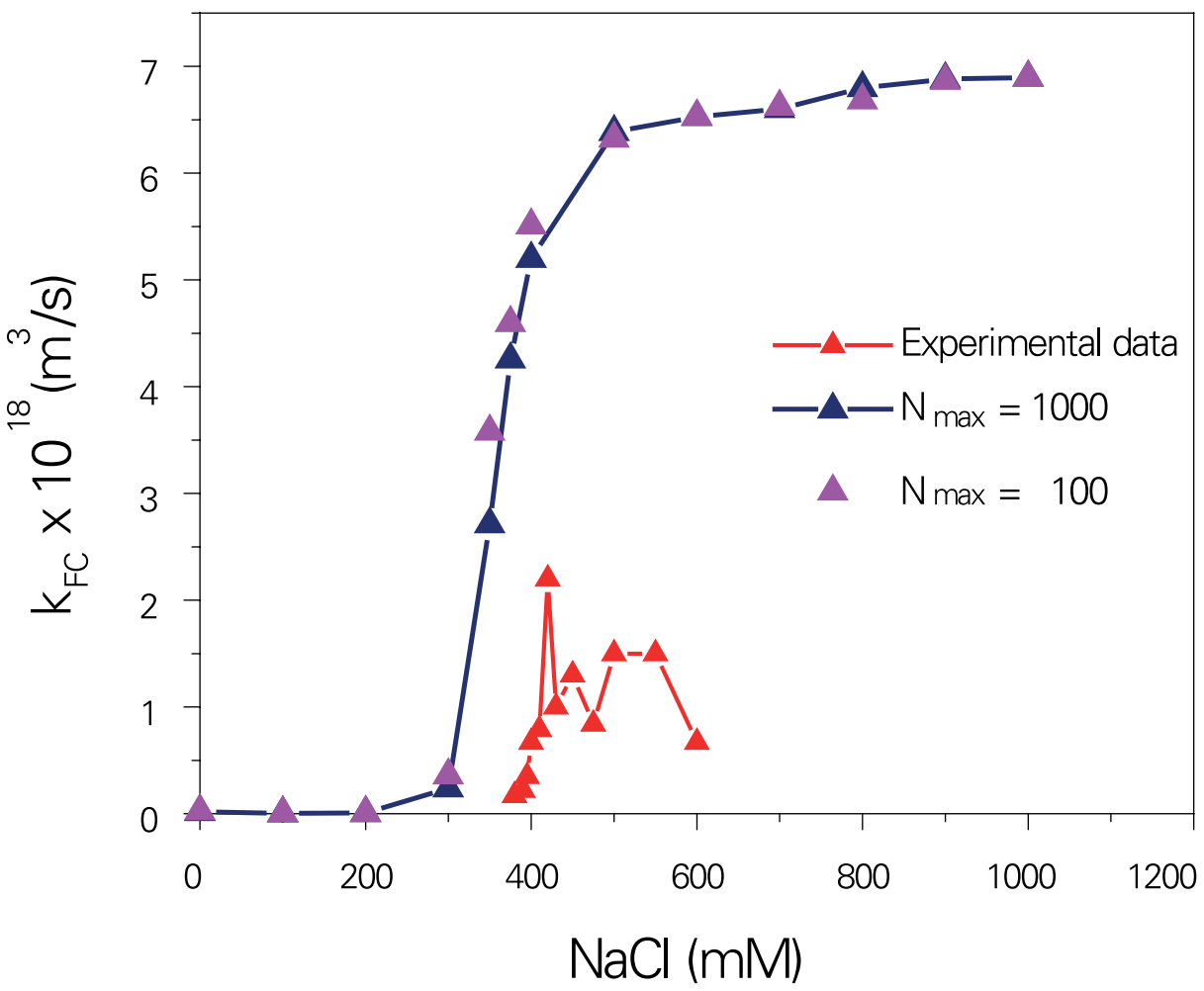

Fig. 24: Prediction of $k_{FC}$ vs. NaCl according to potential Type I in the case of non-deformable drops (see text for details).

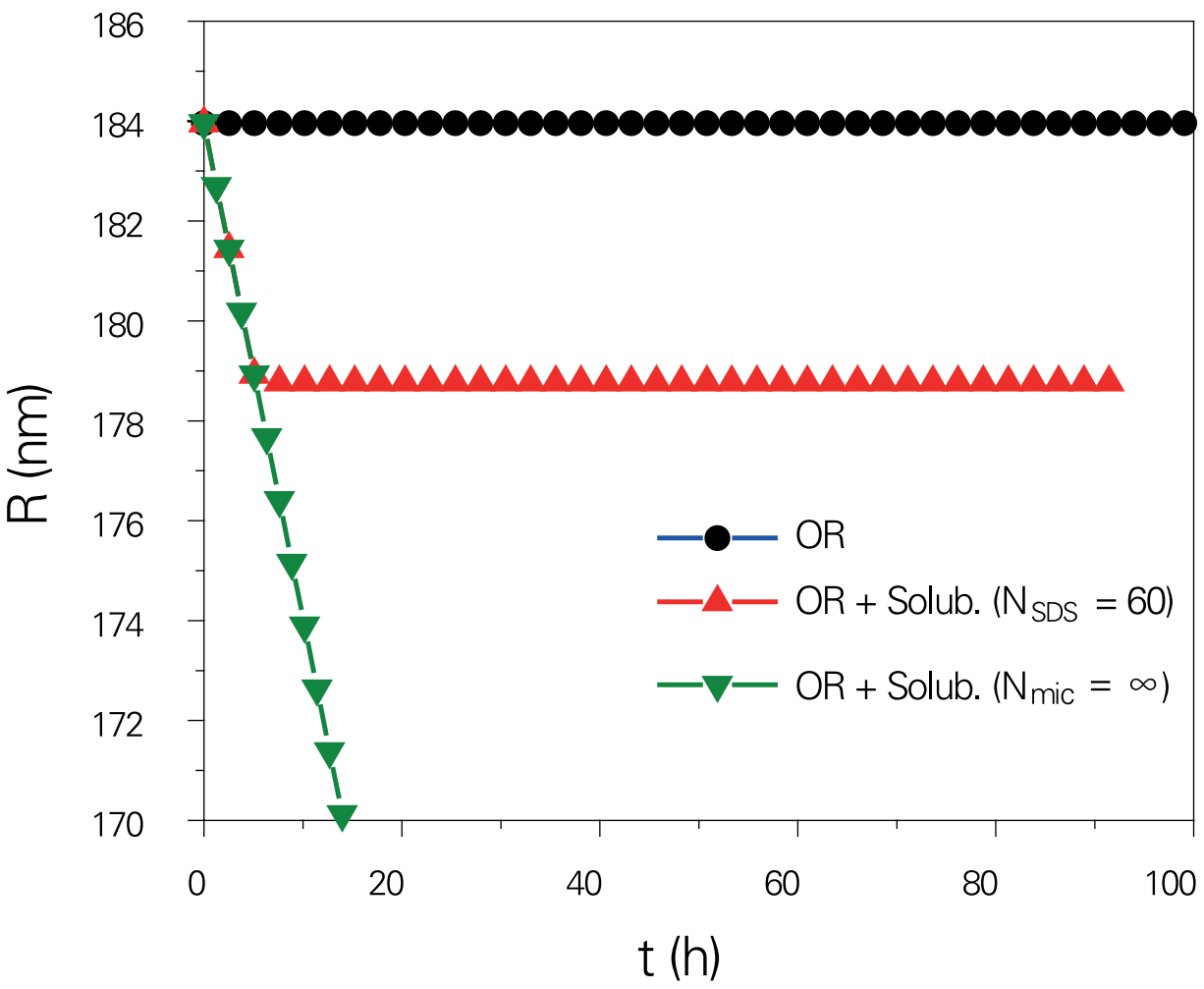

Fig. 25: Change of $R$ vs. $t$ for h/w emulsions. Results from the simulation of 25 immobile drops.

(Eq. (12)). In this case, a high salinity eliminates the potential barrier so there is no difference in calculating either at 1000 mM NaCl or using the van der Waals potential exclusively ($k_{FC}$ = 6.9 x $10^{-18}$ $m^3$/s, $r^2$ = 0.9956). The simulations roughly reproduce the qualitative trend of the experiments, but the values of the flocculation rate are now substantially higher at all salt concentrations. Again, the results do not improve if the time step of the long simulation is lowered ($k_{FC}$ = 7.5 x $10^{-18}$ $m^3$/s, $r^2$ = 0.9977), or the viscosity of the micellar solution is considered ($k_{FC}$ = 7.2 x $10^{-18}$ $m^3$/s, $r^2$ = 0.9947). Contrary to our expectations, the value of $k_{FC}$ increases in the latter case. This remarks the necessity to consider the average error of the calculations which was roughly estimated previously to be approximately one unit of the respective order of magnitude of $k_{FC}$ [Urbina-Villalba, 2005].

Notice that the value of $k_{FC}^{fast}$ could decrease if the rate of solubilization increases. As the size of the drops decreases their diffusion constant increases but their collision frequency decreases.

In summary: the interaction potential corresponding to macroscopic adsorption isotherms (T-I) largely overestimates the flocculation rate, but predicts a qualitative behavior of the systems which is consistent with the experimental evidence. Moreover, it also reproduces the qualitative dependence of $k_{FC}$ vs. [NaCl]. Instead, potential T-II replicates the average value of the aggregation rates, but predicts non-flocculating emulsions at 300 and 500 mM NaCl for at least 300 s. This is incompatible with the variation of the absorbance as a function of time.

It must be remarked at this point that the use of doublets for the appraisal of $k_{FC}$ heavily relies on the long many-particle simulation needed. This calculation spans only a few seconds of actual time. Hence, the new routine of solubilization does not alter significantly the evaluation of this rate. Moreover, it does not change the values of W at all, since the two-particle simulations start from an initial distance of approach of only a few nanometers.

Figure 25 illustrates the rate of change of the average radius predicted for a hexadecane-in-water emulsion. While the radius of a d/w emulsion decreases 4 nm in 2 minutes, the same variation takes 3.8 hours in the case of h/w. In fact, it takes 190 h for the micelle solution to dissolve all the drops under favorable (infinite solubilization) conditions. The solubility of hexadecane is almost three orders of magnitude lower than the one of dodecane (9.6 x $10^{-9}$ vs. 3.5 x $10^{-6}$), and its solubilization rate differs in two orders of magnitude (1.92 x $10^{-13}$ m/s vs. 4.49 x $10^{-11}$ m/s). As a consequence, the solubilization of hexadecane during the time of appraisal of the flocculation rate is expected to be negligible.

Figure 26 and 27 illustrate the evolution of the h/w system (7.5 mM SDS) as a function of the salt concentration. These figures show some features of the Smoluchowskian scheme of aggregation but they do not reproduce it, because only secondary minimum flocculation is possible due to the magnitude of the repulsive barriers (T-II potential). At 300 mM NaCl, there is an eventual formation of doublets (t = 30, 60 and 120 s), and small aggregates of at most four particles, but all of them ultimately break. A similar situation occurs at 350 mM NaCl, except for the fact that larger aggregates of 3 and 4 particles are observed since t = 30 s. At 375 mM NaCl, extended aggregates of 5 – 7 particles are present during the whole simulation, along with a considerable number of primary drops. The aggregates do not necessarily increase their size progressively as predicted by Smoluchowski: from doublets to triplets, etc. At 400 mM NaCl, the number of free particles decreases significantly with respect to the one of previous salinities. Most of them accumulate into large globular clusters after 90 s. In this case, the insets corresponding to t = 30, 60 and 90 s, nicely illustrate a Smoluchowkian type of flocculation: aggregates of 2, 3, and 4 particles are formed and progressively grow mainly due to their collision with primary drops. The same situation occurs for 500 – 900 mM between 30 and 90 seconds. Afterwards, only a few large globular clusters remain. The evolution of these systems between 180 and 240 s nicely illustrates the conclusion of our previous paper [Urbina-Villalba, 2015]: due to the reversibility of the secondary minimum flocculation, aggregates form and decompose at all times. Hence, the rate of change of the total number of aggregates as a function of time is substantially slower than in the case of irreversible aggregation. The system behaves apparently as if there is a repulsive barrier hindering flocculation, but its behavior exemplifies the effect of the secondary minimum on the aggregation rate. Deeper minima favor stable aggregates and faster flocculation rates. For [NaCl] > 400 mM, the number of aggregates existing at 240 s does not decrease monotonically with the ionic strength as expected from DLVO theory.

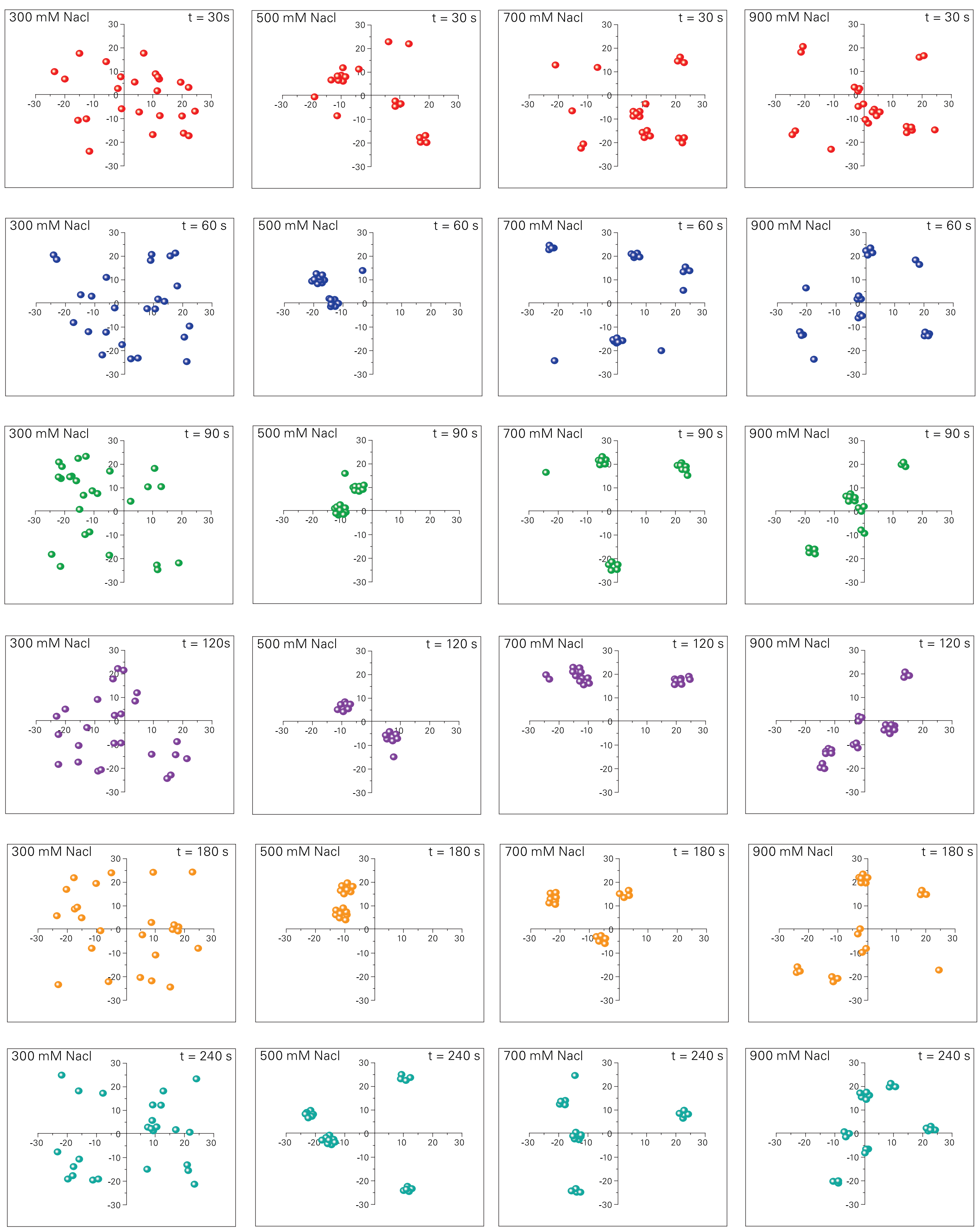

Fig. 26: Evolution of the h/w emulsion stabilized with 7.5 mM SDS for 300, 500, 700, and 900 mM NaCl. Type II potential.

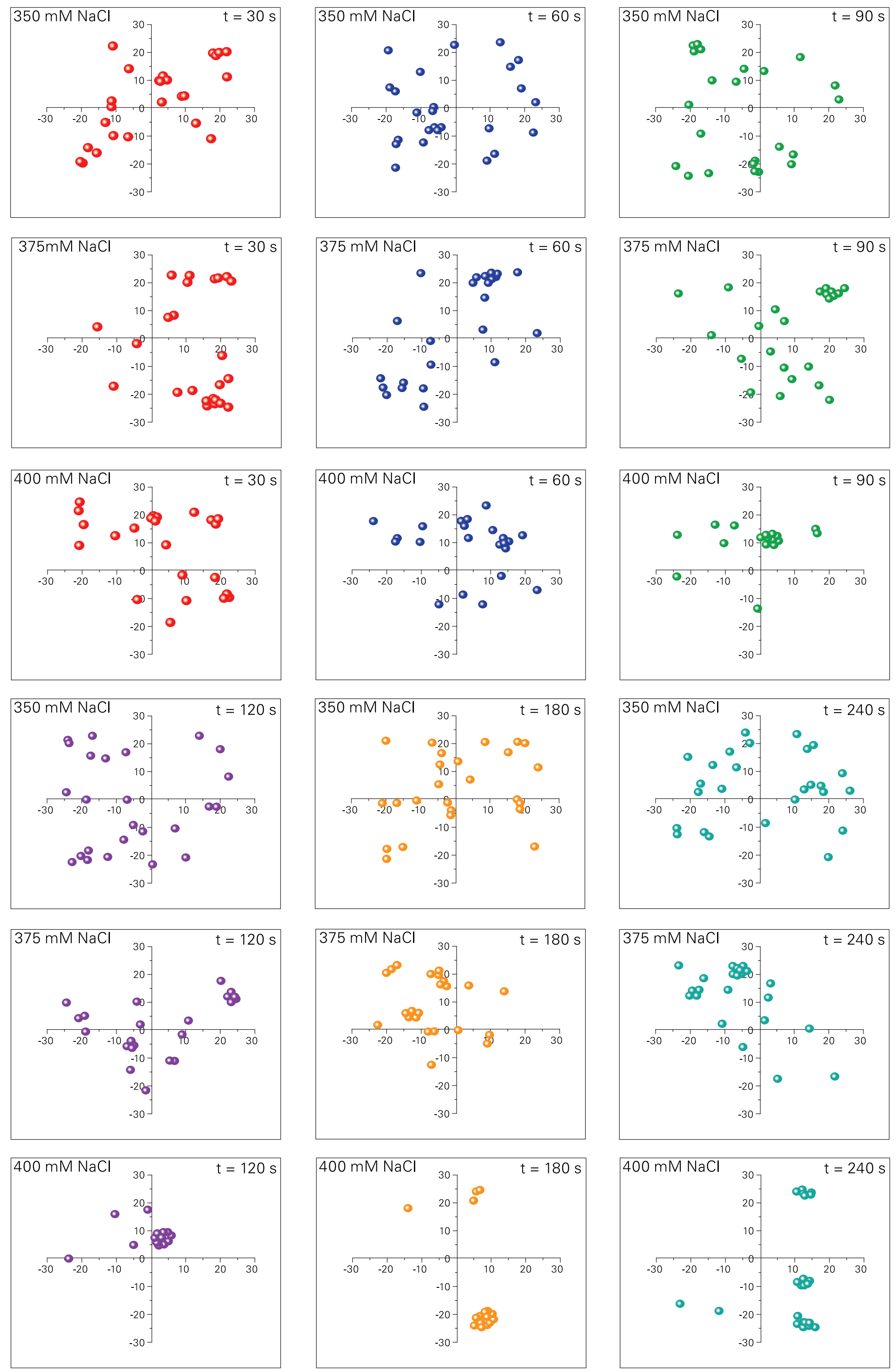

Fig. 27: Evolution of the h/w emulsion stabilized with 7.5 mM SDS for 350, 375, and 400 mM NaCl. Type II potential.

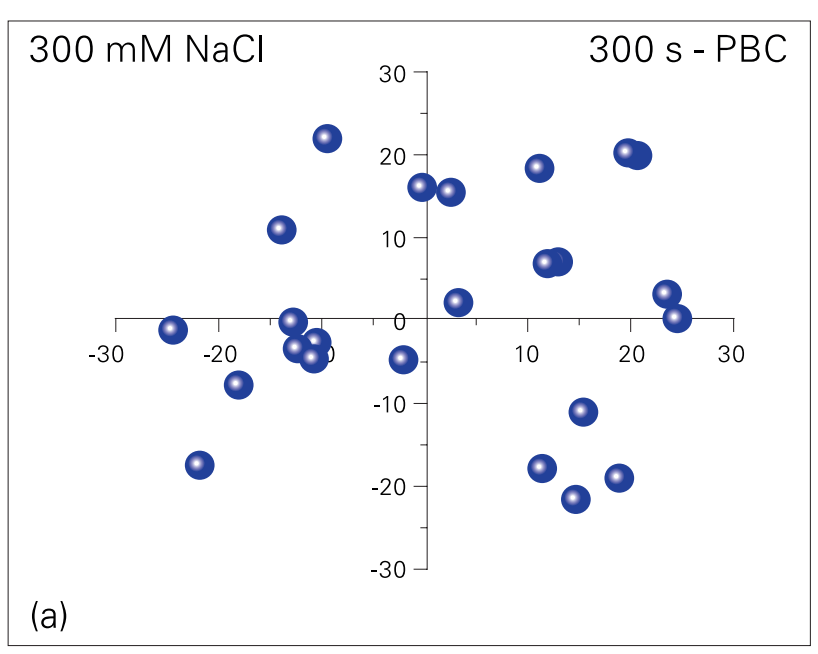

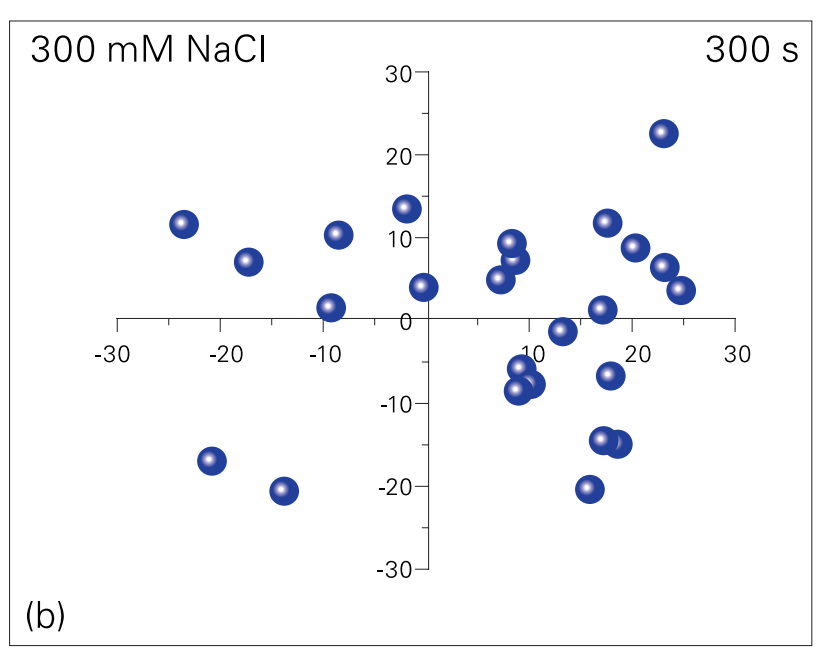

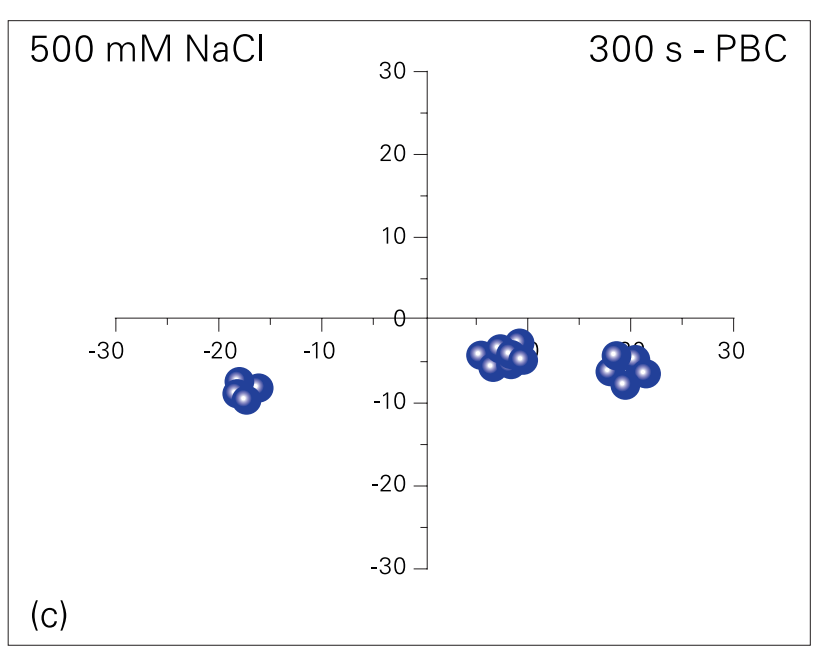

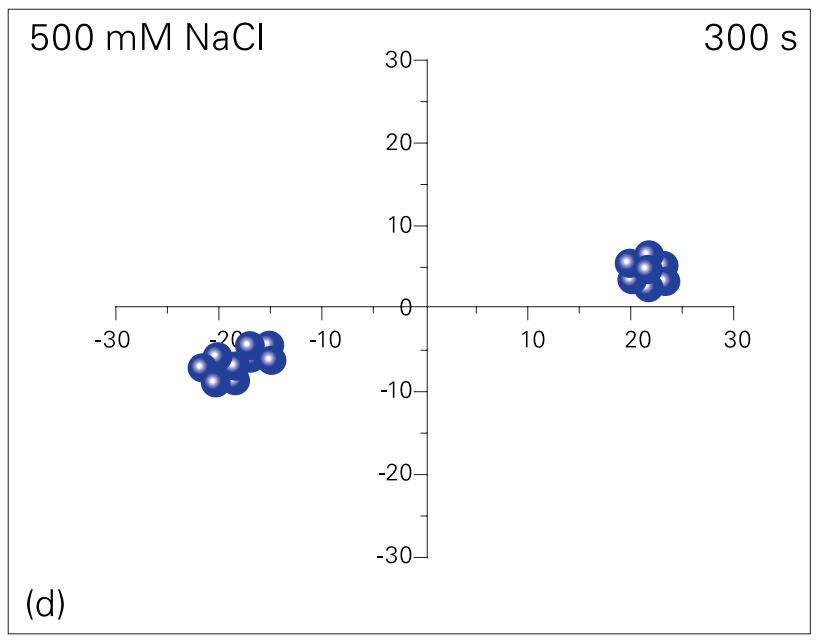

Fig. 28: Effect of gravity on the simulation results. Final state of the systems (t = 300 s) for [NaCl] = 300 and 500 mM, with and without PBC conditions.

While the rest of the figures show XY projections of the simulation cell, Figure 28 shows XZ projections of the final state corresponding to 300 and 500 mM NaCl in 7.5 mM SDS. These additional simulations were run in order to discard a possible decrease of the turbidity due to creaming. Notice that the simulation cell only spans a few nanometers. Hence, it represents a very small volume in the bulk of the solution. Since periodic boundary (PBC) conditions are used, aggregates that leave the cell from the top appear at the bottom and vice versa. However, in two of the simulations we introduced solid walls perpendicular to the z direction (this means reflective boundary conditions at the bottom and at the top of the simulation cell). Aggregates that hit the walls suffer an elastic collision and rebound from them. Therefore, if a sizeable effect of gravity on the movement of the aggregates exists, they should accumulate at the top of the cell. As expected, this did not happen neither for 300 or 500 mM NaCl (Fig. 16).

Figure 29 shows the evolution of the h/w system as a function of the salt concentration for 0.5 mM SDS. The behavior of the systems is very similar to the one observed for 7.5 mM SDS. As in the former case, the T-II potential induces reversible secondary minimum aggregation. Medium size aggregates (5 -7 particles) are already present after 30 s for [NaCl] ≥ 500 mM. They modify their size slightly up to 90 s, and remain stable during the rest of the simulation (t = 300 s). In most cases the number of aggregates is higher than the one corresponding to 7.5 mM SDS at the same salinity. However, the final state of the system is similar in both cases.

The calculations of $k_{FC}$ using the "doublet strategy" (Eq. (12)), did not generate results significantly different than the ones reported in Fig. 10 of Ref. [Urbina-Villalba, 2016], and therefore are not shown. The values of $k_{FC}^{fast}$ for spherical drops subject to van der Waals or DLVO forces at 1000 mM NaCl, were: 8.3 x $10^{-18}$ $m^3/s$ ($r^2$ = 0.9987) and 4.3 x $10^{-18}$ $m^3/s$ ($r^2$ = 0.9776). For deformable droplets with and without capillary oscillations, fast flocculation rates of: 8.5 x $10^{-18}$ $m^3/s$ ($r^2$ = 0.9989) and 7.2 x $10^{-18}$ $m^3/s$ ($r^2$ = 0.9976) were obtained, respectively. It is clear that long many-particle simulations are required in order to appraise the possible effect of solubilization on the aggregation rate.

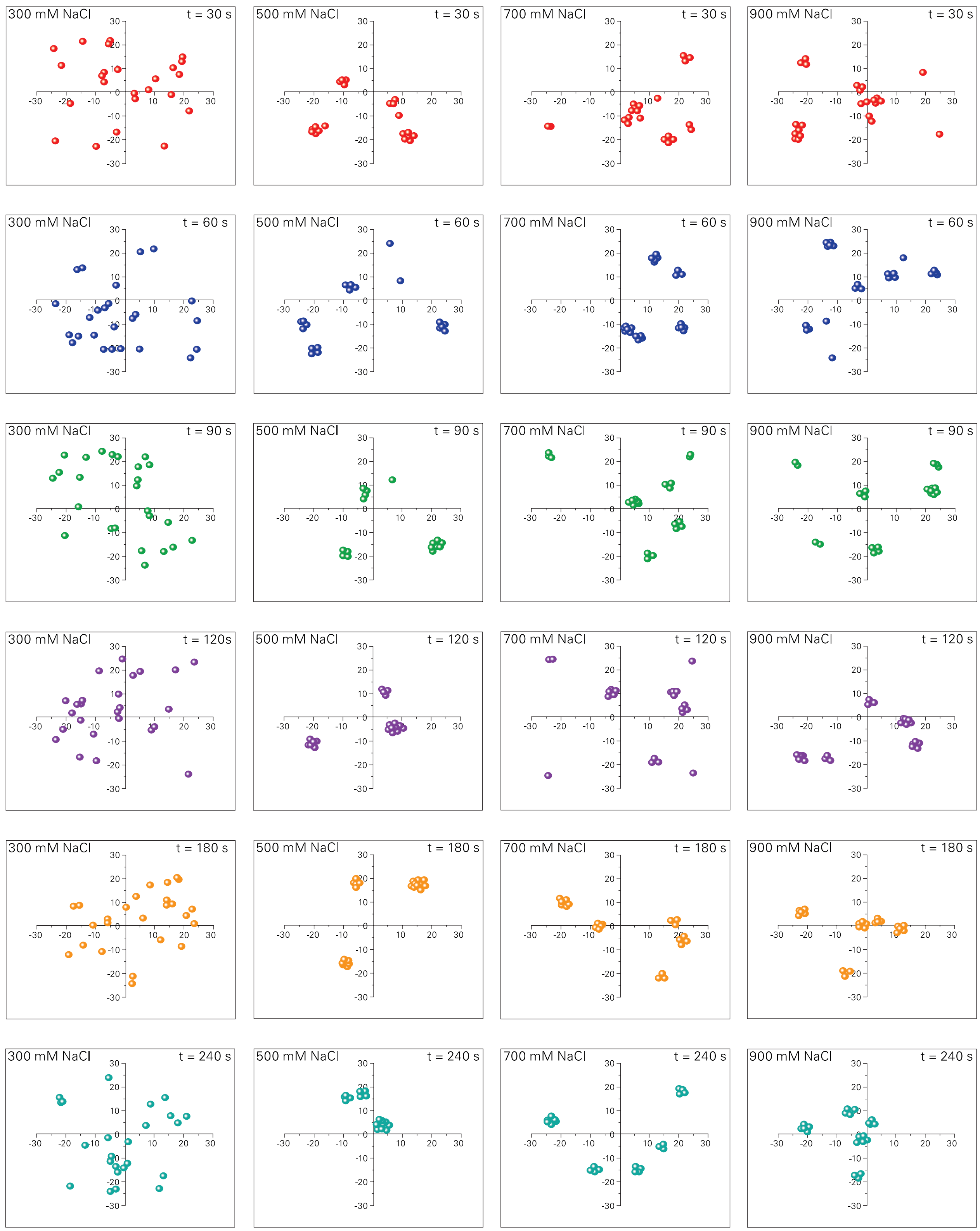

Fig. 29: Evolution of the h/w emulsion stabilized with 0.5 mM SDS for 300, 500, 700, and 900 mM NaCl. Type II potential.

### 6.3 Change of R vs. t

In a previous work [Rahn-Chique, 2015] the independent contributions of flocculation, coalescence and Ostwald ripening were simply added in order to predict the change of the average radius of a set of oil-in-water emulsions. In most cases, the contribution of Ostwald ripening was negligible. Yet the predictions of Eq. (7) were still significantly lower than the experimental findings. Equation (7) was evaluated assuming that average radius of extended aggregates can be approximated to the one of spherical drops: $R_{k,a} \approx R_{k,s}$ (Eq. (8)). This is probably the case if only globular aggregates are formed. As expected, the forecast improved if a distinct size dependence (k-dependence) was used for $R_{k,a}$ and $R_{k,s}$. Similarly, Eq. (9) was able to reproduce the experimental data only if the scattering coefficient $Q$ ($Q = Q_s \approx Q_a$) was used as an adjustable parameter. The reason being the same: $Q_a \neq Q_s$. Since the type of aggregates depends on the salt concentration, an apparent dependence of $Q$ on the physicochemical conditions appeared.

Unfortunately most instruments do not differentiate between aggregates and drops. Moreover, our current equipments do not allow measuring of the average radius with significant precision within the first 80 seconds after the perturbation of the emulsion. Hence, direct comparison with the absorbance data during this lapse is not possible. Only relatively long times can be accessed (Figures 30-34).

In the absence of salt (Fig. 30), the cube average radius of d/w emulsions increases linearly but in a non monotonic (oscillatory) fashion. According to previous studies [Urbina-Villalba, 2019; 2014], Ostwald ripening only augment the critical radius when drops are eliminated either by coalescence and/or complete dissolution. The slopes of the curves corresponding to 0.5 mM SDS (4.4 – 4.5 x $10^{-26}$ $m^3/s$) are at least four times larger than predicted by the theory of Lifshitz, Slyosov and Wagner [Lifshitz 1961, Wagner 1961] for surfactant-less emulsions of dodecane: 1.3 x $10^{-26}$ $m^3/s$ [Sakai, 2002; Urbina-Villalba, 2009; 2014]. According to LSW, the rate of ripening ($V_{OR}$) is equal:

$$V_{OR} = \frac{dR_c^3}{dt} = \frac{4}{9}\,\alpha D_m C_\infty \qquad (13)$$

where $R_c$, $D_m$, $C_\infty$ and $\alpha$ stand for the critical (number average) radius, the diffusion constant of the oil molecules in water, their solubility, and the capillary length of the drops, defined as:

$$\alpha = \frac{2\,\gamma\,V_m}{\tilde{R}\,T} \qquad (14)$$

where: $\tilde{R}$ is the universal gas constant, $\gamma$ the interfacial tension of the drops, and $V_m$ the molar volume of the oil. Using the interfacial tension of a dodecane drop in contact with an aqueous solution of 8 mM SDS (1.1 mN/m), an Ostwald ripening rate of 2.9 x $10^{-28}$ $m^3$ is obtained [Rahn-Chique, 2015]. Since the actual rates are at least two orders of magnitude higher, a substantial degree of flocculation and coalescence is expected. In this event (ripening or flocculation), the rates corresponding to 7.5 mM SDS emulsions should be lower than the ones of 0.5 mM SDS emulsions, as it was experimentally observed (1.5 – 1.6 x $10^{-26}$ $m^3/s$).

If the emulsions would solely flocculate, a short time limit for the rate of change of the cube average radius due to flocculation would be:

$$\left[\frac{dR^3}{dt}\right]_{t\to 0} = (3p)\,[k_F\, n_0\, R_0^3\,] = (d_f)^{-1}\left[\frac{9\,k_F\,\phi}{4\,\pi}\right] \qquad (15)$$

In this case, p = $(d_f)^{-1}$ where $d_f$ is the fractal dimension for either RLCA (~2.1) or DLCA (~1.7) aggregation [Urbina-Villalba, 2009]. Values of 2.7 and 2.6 were found for 0.5 mM SDS at T = 25 and 20 °C. Similar results were obtained for 7.5 mM SDS: 2.8 and 2.1, respectively. These values are close to the one obtained by Lozsan for mixed oil-in-water emulsions of hexadecane and dodecane destabilized with salt: 2.4 [Lozsan, 2012]. They are also similar to the one predicted theoretically for Apollonian spheres [Borkovec, 1994]. These results confirm kinetic-controlled floccula-tion as the most relevant process of destabilization in the mother emulsions.

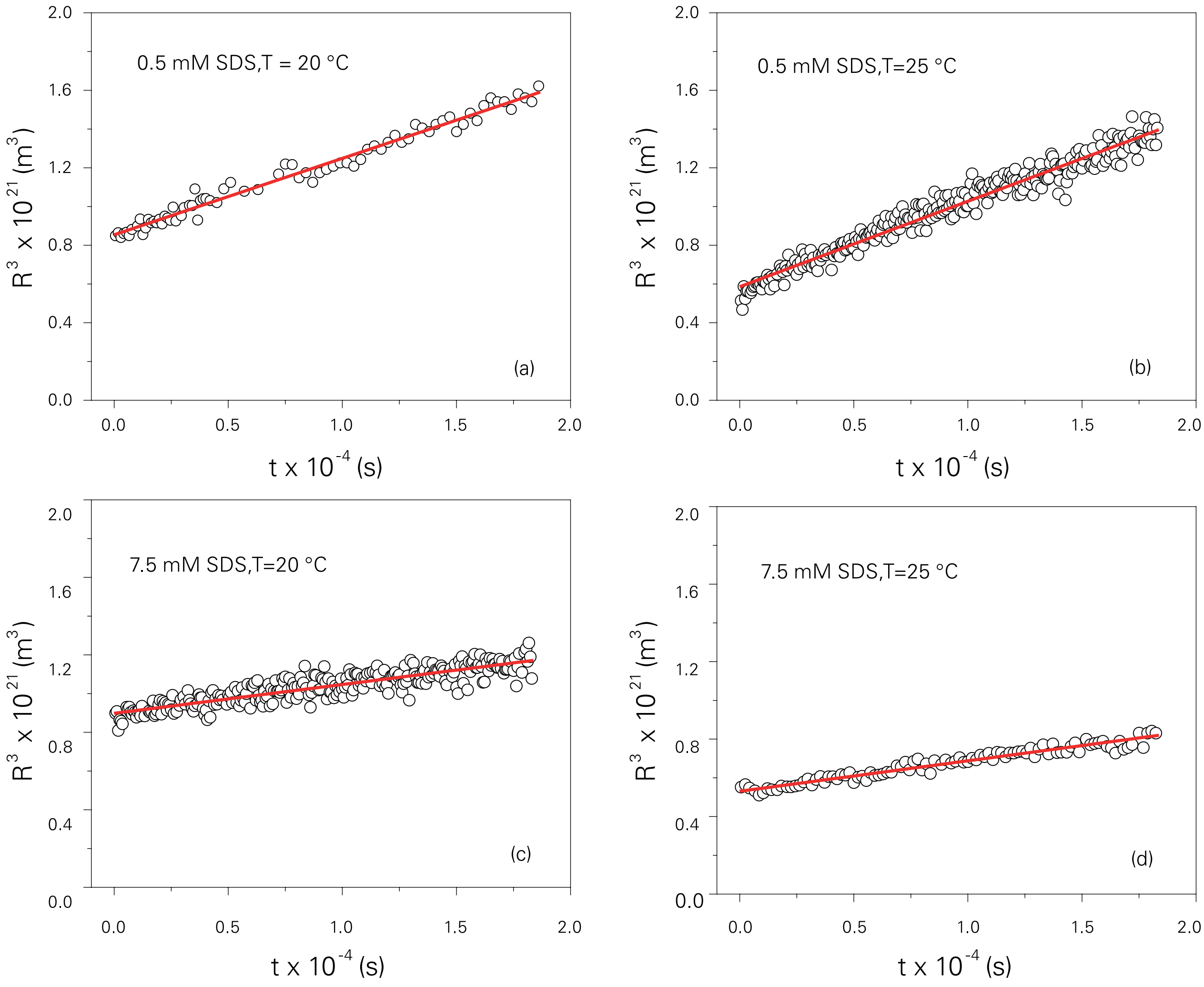

Fig. 30: Experimental variation of $R^3$ vs. t for dodecane in water emulsions in the absence of salt (T = 20 and 25 °C, 0.5 and 7.5 mM SDS).

Figs. 31-34 show the evolution of the average radius of emulsions after the addition of salt ([SDS] = 0.5, 7.5 mM, T = 20, 25 °C). Several curves show a concave downwards increase typical of flocculation. In particular, all 500 mM NaCl systems behave in this way. The radius increases monotonically in 600 – 700 nm during 500 seconds. Emulsions corresponding to 700 mM NaCl behave similarly, except at 20 °C and 7.5 mM SDS, where crystals appear. In this case, the measurements are erratic and not reproducible (significant differences appear in three subsequent measurements). The particle size of the scatterers reaches between 1 and 3 μm.

Emulsions are fairly stable in 300 mM NaCl except at 20 °C and 7.5 mM SDS, in which case the radius increases 100 nm in 500 s (Fig. 34). Curiously emulsions show flocculation at 350 mM NaCl except in the case of 0.5 mM SDS at 25 °C. Under these conditions the radius diminishes linearly between 100 and 200 s, and then stabilizes around 80 nm (Fig. 31). This behavior resembles the effect of solubilization (Fig. 15). However, the rate of "solubilization" calculated from the slope of the decrease ($dR/dt$ = -6.46 x $10^{-10}$ $m^3/s$) is higher than the one reported by Ariyaprakai [2008]: 4.49 x $10^{-11}$ m/s.

If the drops are sufficiently smaller than the wavelength of light (400 nm) it is possible to connect the values of the absorbance to the average radius of the particles assuming Rayleigh scattering. This procedure was used to obtain approximate values of the radius during the first 300 s after the addition of salt. By definition:

$$Abs \;=\; Log \frac{I_0}{I_0 - I_s} \tag{16}$$

where $I_0$ is the intensity of light reaching the sample, and $I_s$, the intensity of light scattered by the particles. According to Rayleigh:

$$I_S \;=\; I_0 \left[ N \left( \frac{1 + Cos^2\,\theta}{2L^2} \right) \left( \frac{2\pi}{\lambda} \right)^4 \left( \frac{n^2 - 1}{n^2 + 2} \right) \right] (R)^6 \;=\; I_0 C\, R^6 \tag{17}$$

where $N$ the number of particles per unit volume, $\theta$ the scattering angle, L the length from the particle to the detector, $\lambda$ the wavelength of incident light, $n$ the relative refractive index between the particles and the aqueous solution. Hence:

$$Abs \;=\; -Log[1 - CR^6] \tag{18}$$

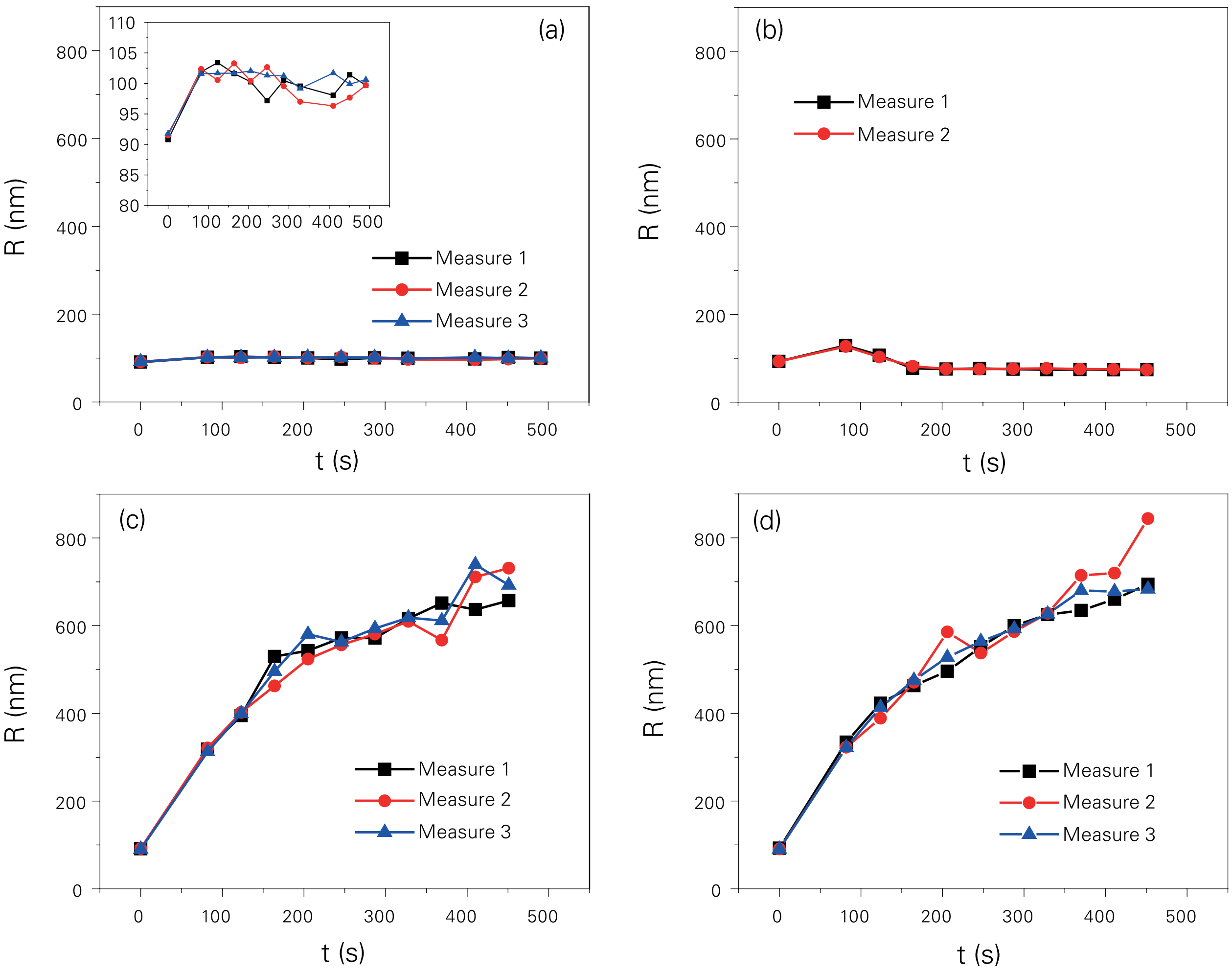

Fig. 31: Experimental variation of $R^3$ vs. t for dodecane in water emulsions in 300, 350, 500 and 700 mM NaCl (T = 25 °C, 0.5 mM SDS).

Constant "C" can be evaluated if the initial radius of the dispersion is known:

$$C \;=\; (1 \;-10^{-Abs(R_0)})/R_0^6 \tag{19}$$

Finally:

$$R(t) \;=\; \left[ \frac{1 - 10^{-Abs}}{C} \right]^{1/6} \tag{20}$$

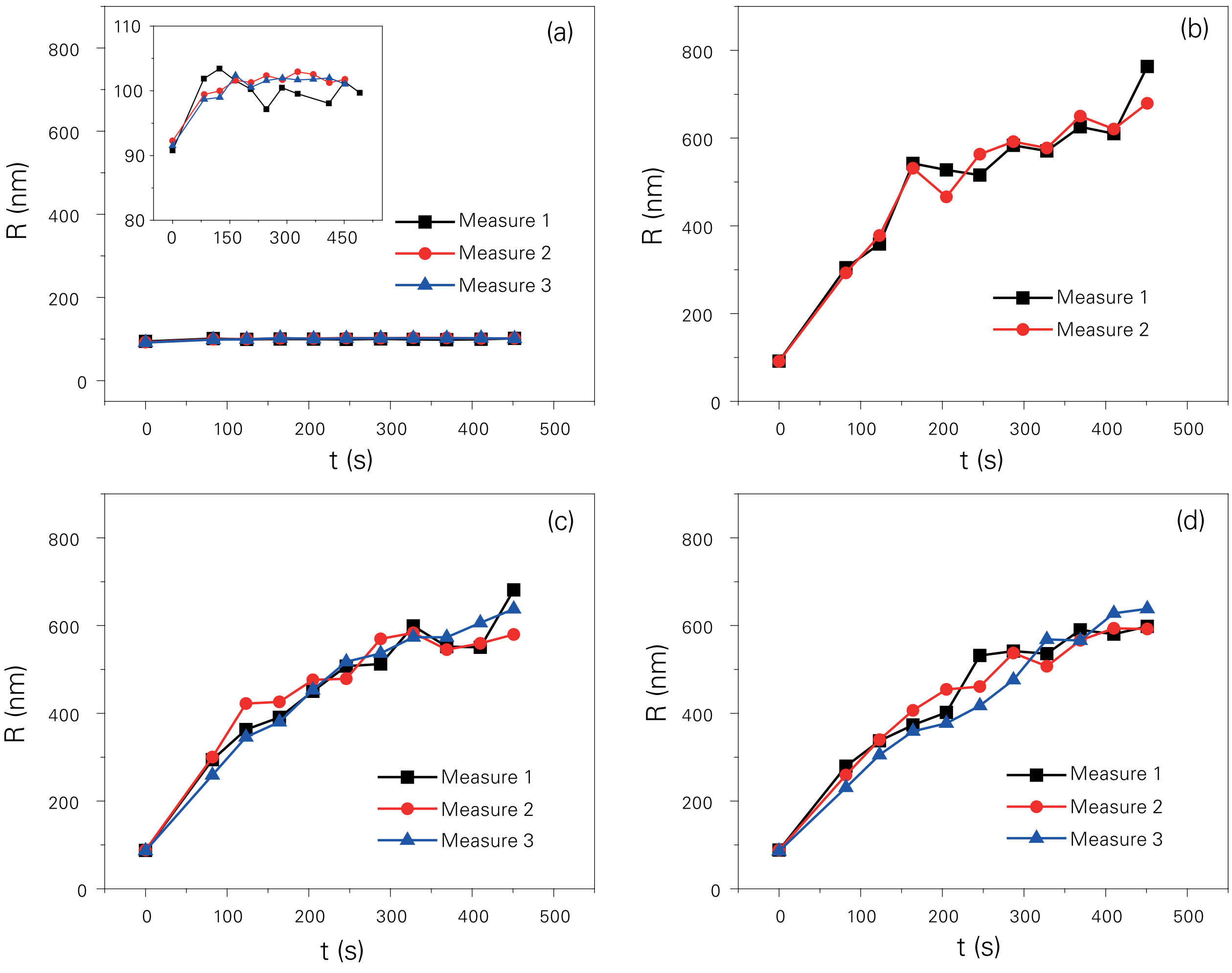

Fig. 32: Experimental variation of $R^3$ vs. t for dodecane in water emulsions in 300, 350, 500 and 700 mM NaCl (T = 20 °C, 0.5 mM SDS).

As shown in Figure 35 the radii deduced from the absorbance and its variation in time ($dR/dt$) are lower than those reported by the Goniometer BI-200SM. Constant C was evaluated assuming $R_0$ = 82 nm for T = 25 °C emulsions. This assumption already gives rise to 20-nm differences for the emulsions preserved at 20 °C. The reason is that the initial radius measured by the Goniometer is close to the theoretical one ($R_0$) in the former emulsions, but it differs considerably for 0.5 and 7.5 mM SDS dispersions at T = 20 °C: 94.7 and 96.5 nm, respectively, suggesting the higher instability of these systems. In any event, the predictions of the average radius obtained from Eq. (20) are reasonably close to the measurements. Moreover, the estimates of Eq. (8) are also satisfactory. This corroborates the results of the fractal coefficients, suggesting that flocculation and coalescence are the main destabilization processes in the mother emulsions.

When the procedure outlined above is applied to a system in which the absorbance monotonically decreases (7.5 mM SDS, 25 °C, 300 mM NaCl), the initial radius also decreases from 82 to 77 nm in 300 seconds; presumably as a consequence of solubilization (Figure 36a). The terminal part of the curve (above 27 s) shows an approximate negative slope of 1.2 x $10^{-11}$ m/s ($r^2$ = 0.9957), the same order of magnitude reported by Ariyaprakai [2008]: 4.49 x $10^{-11}$ m/s. This suggests the occurrence of micelle solubilization.

Fig. 36b shows the results corresponding to one of the emulsions with 0.5 mM SDS, 25 °C, at 350 mM NaCl. The radius rises from 82 to 87 nm, and the progressively diminishes until reaching 80 nm. Using (rough) linear fittings to approximate the variation of ($dR/dt$) before and after the maximum yields: 4.9 x $10^{-10}$ m/s ($r^2$ = 0.9708) and -6.2 x $10^{-11}$ m/s ($r^2$ = 0.9788), respectively. This decrease is similar and of the same order of magnitude than the one reported by Ariyaprakai. A terminal slope of 2.2 x $10^{-12}$ m/s ($r^2$ = 0.9002) was also observed.

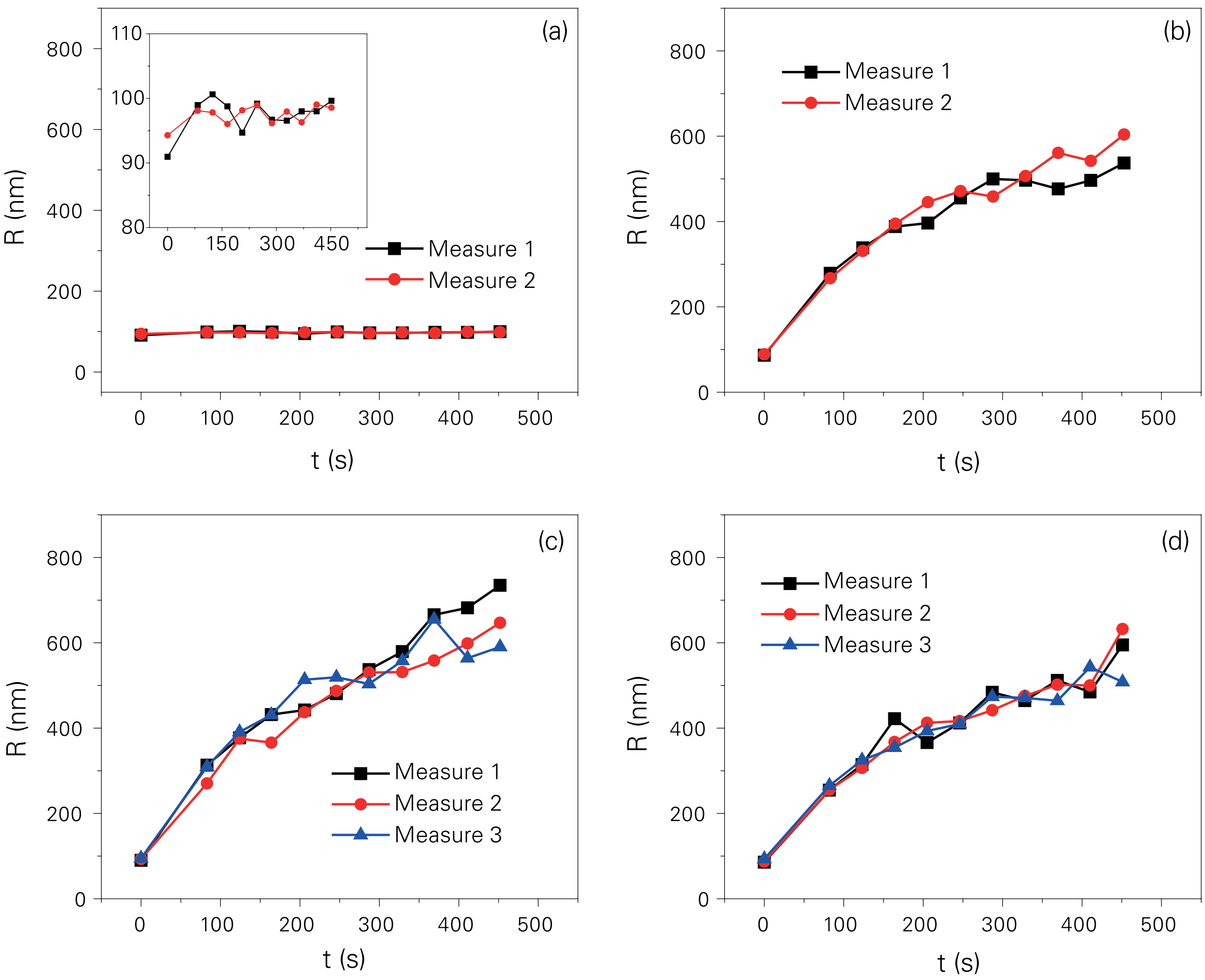

Fig. 33: Experimental variation of $R^3$ vs. t for dodecane in water emulsions in 300, 350, 500 and 700 mM NaCl (T = 25 °C, 7.5 mM SDS).

## 7. CONCLUSIONS

The addition of salt to a dodecane-in-water emulsion lowers the CMC of the surfactant considerably, promoting a substantial solubilization of the drops. For 7.5 mM SDS and below 500 mM NaCl, solubilization prevails over flocculation at all times. In the presence of salt, the maximum volume fraction of oil that can be solubilized ($\phi_c$) is significantly larger than the one expected from the dependence of $\phi_c$ on the total surfactant concentration in the absence of electrolytes. However, the rate of solubilization is much slower. Above 500 mM, a significant degree of aggregation occurs during a transient period, and only partial solubilization is attained afterwards.

The behavior of the absorbance with respect to the salt concentration depends on the competition between flocculation, crystal formation and solubilization. In 0.5 mM SDS, systems with 300 mM NaCl are stable at 20 and 25 degrees (the absorbance remains constant as a function of time). But in 7.5 mM SDS, a significant solubilization occurs at both temperatures.

Higher salt concentrations promote an initial increase of the absorbance due to flocculation and possibly crystal formation. Macroscopic crystals are only observed at times substantially longer than 60 seconds. Therefore, their possible influence during the initial increase of the absorbance is more a hypothesis rather than an experimental fact. Following the initial increase, it is observed that the absorbance either remains constant if the Krafft temperature is attained, or otherwise decreases due to the solubilization of the drops. This is consistent with the fact that micelles and crystals do not coexist. Crystals are observed at T = 20 °C for [NaCl] at 500 mM, and at T = 25 °C for [NaCl] = 700 mM.

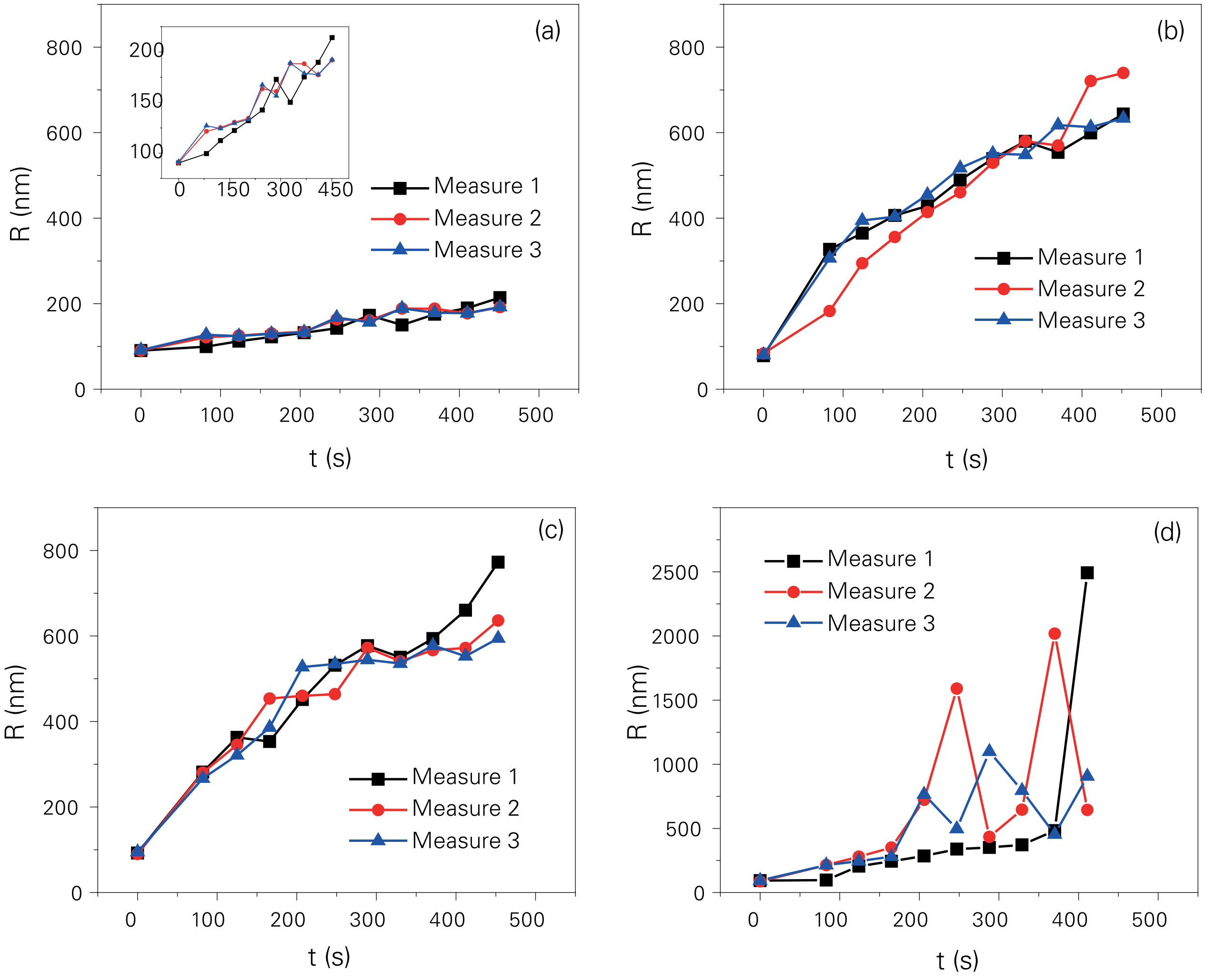

Fig. 34: Experimental variation of $R^3$ vs. t for dodecane in water emulsions in 300, 350, 500 and 700 mM NaCl (T = 20 °C, 7.5 mM SDS).

Between 300 and 500 mM NaCl, a maximum of absorbance is observed. At T = 25 °C, this occurs around 350 mM NaCl for both 0.5 and 7.5 mM SDS. This behavior is likely to result from the competition between flocculation and solubilization. However, the mechanism of solubilization in the case of 0.5 mM SDS is unclear. The influence of gravity and/or Ostwald ripening was discarded based on the variation of the turbidity as a function of height, and the behavior of emulsions containing squalene. According to 25-particle simulations a maximum might be caused by the coalescence of the drops. As the size of the drops grows the scattering of light initially increases, but the number of particles per unit volume diminishes promoting a subsequent decrease of the absorbance.

Since the formation of SDS crystals is slower than the aggregation of the drops, the use of the absorbance for the appraisal of the flocculation rate under this condition induces slower rates. Hence, the values of $k_{FC}$ corresponding to 500 and 700 mM SDS might be lower for 7.5 than for 0.5 mM SDS (Table 4). However, the formation of crystals is only observed for t >> 60 s. We do not have the means to study the nucleation of the crystals during short periods of time.

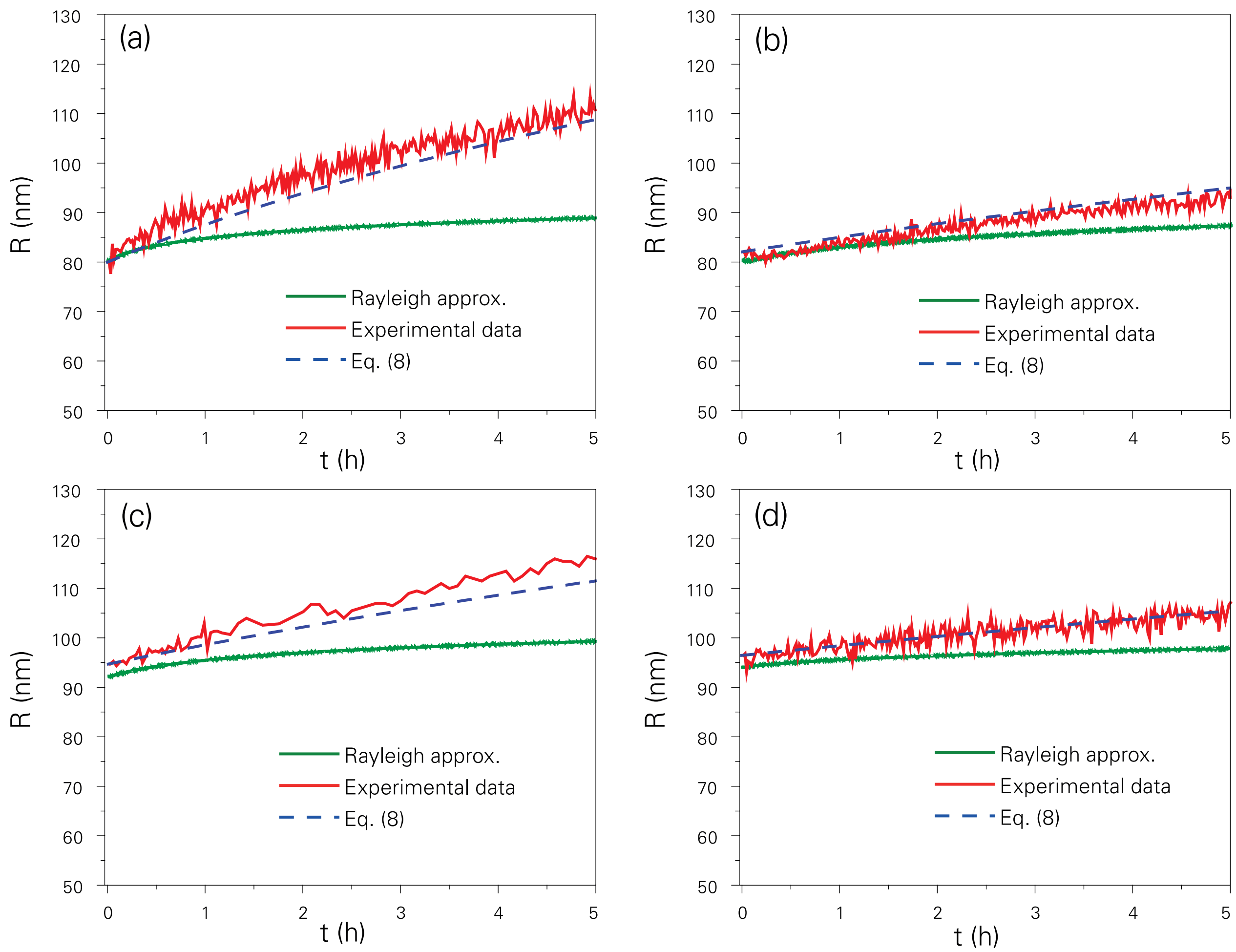

Fig. 35: Emulsions without salt. Comparison between the average radius evaluated from a Brookhaven BI-200SM Goniometer, and the values deduced from the Absorbance of the dispersions assuming Rayleigh scattering. The predictions of Eq. 8 are also shown. a) 0.5 mM SDS T = 25 °C; (b) 7.5 mM SDS T = 25 °C; c) 0.5 mM SDS T = 20 °C, (d) 7.5 mM SDS T = 20 °C.

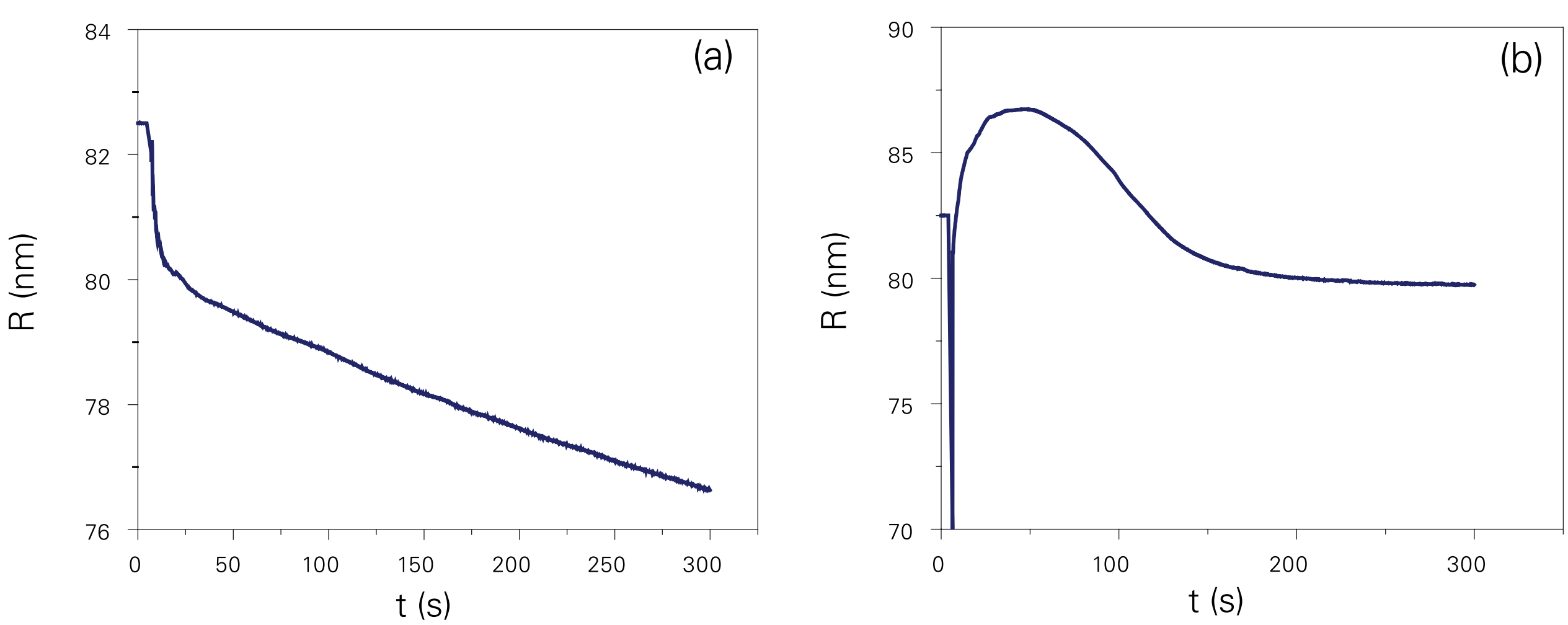

Fig. 36: Variation of the average radius deduced from the Absorbance of the dispersions assuming Rayleigh scattering. (a) 300 mM NaCl, 7.5 mM SDS, T = 25 °C; (b) 350 mM NaCl, 0.5 mM SDS, T = 20 °C.

Whenever solubilization occurs, the use of doublets for the estimation of the stability ratio is questionable: the simulation time is insufficient to appraise the effects of solubilization. Still, the qualitative behavior of the emulsions can be reproduced using the interaction potentials of Fig. 4 (T-I parameterization), but substantially faster aggregation rates are predicted. Nevertheless, Type II potentials (Fig. 5) reproduce the average value of the flocculation rates, but the qualitative behavior forecasted is inconsistent with the observations.

In the case of h/w emulsions, 25-particle simulations nicely illustrate the effect of the secondary minimum on the dynamics of the flocculation process. The reversibility of the aggregation process leads to a lower aggregation rate due to a slower decrease of the total number of aggregates per unit volume as a function of time. The influence of solubilization is negligible because the molecular solubility of hexadecane is very low. The estimation of W using doublets, reproduces the results of former simulations. Hence, the h/w simulations do not provide an explanation to the terminal decrease of the curve $k_{FC}$ vs. [NaCl] experimentally observed for [NaCl] > 500 mM at 7.5 mM SDS.

In the case of nanoemulsions, the use of the formulae for Rayleigh scattering provides a mean to relate the values of the absorbance with an average radius. Hence, the same absorbance data might be used to determine the average radius and the aggregation rate. This approximate procedure is useful to understand the qualitative behavior of the absorbance in terms of the size of the particles.